\documentclass[pra,twocolumn,preprintnumbers,amsmath,amssymb,showpacs,
nofootinbib,floatfix]{revtex4}

\usepackage{graphicx,bm}

\makeatletter
\def\graphicscale{\twocolumn@sw{0.33}{0.4}}
\def\graphicthreescale{\twocolumn@sw{0.33}{0.4}}
\makeatother

\begin{document}

\title{Equilibrium and off-equilibrium trap-size scaling in 1D ultracold
  bosonic gases}

 \author{Massimo Campostrini and Ettore Vicari} \affiliation{Dipartimento di
   Fisica dell'Universit\`a di Pisa and I.N.F.N., Sezione di Pisa, Largo Bruno
   Pontecorvo 2, I-56127 Pisa, Italy} 
\date{October 9, 2010}

\begin{abstract}
  
  We study some aspects of equilibrium and off equilibrium quantum dynamics of
  dilute bosonic gases in the presence of a trapping potential.  We consider
  systems with a fixed number of particles and study their scaling behavior
  with increasing the trap size.
  
  We focus on one-dimensional (1D) bosonic systems, such as gases described by
  the Lieb-Liniger model and its Tonks-Girardeau limit of impenetrable bosons,
  and gases constrained in optical lattices as described by the Bose-Hubbard
  model.  We study their quantum (zero-temperature) behavior at equilibrium
  and off equilibrium during the unitary time evolution arising from changes
  of the trapping potential, which may be instantaneous or described by a
  power-law time dependence, starting from the equilibrium ground state for an
  initial trap size.
  
  Renormalization-group scaling arguments, analytical and numerical
  calculations show that the trap-size dependence of the equilibrium and
  off-equilibrium dynamics can be cast in the form of a trap-size scaling in
  the low-density regime, characterized by universal power laws of the trap
  size, in dilute gases with repulsive contact interactions and lattice
  systems described by the Bose-Hubbard model.  The scaling functions
  corresponding to several physically interesting observables are computed.

  Our results are of experimental relevance for systems of cold atomic gases
  trapped by tunable confining potentials.

\end{abstract}



\maketitle

\section{Introduction}
\label{intro}

The dynamics of strongly correlated quantum systems is a fundamental physical
issue which has attracted much theoretical interest.  The achievement of
Bose-Einstein condensation in dilute atomic vapors~\cite{CWK-02} and the
impressive progress in the experimental manipulation of cold atoms in optical
lattices, see, e.g., Ref.~\cite{BDZ-08} and references therein, have provided
a great opportunity to investigate the interplay between quantum and
statistical behaviors in particle systems, and issues related to the unitary
quantum evolution of closed many-body systems, exploiting their low
dissipation rate which allows to maintain phase coherence for a long time.
Transitions between different quantum phases have been experimentally
observed, such as those related to the formation of a Bose-Einstein condensate
in interacting Bose gases~\cite{HSBBD-06,DRBOKS-07,CRRHP-09} and quantum
Mott-insulator to superfluid transitions in atomic systems constrained in
optical lattices, see, e.g.,
Refs.~\cite{GBMHS-02,KWW-05,FWMGB-06,SPP-07,CFFFI-09,MCA-10}.  Accurate
experimental studies of nonequilibrium properties of quantum many-body systems
of ultracold atoms have also been reported, see, e.g.,
Refs.~\cite{Greiner-etal-02,KWW-06,Sadler-etal-06}.

An important feature of these experiments is the presence of a confining
potential which traps the particles within a limited spatial region.  The
capability of varying the confining potential, which may also depend on the
spatial directions, allows to vary the effective spatial geometry of the
particle systems, including quasi-1D geometries, see, e.g.,
Refs.~\cite{KWW-06,KWW-04,SMSKE-04,PWMMFCSHB-04,THHPRP-04,HLFSS-07}.

In the presence of an optical lattice with lattice spacing $a$, created by
laser-induced standing waves which constrains the particle to stay at the
sites of a lattice, the theoretical framework~\cite{JBCGZ-98} is provided by
the Bose-Hubbard (BH) model~\cite{FWGF-89} with a confining
potential $V(r)$ coupled to the particle density, defined by the
Hamiltonian~\footnote{The BH Hamiltonian for $N$ particles is usually written
  with the kinetic term ${\cal H}_{\rm kin} = - (J/2) \sum_{\langle ij\rangle}
  (b_j^\dagger b_i + b_i^\dagger b_j)$. The difference from Eq.~(\ref{bhmN})
  is a $N$-dependent constant.  }
\begin{eqnarray}
{\cal H}_{\rm BH} &=& {J\over 2}
\sum_{\langle ij\rangle} (b_j-b_i)^\dagger (b_j-b_i)
\label{bhmN}\\
&&+ {U\over2} \sum_i n_i(n_i-1) + \sum_i V(r_i) n_i ,
\nonumber
\end{eqnarray}
where $\langle ij\rangle$ is the set of nearest-neighbor sites, $b_i$ are
bosonic operators, $n_i\equiv b_i^\dagger b_i$ is the particle density
operator, and $N = \langle \sum_i n_i \rangle$ is the particle number.  We
consider a power-law spatial dependence for the trapping potential,
\begin{equation}
V(r) = {1\over p} v^p r^p, 
\label{potential}
\end{equation}
where $r\equiv |\vec{x}|$ is the distance from the center of the trap, $v$ is
a positive constant and $p$ an even integer number.  Experiments are usually
set up with a harmonic potential, i.e., $p=2$. Examples of experimental traps
described by quartic potentials are reported in Ref.~\cite{BSSD-04}.  The
hard-core (HC) limit $U\to\infty$ of the BH model implies that the particle
number $n_i$ per site is restricted to the values $n_i=0,1$.  In one dimension
the HC limit can be exactly mapped into a lattice model of spinless fermions,
see, e.g., Ref.~\cite{Sachdev-book}.

We consider
\begin{equation}
l \equiv {J^{1/p}\over v}
\label{trapsizedef}
\end{equation}
as the size of the trap within the confined BH model~\cite{RM-04,CV-10}.  In
the case of harmonic traps, $l\sim \omega^{-1}$ where $\omega$ is the trap
frequency.  The definition (\ref{trapsizedef}) of trap size naturally arises
when we consider the {\em thermodynamic} limit, which is generally defined 
as $N,l\to\infty$ keeping $N/l^d$ fixed~\cite{BDZ-08,PGS-04}, and it is
equivalent to introducing a chemical potential $\mu$, adding the
term $\mu\sum_i n_i$ to the Hamiltonian (\ref{bhmN}).

In the absence of a lattice structure, the basic model to describe the
many-body features of a boson gas confined to an effective 1D geometry is the
Lieb-Liniger (LL) model with an effective two-particle repulsive
contact interaction~\cite{LL-63},
\begin{equation}
{\cal H}_{\rm LL} = \sum_{i=1}^N \left[ {p_i^2\over 2m} + V(x_i)\right] +
g \sum_{i\ne j} \delta(x_i-x_j)
\label{LLHam}
\end{equation}
where $N$ is the number of particles and $V(x)$ is the confining potential.
The limit of infinitely strong repulsive interactions corresponds to a 1D gas
of impenetrable bosons~\cite{Girardeau-60,Girardeau-65}, the Tonks-Girardeau
(TG) gas.  1D Bose gases with repulsive two-particle short-ranged interactions
become more and more nonideal with decreasing the particle density, acquiring
fermion-like properties, so that the 1D gas of impenetrable bosons is expected
to provide an effective description of the low-density regime~\cite{PSW-00}.
1D Bose gases of cold atoms have been realized 
experimentally~\cite{KWW-04,PWMMFCSHB-04,THHPRP-04,KWW-06,HLFSS-07}.

In this paper we address issues related to the equilibrium and off equilibrium
quantum dynamics of dilute bosonic gases in the presence of a confining
potential.  We consider 1D trapped gases constituted by $N$ bosonic particles,
at the equilibrium and off equilibrium during the unitary time evolution
arising from changes of the trapping potential. In the latter case we consider
a system of $N$ particles in a trap of size $l_0$, which is prepared in its
ground state at $t=0$, the trapping potential is then varied as
\begin{eqnarray}
V(r,t) = {1\over p} u^p K(t/t_q) r^p,\label{tidpo}
\end{eqnarray}
with $K(0)=1$ and the parameter $t_q$ providing the time rate of the variation
of the trapping. Since the particle number $\hat{N}\equiv \sum_x n_x$ commutes
with the time-dependent Hamiltonian even when the trapping potential depends
on the time, the particle number $N$ remains unchanged during the dynamical
process.  Istantaneous changes from the initial trap size $l_0\equiv 1/u$ to a
final trap size $l_f$, or a complete drop of the trap (corresponding to
$l_f\to\infty$), give rise to interesting cases of off-equilibrium evolutions.
Moreover, we also consider a power-law time dependence of the confining
potential such as
\begin{eqnarray}
K(t/t_q) = \tau^q,\quad \tau\equiv 1+t/t_q .
\label{ktaudef}
\end{eqnarray}
Adiabatic time evolutions apply when the change of the external potential is
very slow, thus it requires a large time rate $|t_q|$.  

Several theoretical studies have already been dedicated to issues related to
the quantum behavior of trapped bosonic gases, in the continuum and on the
lattice (i.e., the BH model), in equilibrium conditions and off equilibrium
due to time-variations of the trapping potential, see, e.g.,
Refs.~\cite{BDZ-08,PGS-04,KSS-96,GW-00,KNSX-00,DLO-01,
GWT-01,LGW-02,BRSRMDT-02,
MS-02,PSG-02,FFGW-02,KB-02,Papenbrock-03,FFGW-03,GS-03,RM-04,
Gangardt-04,KSDZ-04,PRHD-04,WATB-04,MG-05,GFF-05,CDEO-08,PG-08,
BKMRS-08,RBRS-09,CV-10,CV-10-2,CK-10,Roux-10,PPS-10}.

In this paper we further investigate these issues within the framework of the
trap-size scaling (TSS) theory~\cite{CV-10,CV-09}.  We consider systems with a
fixed number of particles $N$ and study their scaling behavior with increasing
the trap size $l$.  We study the asymptotic trap-size dependence in the
low-density regime which is characterized by universal power laws, as we shall
see.

Using renormalization-group (RG) scaling arguments, we derive the power-law
behaviors which describe the asymptotic TSS in the low-density regime at
equilibrium. They are determined by the continuum {\em nonrelativistic}
bosonic $\Phi^4$ theory, which is the same continuum theory describing the
quantum critical behavior at Mott transitions driven by the chemical
potential~\cite{FWGF-89}.  This implies that in one and two dimensions the
power-law TSS is characterized by the dynamic exponent $z=2$, the RG
dimensions $y_b=d/2$ and $y_n=d$ of the bosonic and particle density operators
respectively~\cite{FWGF-89,Sachdev-book}, and the trap exponent
$\theta=p/(p+2)$~\cite{CV-10}.  This general TSS scenario is supported by
analytical and numerical calculations within 1D systems.  We show that the TSS
is universal for diluted gases with repulsive contact interactions such as the
LL model (\ref{LLHam}) and lattice systems described by the BH model
(\ref{bhmN}).

We consider the off-equilibrium behavior arising from time variations of the
confining potentials, which may be instantaneous or described by a power-law
time dependence, such as Eq.~(\ref{ktaudef}), starting from the equilibrium
ground state for an initial trap size $l_0$.  We put forward scaling Ansatz
for the asymptotic TSS with respect to the initial trap size $l_0$ in the
large-$l_0$ limit. Then we study the trap-size dependence of the
off-equilibrium dynamics of 1D bosonic gases of $N$ particles, by assuming an
adabiatic evolution in the case of slow changes of the Hamiltonian parameters,
and by analyzing the solutions of the Sch\"rodinger equation of $N$
impenetrable bosons in the presence of time-dependent harmonic potentials.

Our results are of experimental relevance for systems of cold atomic gases
trapped by tunable confining potentials. Indeed, the long characteristic time
scales of these systems may allow a scaling study of the trap-size dependence
of the zero-temperature properties of $N$-particle boson gases in the
low-density regime, at equilibrium and off equilibrium during the time
evolution of the confining potential.

The paper is organized as follows.  In Sec.~\ref{TSSeq} we derive the
asymptotic TSS of bosonic systems of $N$ particles at equilibrium in
the large trap-size limit, using general scaling arguments.  In
Sec.~\ref{static} we focus on the 1D BH model, we present analytical
and numerical calculations in the HC limit and at finite $U$, which
support the RG scaling predictions, and provide the universal scaling
functions at equilibrium.  In Sec.~\ref{bosgas} we consider 1D bosonic
gases at low density, and, in particular, in the TG limit which
describes the low-density behavior of the LL model: we show that the
trap-size dependence of $N$-particle systems is identical to the
asymptotic TSS of the 1D BH model with the same number of
particles. In Sec.~\ref{dynTSS} we turn to off-equilibrium dynamics,
considering time-dependent confining potentials or instantaneous
changes of its parameters; using scaling arguments we extend the
equilibrium TSS Ansatz to the case of the off-equilibrium evolutions,
distinguishing the cases of instantaneous variations and power-law
time dependences of the confining potential.  In Sec.~\ref{slowdyn} we
discuss the case of slow time variations of the trapping potential,
assuming an adiabatic or quasi-adiabatic approximation.
Sec.~\ref{bosgasdyn} is dedicated to the study of the quantum unitary
evolution of 1D impenetrable bosons in time-dependent harmonic traps,
where the off-equilibrium trap-size dependence can be analytically
determined, in particular, for a linear time-dependence of the
confining potential and for instantaneous quenches.  Finally, in
Sec.~\ref{conclusions} we summarize our main results and draw our
conclusions.  In App.~\ref{tdhp} we discuss the case of particle systems
confined by a spatial dependence of the hopping
parameter.  In App.~\ref{lnbeh} we study the asymptotic behavior of
the TSS functions for a large number of particles, showing that they
have a nontrivial large-$N$ power-law scaling.  App.~\ref{oscill}
presents a detailed analysis of the off-equilibrium dynamics of a
quantum oscillator with a time-dependent frequency.

\section{Trap-size scaling of $N$ particles at equilibrium}
\label{TSSeq}

Before discussing the TSS of $N$ particles in the large-$l$ limit, let
us note that the large-$l$ limit keeping $N$ fixed differs from that
performed at fixed chemical potential $\mu$, i.e., considering the BH
Hamiltonian
\begin{eqnarray}
{\cal H}_{\rm \mu} =  {\cal H}_{\rm BH}  + (\mu-J) \sum_i n_i .
\label{bhm}
\end{eqnarray}
Indeed, the large trap-size limit, keeping $\mu$ fixed, implies an
increase of the particle number so that
\begin{equation}
N/l^d = \tilde{\rho}(\mu/J)
\label{rhot}
\end{equation}
asymptotically, where $d$ is the spatial dimension and
$\tilde{\rho}(\mu)$ is a finite function of $\mu$.  This {\em
thermodynamic} limit is usually considered when quantum transitions
are studied in confined particle systems, see, e.g.,
Ref.~\cite{CV-10-2}.

We are interested in the low-density regime, which is related to the
limit $\mu\to\mu_c$ where $\tilde{\rho}(\mu_c)=0$, which corresponds
to a low-density to empty-state transition, which may be considered as
a $n=0$ Mott transition.  In the homogeneous BH model without trap,
the low-energy properties at Mott transitions driven by the chemical
potential $\mu$ are described by a {\em nonrelativistic\/}
U(1)-symmetric bosonic $\Phi^4$ field theory~\cite{FWGF-89}, whose
partition function is given by
\begin{eqnarray}
&&Z = \int [D\phi] \exp \left( - \int_0^{1/T} dt \, d^dx
\,{\cal L} \right), \nonumber 
\\
&&{\cal L} = \phi^* \partial_t \phi + {1\over 2 m}
|\nabla \phi|^2 + r |\phi|^2  + u |\phi|^4,
\label{lb}
\end{eqnarray}
where $r \sim \mu-\mu_c$.  The upper critical dimension of this bosonic theory
is $d=2$, thus its critical behavior is of mean-field type
for $d>2$.  For $d=2$ the
field theory is essentially free (apart from logarithmic corrections), thus
the dynamic critical exponent is $z=2$ and the RG dimension dimension of the
coupling $\mu$ is $y_\mu = 2$.  In $d=1$ the theory turns out to be equivalent
to a free field theory of nonrelativistic spinless fermions, thus $z=2$ and
$y_\mu=2$ as well, see, e.g., Ref.~\cite{Sachdev-book}

The quantum critical behaviors in the presence of the trapping potential can
be described in the theoretical framework of the TSS theory~\cite{CV-10}, 
which introduces 
a trap critical exponent $\theta$ which describes how the length scale at the
quantum critical point diverges with increasing the
trap size $l$, i.e., $\xi\sim
l^{\theta}$~\cite{CV-09}.  The trap exponent at the Mott transitions of 1D and
2D BH models is
\begin{equation}
\theta=p/(p+2). 
\label{thetaexp}
\end{equation}
At the low-density to empty-state transition the TSS of the free-energy
density in the presence of a confining potential (\ref{potential}) is given
by~\cite{CV-10,CV-10-2}
\begin{equation}
F(\mu,T,l,x) = l^{-\theta (d+z)} 
{\cal F}(\bar{\mu} l^{\theta/\nu},Tl^{\theta z},xl^{-\theta}),
\label{freee}
\end{equation}
where $x$ is the distance from the middle of the trap, $T$ is the
temperature, $\bar{\mu}\equiv \mu-\mu_c$, and $\nu\equiv 1/y_\mu$.
The zero-temperature TSS of a generic observable, whose low-density
critical behavior is described by the RG dimension $y_o$ in the
homogeneous system, is given by
\begin{equation}
\langle O\rangle(\mu,l,x) \sim  l^{-y_o \theta} 
{\cal O}(\bar{\mu} l^{\theta/\nu}, x l^{-\theta}).
\label{scalbeheq}
\end{equation}
For example, any low-energy scale at $T=0$ is expected to behave as $E =
l^{-z\theta} {\cal E}(\bar{\mu} l^{\theta/\nu})$; the particle density as
$\langle n_x \rangle = l^{-d\theta} {\cal D}(\bar{\mu}
l^{\theta/\nu},xl^{-\theta})$ (using the fact that the RG dimension of the
density operator is $y_n=d$); the one-particle density matrix as
\begin{equation}
\rho_1(x,y)=\langle b^\dagger_x b_y\rangle = l^{-d\theta} {\cal
  M}(\bar{\mu}l^{\theta/\nu},xl^{-\theta},yl^{-\theta}),
\label{rho1mu}
\end{equation}
using the fact that $y_b=d/2$; etc....

The limit $\bar{\mu}\to 0$ corresponds to the low-density regime $Na^d/l^d\to
0$ where $a$ is the lattice spacing.  A TSS Ansatz for the large-$l$ trap-size
dependence at fixed particle number $N$ can be derived by replacing the
dependence on $\bar{\mu} l^{\theta/\nu}$ with that on $N$. See also 
the next
section for an explicit derivation in 1D systems.  
Therefore, for a generic observable we expect
\begin{equation}
\langle O\rangle(N,l,x) \sim  l^{-y_o \theta} {\cal O}_N(x l^{-\theta}).
\label{scalbehNeq}
\end{equation}
In particular, the gap, i.e., the energy difference of the lowest states,
behaves as
\begin{equation}
\Delta_N \approx A_N l^{-z\theta} , 
\label{deltaNtss}
\end{equation}
where $A_N$ is a $N$-dependent amplitude.  Since the RG dimension of the
particle density operator $n_i$ is given by $y_n=d+z-y_\mu=d$, we expect that
\begin{eqnarray}
\rho(x)\equiv \langle n_x \rangle \approx l^{-d\theta} {\cal D}_N(X), 
\label{dnphitss}
\end{eqnarray}
and
\begin{eqnarray}
G_n(x,y)\equiv 
\langle n_{x} n_{y}\rangle-\langle n_{x}\rangle \langle n_{y}\rangle 
\approx  l^{-2d\theta} {\cal G}_N(X,Y),\label{gncotss}
\end{eqnarray}
where $X=x/l^{\theta}$ and $Y=y/l^{\theta}$, and the scaling functions ${\cal
  D}_N$ and ${\cal G}_N$ depend on $N$.  The RG dimension of the boson
operator is $y_b=d/2$, thus the low-density TSS of the one-particle density
matrix is
\begin{equation}
\rho_1(x,y) \equiv \langle b_{x}^\dagger b_{y} \rangle
\approx l^{-d\theta} {\cal M}_N(X,Y).
\label{gbscal}
\end{equation}
Finally, we consider the momentum distribution,  defined as
\begin{equation}
n_k \equiv {1\over N} \sum_{x,y} e^{ik(x-y)} \rho_1(x,y),
\label{nkdef}
\end{equation}
normalized so that $\int {dk\over 2\pi} n_k  =1$. 
$n_k$ is usually accessible experimentally from the intereference patterns of
absorption images taken after the drop of the trap and the expansion of the
atomic gas.  Eq.~(\ref{gbscal}) implies
\begin{equation}
n_k \approx l^{\theta(2-d)} {\cal N}_N(K), \quad K=l^\theta k.
\label{nkscal}
\end{equation}

In the next sections we report analytic and numerical calculations for the BH
model and a gas of impenetrable bosons, showing that they share the same
asymptotic TSS behavior, with the same scaling functions.

\section{Trapped particles at equilibrium on a 1D lattice}
\label{static}

In this section we address issues related to the equilibrium properties of a
1D lattice system of $N$ bosonic particles confined by a trapping potential,
described by the 1D BH model (\ref{bhmN}).  These results will also be
relevant for the off-equilibrium evolution of the system under variations of
the confining potential, in particular when the dynamics is so slow to admit
the adiabatic approximation. In the following we set the hopping parameter
$J=1$, thus the trap size (\ref{trapsizedef}) simply becomes $l=1/v$.

To complete the definition of the BH model in the presence of the
confining potential, we consider traps whose center coincides with a
site of the lattice, so that the lattice model has a reflection
symmetry with respect to the center of the trap.  However, any other
particular choice, i.e., centering the trap anywhere between two
sites, does not change the asymptotic low-density TSS behavior,
leading to an effective asymptotic reflection symmetry.

\subsection{The hard-core limit}
\label{HClimit}

We first consider the hard-core (HC) limit $U\to\infty$ of the 1D BH model,
which allows us to study the effects of the confining potential by exact and
very accurate numerical results. The HC limit implies that the particle number
$n_i$ per site is restricted to the values $n_i=0,1$.  In this limit the 1D BH
model (\ref{bhm}) can be mapped into the XX chain model with lattice spacing
$a$ and a space-dependent transverse external field,
\begin{eqnarray}
H_{\rm XX} &=& - \sum_i \left( S^x_i S^x_{i+1} + S^y_i S^y_{i+1} \right)
\nonumber \\&-&
\sum_i [\mu+V(x_i)] S^z_i,
\label{XX}
\end{eqnarray}
where $S^a_i=\sigma^a_i/2$ and $\sigma^a$ are the Pauli matrices, which are
related to the boson operators $b_i$ by $\sigma^x_i = b_i^\dagger + b_i$,
$\sigma^y_i = i(b_i^\dagger - b_i)$, $\sigma^z_i = 1-2b_i^\dagger b_i$.  
Then, by a Jordan-Wigner transformation, one can
further map it into a model of spinless fermions, see, e.g.,
Ref.~\cite{Sachdev-book}, with a quadratic Hamiltonian
\begin{eqnarray}
H_q = \sum_{ij} c_{i}^\dagger h_{ij} c_{j} + (\mu-1) \sum_i c_i^\dagger c_i,
\label{fermmod}
\end{eqnarray}
where $c_i$ is a spinless fermion operator,
and
\begin{eqnarray}
h_{ij} = \delta_{ij} - {1\over 2} \delta_{i,j-1} - {1\over 2} \delta_{i,j+1}
+ V(x_i) \delta_{ij}. 
\label{hijdef}
\end{eqnarray}

Issues related to quantum transitions in particle systems are best
discussed in the presence of the chemical potential $\mu$.  In the
absence of the trap, the 1D HC-BH model with a chemical potential
$\mu$ has three phases: two Mott insulator phases, for $\mu<-1$ with
$\langle n_i\rangle=1$ and for $\mu>1$ with $\langle n_i\rangle=0$,
separated by a gapless superfluid phase for $|\mu|<1$.  Therefore,
there are two Mott insulator to superfluid transitions at $\mu=-1$ and
$\mu=1$.  These transitions are characterized by the dynamic exponent
$z=2$ and the RG dimension of the chemical potential
$y_\mu=2$~\cite{FWGF-89}.  The gapless superfluid phase is instead
described by a free massless bosonic field theory with dynamic
exponent $z=1$, see, e.g., Ref.~\cite{Sachdev-book}.

In 1D particle systems, the {\em thermodynamic} limit at fixed $\mu$
corresponds to $N,l\to\infty$ keeping the ratio $N/l$ fixed.
Indeed, we have
\begin{equation}
N \equiv \langle \sum_i b_i^\dagger b_i  
\rangle = \tilde{\rho}(\mu) l + O(1)\label{cmudef}
\end{equation}
The function $\tilde{\rho}(\mu)$ can be computed in the HC limit.  The
particle density in the large-$l$ limit turns out to approach its local
density approximation (LDA), with corrections that are suppressed by powers of
the trap size and present a nontrivial TSS behaviour~\cite{CV-10-2}.  Within
the LDA, the particle density at the spatial coordinate $x$ equals
the particle density of the homogeneous system at the
effective chemical potential
\begin{equation}
\mu_{\rm eff}(x) \equiv \mu + {1\over p} \left({x\over l}\right)^p.
\label{mueff}
\end{equation}
The LDA of the particle density reads
$\langle n_x \rangle_{\rm lda} \equiv \rho_{\rm lda}(x/l)$, where
\begin{equation}
\rho_{\rm lda}(x/l) = 
\kern-10pt \quad\left\{
\begin{array}{l@{\ \ }l@{\ \ }l}
0 & {\rm for} & \mu_{\rm eff}(x) > 1, \\
(1/\pi)\arccos\mu_{\rm eff}(x) &
    {\rm for} & -1 \le \mu_{\rm eff}(x) \le 1, \\
1 & {\rm for} & \mu_{\rm eff}(x) < -1. \\
\end{array} \right.
\label{nxlda}
\end{equation}
Asymptotically, the total particle number is obtained by integrating 
the LDA of the particle density $\rho_{\rm lda}$, obtaining
\begin{eqnarray}
\tilde{\rho}(\mu) = 2  \int_0^\infty \rho_{\rm lda}(y) \,{\rm d} y.
\label{cmu}
\end{eqnarray}
In the low-density regime, $\bar\mu\equiv\mu-1\to 0$,
\begin{equation}
\tilde{\rho}(\mu) = c |\bar\mu|^{(2+p)/(2p)} [1 + O(\bar\mu)],
\label{tilderho}
\end{equation}
where $c$ is a $p$-dependent constant; $c=1$ for $p=2$.  

Eq.~(\ref{cmudef}) provides the correspondence between the ratio $N/l$ and
$\mu$.  In particular, when $0<N/l<\tilde{\rho}(-1)$ the system is effectively
in the superfluid phase, while for $N/l>\tilde{\rho}(-1)$ the $n=1$ Mott phase
appears around the center of the trap.  Eq.~(\ref{cmu}) gives
$\tilde{\rho}(-1)=2.54648$ for $p=2$ and $\tilde{\rho}(-1)=2.56561$ for $p=4$.
Eqs.~(\ref{cmudef}), (\ref{nxlda}), 
and (\ref{tilderho}) imply that the particle
density at the origin scales as a nontrivial power law of the ratio $N/l$ in
the low-density regime,
\begin{equation}
\rho(0) \sim (N/l)^{\theta}.
\label{asybehrho0}
\end{equation}

At the low-density Mott transition, around $\mu=1$, the TSS limit can be
analytically derived within the quadratic spinless fermion
representation~\cite{CV-10-2}. This is obtained by rescaling the spatial
distance from the origin as 
\begin{equation}
x = l^{\theta} X,
\label{xresc1}
\end{equation}
and the difference
$\bar{\mu}\equiv 1-\mu$ as 
\begin{equation}
\bar{\mu} = l^{-2\theta} \mu_r.
\label{muresc1}
\end{equation}
Any low-energy scale turns out to behave as $E \approx l^{-2\theta}
{\cal E}_\Delta(\mu_r)$, the particle density behaves as $\langle n_x
\rangle \approx l^{-\theta} {\cal D}(\mu_r,X)$, etc..., in agreement
with the scaling Ansatz reported in the previous section,
cf. Eq.~(\ref{scalbeheq}).  Analytic and numerical calculations of the
above scaling functions are reported in
Ref.~\cite{CV-10-2}.~\footnote{Note that here the definition of the
trap size, cf.  Eq.~(\ref{potential}), differs from that of
Refs.~\cite{CV-10,CV-10-2} by a factor $p^{-1/p}$.}

\subsection{TSS of a system of $N$ particles}
\label{TSSn}

We now derive the TSS as a function of the particle number $N$ in the low
density regime $Na/l\ll 1$.  This is worth being discussed in some detail,
because the corresponding TSS functions are not trivially derived from the TSS
in the presence of a chemical potential, which were already computed in
Ref.~\cite{CV-10-2}.

\subsubsection{TSS limit}
\label{TSSlim}

In the fermion representation the Hamiltonian 
\begin{equation}
H_c = \sum_{ij} c_{i}^\dagger h_{ij} c_{j}
\label{hcn}
\end{equation}
can be diagonalized by introducing new
canonical fermionic variables 
$\eta_k=\sum_i \phi_{ki} c_i$,
where $\phi$ satisfies the equation
\begin{equation}
h_{ij}\phi_{kj} = \omega_k \phi_{ki},
\label{hijeq}
\end{equation}
so that
\begin{equation}
H_c = \sum_k \omega_k \eta_k^\dagger \eta_k.  
\label{hcdia}
\end{equation}
The ground state of a system of $N$ particles is then given by the 
$\eta$-fermions filling the $N$ lowest one-particle levels.  For the lowest
states, assuming smoothness, we may consider the continuum limit of
Eq.~(\ref{hijeq}), by replacing $\phi_{kx}\to \phi_k(x)$ and rewriting the
discrete differences as
\begin{equation}
\phi(x+a)-\phi(x)= a {d\phi(x)\over dx} + {1\over 2} a^2  
{d^2\phi(x)\over dx^2} + ...,
\label{discr}
\end{equation}
where $a$ is the lattice spacing.  Then, rewriting the resulting equation in
terms of the rescaled quantities
\begin{eqnarray}
&&X\equiv a^{-2\theta/p} l^{-\theta} x, \label{xresc} \\
&&e_k \equiv  a^{-2\theta} l^{2\theta}\omega_k, \label{eresc} \\
&&\varphi_k(X)\equiv a^{\theta/p} 
l^{\theta/2}\phi_k(a^{-2\theta/p} l^\theta X), \label{phiresc} 
\end{eqnarray}
with $\theta$ given by Eq.~(\ref{thetaexp}), and neglecting terms which are
suppressed by powers of the trap size, one arrives at a 
Schr\"odinger-like
equation 
\begin{equation}
\left( - {1\over 2} {d^2 \over dX^2} + {1\over p} X^p \right)\varphi_k(X) 
= e_k \varphi_k(X).
\label{trapscaleqxx}
\end{equation}
This equation describes the TSS limit at fixed $N$, i.e., $l\to
\infty$, $x\to \infty$, keeping the scaling variable $X$ fixed.  
The next-to-leading terms in the large-$l$ limit, arising from
the higher order terms in the expansion (\ref{discr}), give rise to
$O(l^{-2\theta})$ scaling corrections~\cite{CV-10-2}. Moreover, one can easily
check that a shift of the center of the trap, by $\delta < a$, generally
induces $O(l^{-\theta})$ subleading corrections.

Solving Eq.~(\ref{trapscaleqxx}) for $p=2$, we obtain
\begin{eqnarray}
&&e_k = k + 1/2, \quad k\ge 0, \label{eq:p2eig}\\
&&\varphi_k(X) = {H_k(X)\over
 \pi^{1/4} 2^{k/2} (k!)^{1/2}} \, \exp(-X^2/2), 
 \nonumber 
\end{eqnarray}
where $X\equiv x/(al)^{1/2}$ and $H_k$ are Hermite's polynomials.
For $p=4$, Eq.\ (\ref{trapscaleqxx}) can be solved numerically by Numerov's
method, see, e.g., Ref.~\cite{book-numerov}; the resulting energy levels are 
$e_0= 0.420805$,
$e_1= 1.50790$,
$e_2= 2.95880$,
$e_3= 4.62122$,
$e_4= 6.45350$,
$e_5= 8.42843$, etc....
The Bohr-Sommerfield quantization formula, see, e.g., Ref.~\cite{Landau-book}, 
gives the asymptotic
large-$k$ behavior 
\begin{eqnarray}
e_k &\approx& b_4 (k+1/2)^{4/3},
\nonumber \\
b_4 &=& {\pi^{2/3} \Gamma(7/4)^{4/3}\over
2^{4/3}\Gamma(5/4)^{4/3}}\cong 0.867145.
\label{eq:p4eig}
\end{eqnarray}
This formula provides a good approximation for
relatively low levels already: it is accurate to $0.1\%$ already for $e_5$.
For $p\to\infty$, Eq.\ (\ref{trapscaleqxx}) becomes equivalent to the
Schr\"odinger equation of a free particle in a box of size $L=2l$ with
boundary conditions $\varphi(-1)=\varphi(1)=0$, leading to
\begin{eqnarray}
&&e_k = {\pi^2\over 8} (k+1)^2, \qquad k\ge 0,
\label{eq:pinfeig}\\
&&\varphi_k(X) = \sin\left[{\pi\over 2} (k+1) (X+1)\right],
\nonumber
\end{eqnarray}
where $X\equiv x/l$. 

In App.~\ref{tdhp} we show that the same TSS limit is obtained when
the trap is induced by a spatial dependence of the hopping parameter.

\subsubsection{TSS of observables}
\label{tssobs}

In the following we set the lattice spacing $a=1$ to simplify the
expressions.  The dependence on $a$ and $J$ can be easily recovered by
a dimensional analysis.

The TSS limit leading to Eq.~(\ref{trapscaleqxx}) allows us to compute the
low-density trap-size dependence of the observables.  For example, the gap,
i.e., the difference of the energy of the lowest states, behaves as
\begin{equation}
\Delta_N = A_N l^{-2\theta} \left[ 1 + O(l^{-2\theta})\right],
\label{deltaN}
\end{equation}
with $A_2=1$ for $p=2$, and 
\begin{equation}
A_N = a_p N^{2\theta-1} \left[ 1 + O(1/N)\right]
\label{apn}
\end{equation}
for a generic power law $p$: $a_4\approx 1.15619$ and $a_\infty =
\pi^2/2$.  Note that the gap at fixed $N$ differs from the difference of the
energy of the lowest states at fixed chemical potential $\mu$, which behaves
as $\Delta_\mu=l^{-2\theta}{\cal E}_\Delta(\mu_r)$, because the latter
involves states of subsequent particle number sectors, giving rise to zeroes
in the scaling function ${\cal E}_\Delta(\mu_r)$ for $\mu_r<0$, see
Ref.~\cite{CV-10-2}.

In the low-density regime the particle density behaves as
\begin{eqnarray}
&&\rho(x)\equiv \langle n_x \rangle = 
l^{-\theta} {\cal D}_N(X) \left[1+ O(l^{-2\theta})\right],
\label{nxn}\\
&& {\cal D}_N(X) = \sum_{k=0}^{N-1} \varphi^2_k(X).
\label{dnphi}
\end{eqnarray}

\begin{figure}[tbp]
\includegraphics*[scale=\graphicscale]{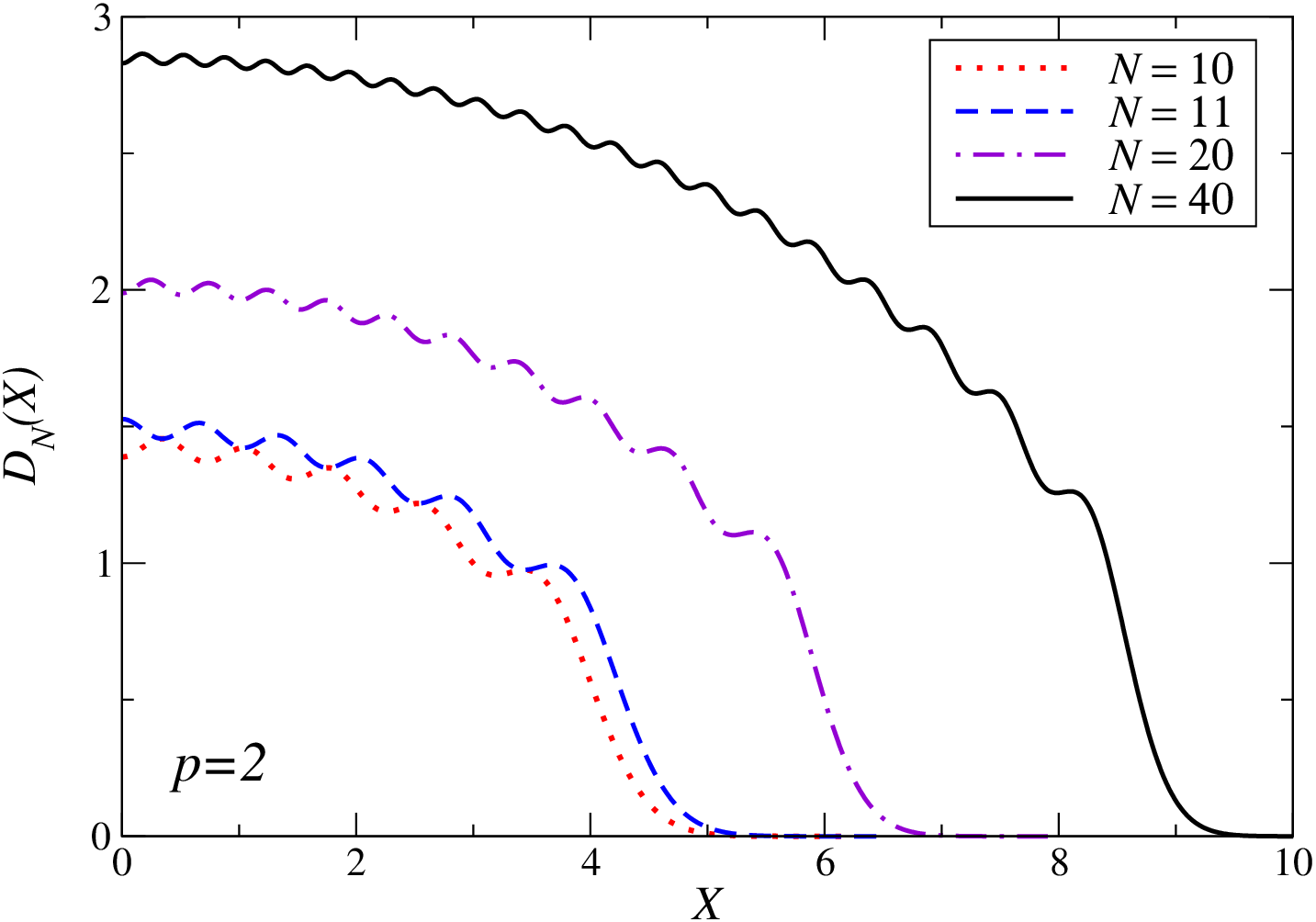}
\caption{ (Color online)
  The scaling function ${\cal D}_N(X)$, cf. Eq.~(\ref{dnphi}), for $p=2$.
  Since ${\cal D}_N(X)={\cal D}_N(-X)$ due to the reflection symmetry with
  respect to the center of the trap, we show only the curves for $X\ge 0$.  }
\label{dnxp2}
\end{figure}

\begin{figure}[tbp]
\includegraphics*[scale=\graphicscale]{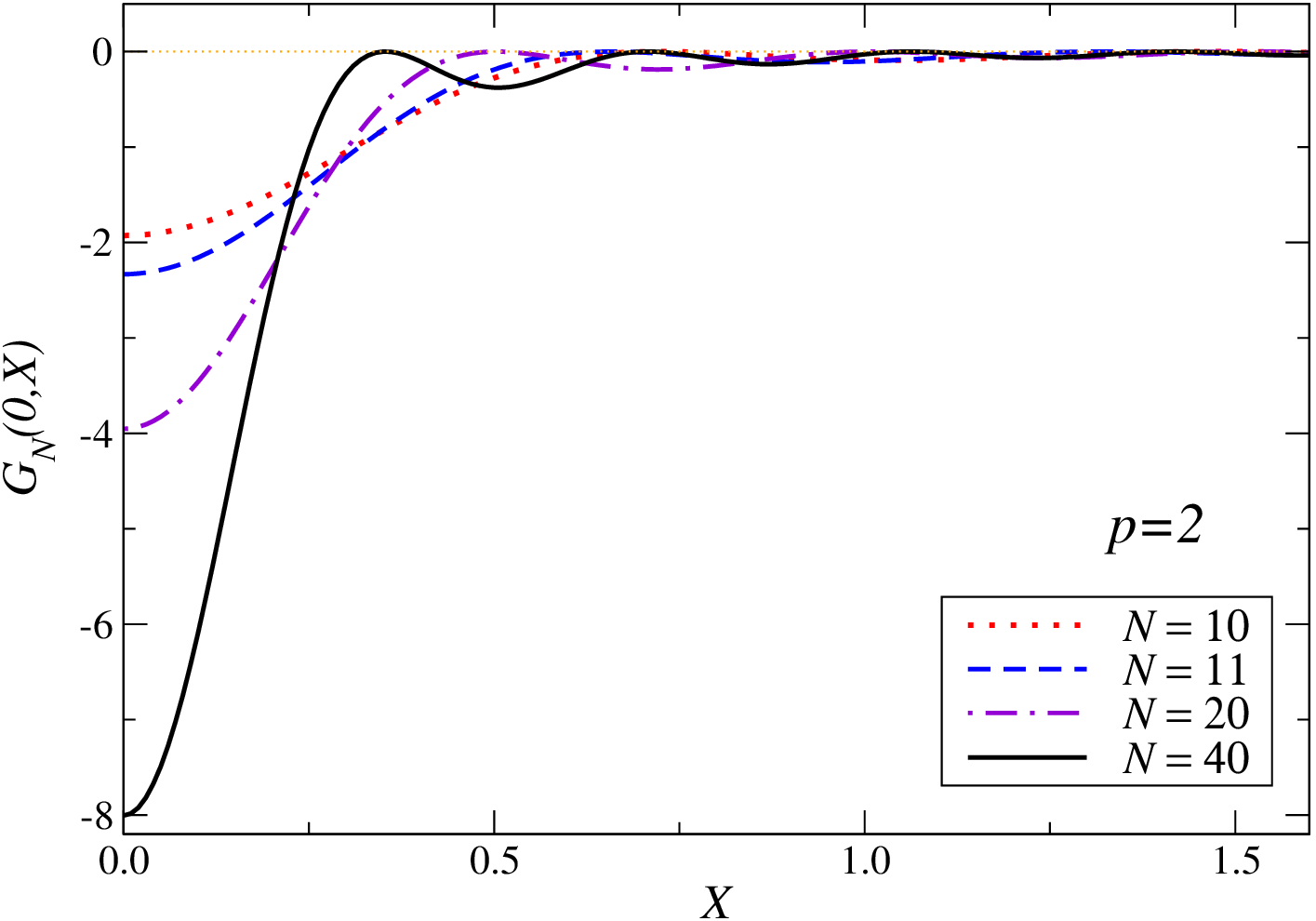}
\caption{ (Color online)
The scaling function 
${\cal G}_N(0,X)$, cf. Eq.~(\ref{eq:GnTSS}), for $p=2$.
}
\label{gnxp2}
\end{figure}

Fig.~\ref{dnxp2} shows results for the spatial dependence of ${\cal
D}_N(X)$ for $p=2$ and several values of $N$.  Note the peculiar
structure of ${\cal D}_N(X)$ characterized by $N$ local maxima, which
get suppressed at large $N$ by powers of $1/N$.  Due to the parity of
the Hermite polynomials, ${\cal D}_{2j-1}(0)= {\cal D}_{2j}(0)$.

Straightforward calculations show that the density-density correlator
behaves as
\begin{eqnarray}
G_n(x,y)\equiv 
\langle n_x n_y\rangle_c
\approx  l^{-2\theta} {\cal G}_N(X,Y),\label{gnco}
\end{eqnarray}
where $X=x/l^{\theta}$, $Y=y/l^{\theta}$, and
\begin{eqnarray}
{\cal G}_N(X,Y) = 
-\Bigl[\textstyle\sum_{k=0}^{N-1} \varphi_k(X) \varphi_k(Y)\Bigr]^2.
\label{eq:GnTSS}
\end{eqnarray}
Fig.~\ref{gnxp2} shows plots of ${\cal G}_N(0,X)$ for the harmonic potential.

\begin{figure}[tbp]
\includegraphics*[scale=\graphicscale]{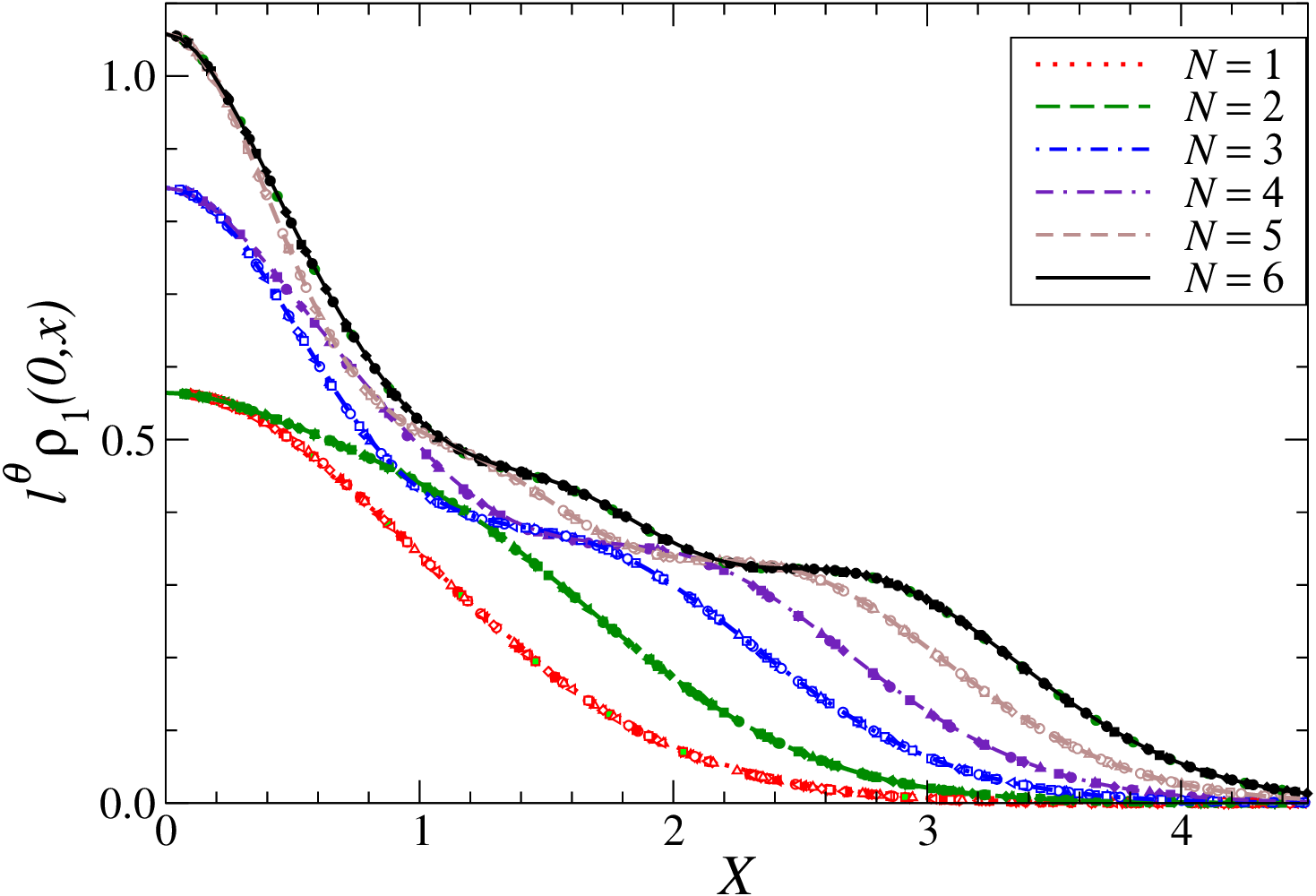}
\caption{(Color online) Results for the one-particle density matrix: we plot
  $l^\theta \rho_1(0,x)$ vs $X\equiv x/\l^\theta$ for $p=2$, thus
  $\theta=1/2$, for several values of $N$.  The data points are numerical
  results for the HC-BH model at fixed $N$ and trap size $l$, with $10\lesssim
  l\lesssim 10^3$. The continuous lines are the curves for systems of $N$
  impenetrable bosons.  The data of the HC-BH model clearly approach these
  curves in the large trap-size limit.  }
\label{gbp2}
\end{figure}

The one-particle density matrix, cf. Eq.~(\ref{gbscal}), cannot be easily
derived from the solutions of Eq.~(\ref{trapscaleqxx}), because the
fermion-boson map exploited in the HC limit is not trivial, and, in
particular, it is non local.  However, as we shall show in Sec.~\ref{bosgas},
the asymptotic TSS of the BH model in the low-density regime coincides with
the trap-size dependence of a 1D gas of impenetrable bosons, whose
one-particle density matrix can be computed using the known ground-state wave
function.  Some results for the harmonic potential are shown in
Fig.~\ref{gbp2}.

The TSS functions of the observables considered above show nontrivial
power-law scalings with respect to the particle number $N$ at large
$N$.  Their large-$N$ behaviors are reported in
App.~\ref{lnbeh}.

\subsection{Numerical results}
\label{numres}

Beside deriving the asymptotic behaviors in the low-density region, we present
numerical calculations at fixed particle number $N$ and trap size $l$.  We
exploit the quadratic spinless fermion representation (\ref{fermmod}) of the
1D HC-BH model, which allows us to perform computations for very large
systems, since they only require the diagonalization of a $L\times L$ matrix
where $L$ is the number of lattice sites.  We obtain numerical results for
chains of size $L$, with a trap of size $l$ centered at the middle site (we
consider odd $L$); we choose $L$ large enough to have negligible finite-$L$
effects.  This can been accurately checked by comparing results at fixed $l$
and increasing values of $L$.  Thus, the results at fixed $N$ and $l$ that we
shall present, respectively up to $N\approx 10^2$ and $l=O(10^3)$, are the
infinite chain size limit (keeping $N$ and $l$ fixed) with great accuracy.
For more details see Ref.~\cite{CV-10-2}, where analogous calculations at
fixed chemical potential were presented.

\begin{figure}[tbp]
\includegraphics*[scale=\graphicscale]{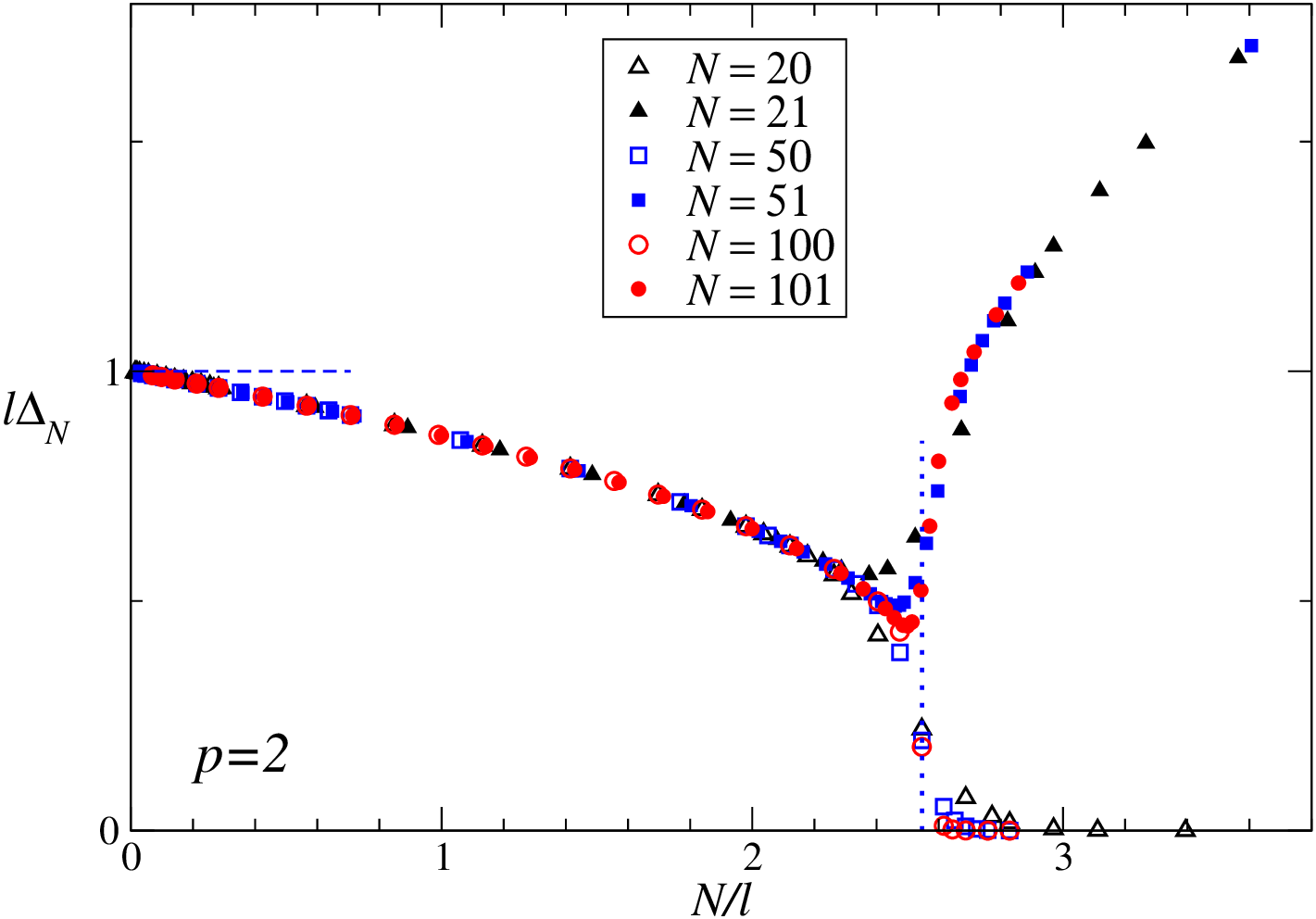}
\caption{ (Color online) $l\Delta_N$ versus $N/l$ for $p=2$. The dashed
  horizontal line indicates the constant value computed in the low-density
  regime, i.e., $N/l\ll 1$.  The vertical dotted line shows the asymptotic
  value of the ratio $N/l$ corresponding to the $n=1$ Mott transition.  }
\label{gapNp2}
\end{figure}

\begin{figure}[tbp]
\includegraphics*[scale=\graphicscale]{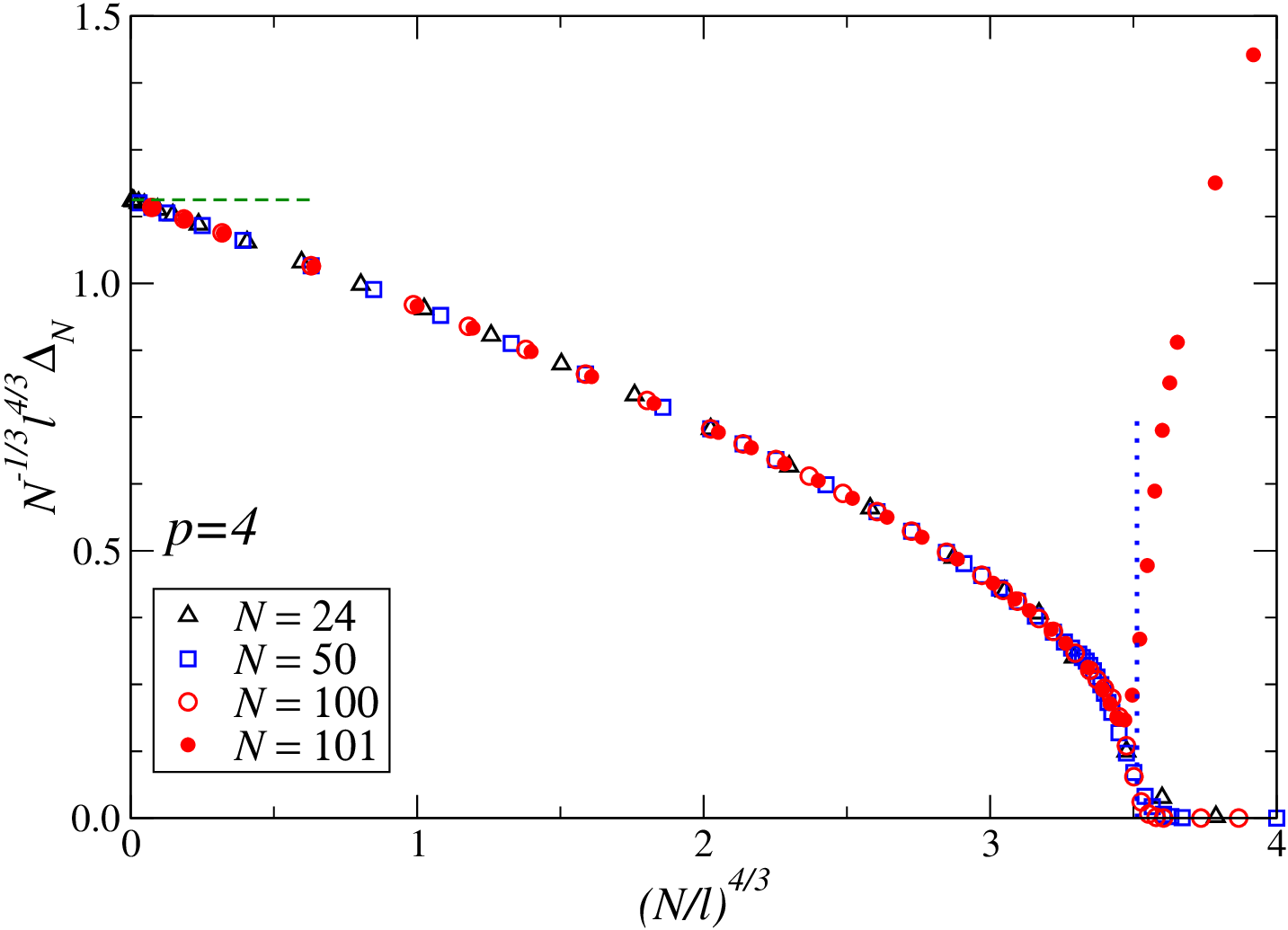}
\caption{ (Color online) Plot of $N^{-1/3} l^{4/3} l\Delta_N$ versus
  $(N/l)^{4/3}$ for $p=4$. The dashed horizontal line indicates the constant
  value computed in the low-density regime, i.e., $N/l\ll 1$.  The vertical
  dotted line shows the asymptotic value of $(N/l)^{4/3}$ corresponding to the
  $n=1$ Mott transition.  }
\label{gapNp4}
\end{figure}

Bosonic particle systems confined to 1D lattices have already been the subject
of several numerical 
investigations~\cite{BRSRMDT-02,KSDZ-04,PRHD-04,WATB-04,RM-04,BKMRS-08,RBRS-09}.  We
study the dependence of some physically interesting observables on the
particle number $N$ and the trap size $l$.

Figs.~\ref{gapNp2} and \ref{gapNp4} show results of the gap for $p=2$
and $p=4$ respectively.  Note that the gap of a system of $N$
impenetrable bosons is identical to the gap of $N$ free fermion
particles in a trap.  Guided by the low-density scaling behavior
(\ref{deltaN}), we plot the quantity
$N^{1-2\theta}l^{2\theta}\Delta_N$ versus $(N/l)^{2\theta}$ (which is
just $l\Delta_N$ vs. $N/l$ for $p=2$).

\begin{figure}[tbp]
\includegraphics*[scale=\graphicscale]{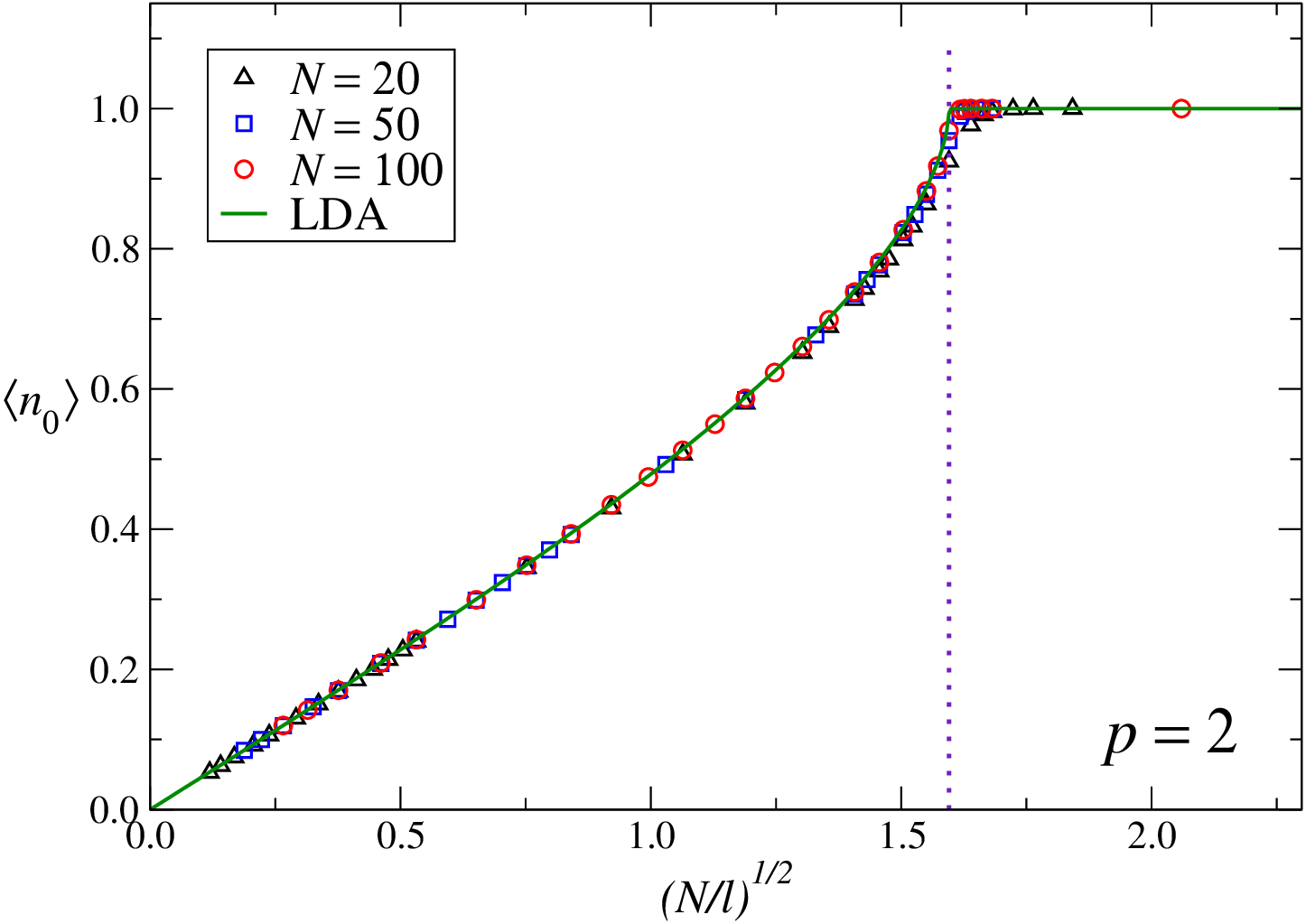}
\caption{ (Color online)
  The particle density at the origin versus $(N/l)^{1/2}$, for $p=2$ for
  several values of $N$ and $l$.  The data points show results obtained
  by solving the HC-BH model at fixed $N$ and $l$, while the continuous
  line shows the LDA.  The vertical dotted line shows the
  asymptotic value of $(N/l)^{1/2}$ corresponding to the $n=1$ Mott
  transition.  }
\label{rho0Np2}
\end{figure}

\begin{figure}[tbp]
\includegraphics*[scale=\graphicscale]{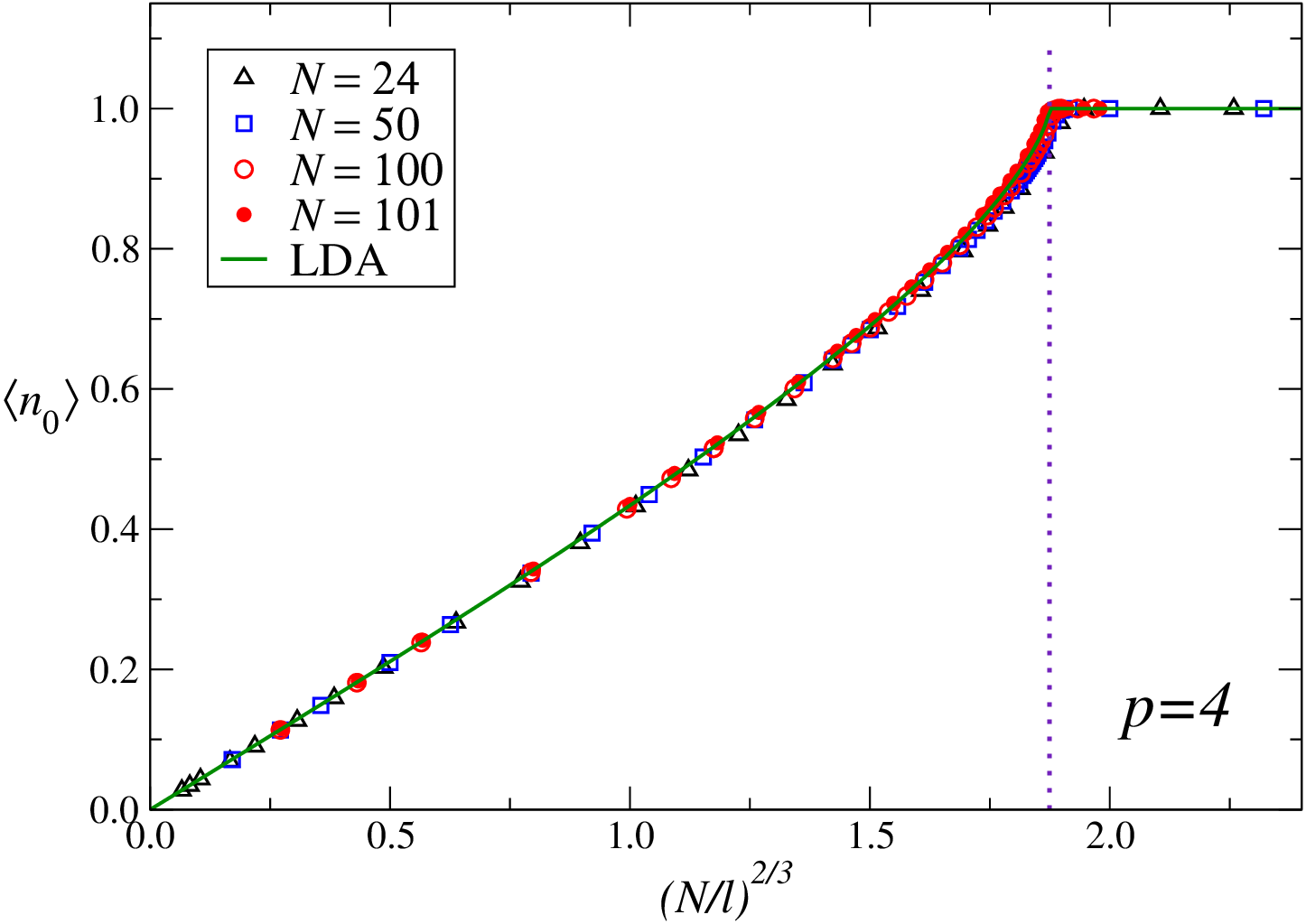}
\caption{ (Color online)
  The particle density at the origin versus $(N/l)^{2/3}$, for $p=4$.  The
  data points show results obtained by solving the HC-BH model at fixed
  $N$ and $l$, while the continuous line shows the LDA. The
  vertical dotted line shows the asymptotic value of $(N/l)^{2/3}$
  corresponding to the $n=1$ Mott transition.  }
\label{rho0Np4}
\end{figure}

In the region of values of $N/l$ corresponding to $1>\mu>-1$, the data
show the asymptotic behavior
\begin{equation}
N \Delta_N = g(N/l),
\label{deltasu}
\end{equation}
in the limit $l\to\infty$, $N\to\infty$ keeping $N/l$ fixed.  The
low-density behavior (\ref{deltaN}) is recovered for $N/l\ll 1$,
because
\begin{equation}
g(x) = c x^{2\theta}\left[ 1 + O(x^{2\theta})\right].
\label{gxbeh}
\end{equation}
Around $N/l=\tilde{\rho}(-1)$, i.e., the value corresponding to the
$n=1$ Mott transition, the behavior for even and odd $N$ begins
differing significantly.  In particular, the data for even $N$ appear
suppressed for $N/l\gtrsim \tilde{\rho}(-1)$.  This is essentially
related to the fact that the trap is centered at the middle site of
the chain. When the region around the center of the trap shows the
$n=1$ Mott phase, we have two degenerate lowest states for even $N$,
differing for a reflection with respect to the middle site; while for
odd $N$ the ground state is unique, and the gap is expected to behave
as $\Delta_N\sim N^{p-1}/l^p$ for $N/l$ sufficiently larger than
$\tilde{\rho}(-1)$, as also shown by Fig.~\ref{gapNp2} for $p=2$,
where the corresponding asymptotic behavior $l\Delta_N\sim N/l$ can be
already seen for $N/l\gtrsim 3$.

Figs.~\ref{rho0Np2} and \ref{rho0Np4} show results of the particle
density at the origin, for $p=2$ and $p=4$ respectively, for some
values of $N$ in the range $20\lesssim N\lesssim 100$, and $l$ up to
$O(10^3)$.  The data appear to follow a unique function of
$\tilde\rho\equiv N/l$, given by the LDA obtained using
Eqs.~(\ref{nxlda}) and (\ref{cmu}), and corrections are hardly
visible.  This fact was already observed by other numerical works,
see, e.g., Refs.~\cite{BKMRS-08,RBRS-09}.  The agreement is already
good for relatively small values of $N$, i.e., $N\gtrsim 20$.  Note
that the LDA reproduces the low-density behavior (\ref{nxn}) for $N\gg
1$, and in particular the leading large-$N$
term of Eq.~(\ref{dnto1on}).

\begin{figure}[tbp]
\includegraphics*[scale=\graphicscale]{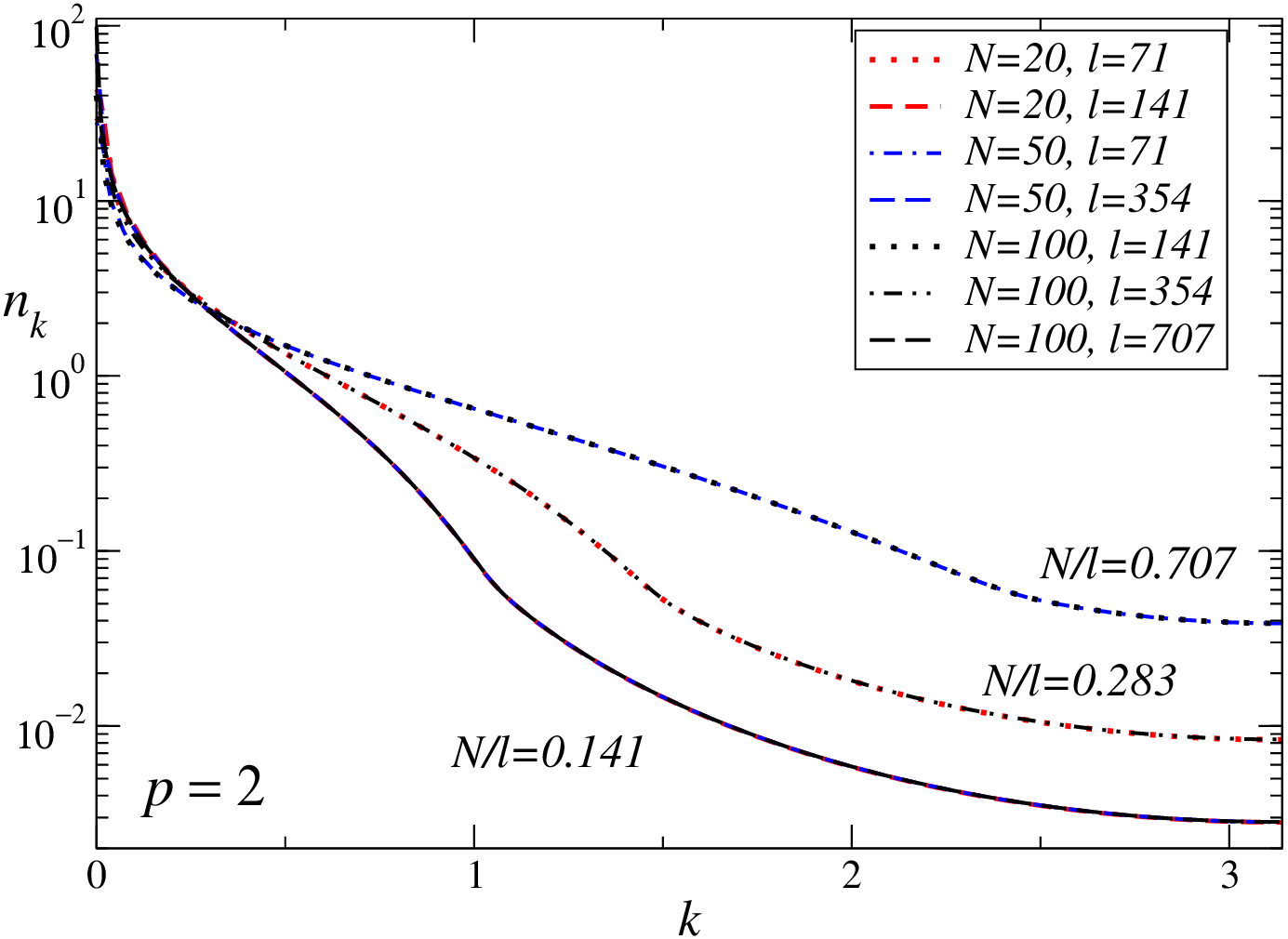}
\includegraphics*[scale=\graphicscale]{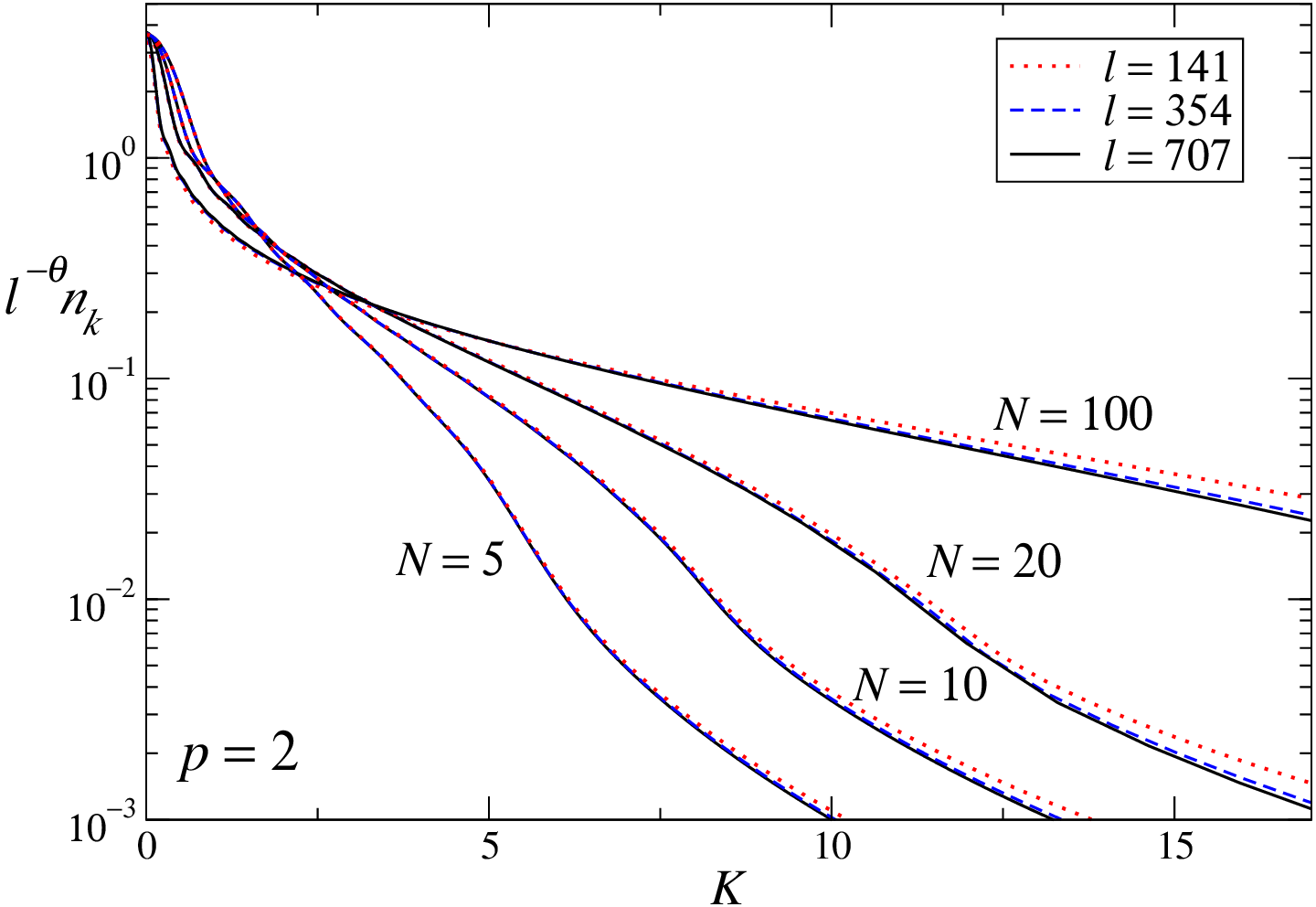}
\includegraphics*[scale=\graphicscale]{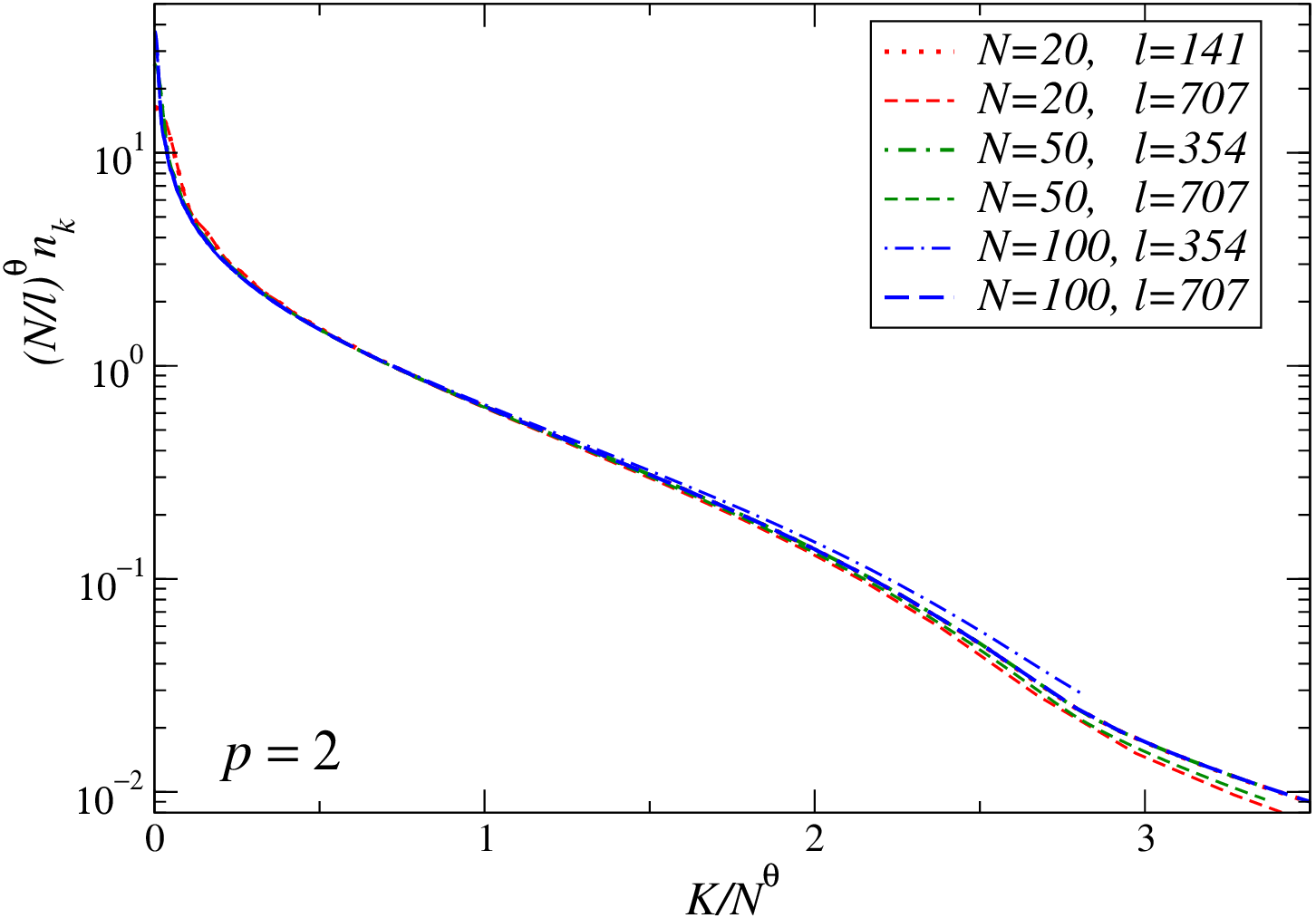}
\caption{(Color online) Results for the momentum distribution $n_k$
for $p=2$, and several values of $N$ and $l$.  We plot $n_k$ vs $k$
(above), $l^{-\theta}n_k$ versus $K\equiv l^{\theta} k$ (middle), and
$(N/l)^{\theta}n_k$ vs $(N/l)^{-\theta}k$ (below).  We recall that
$\theta=1/2$ for $p=2$. }
\label{nkp2}
\end{figure}

\begin{figure}[tbp]
\includegraphics*[scale=\graphicscale]{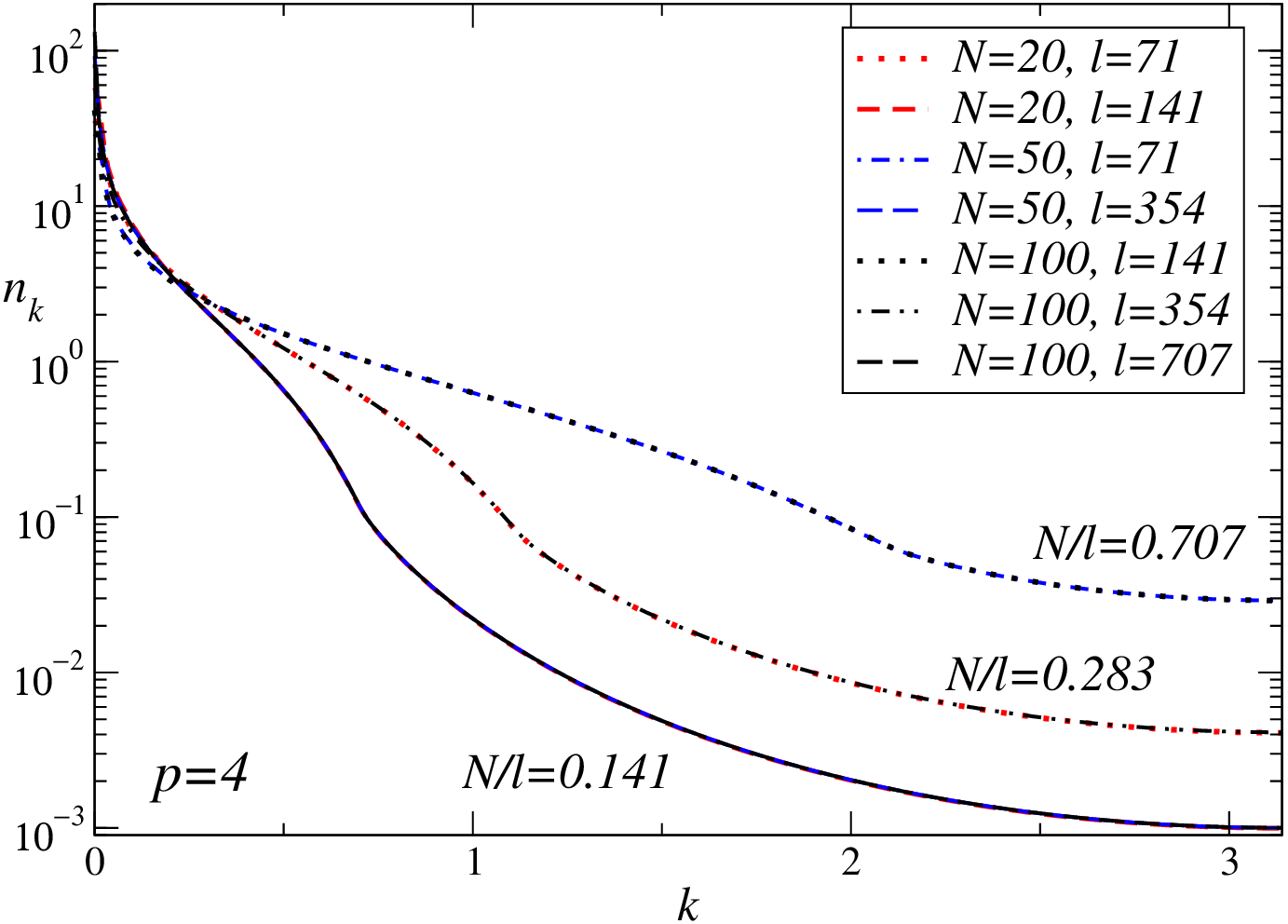}
\includegraphics*[scale=\graphicscale]{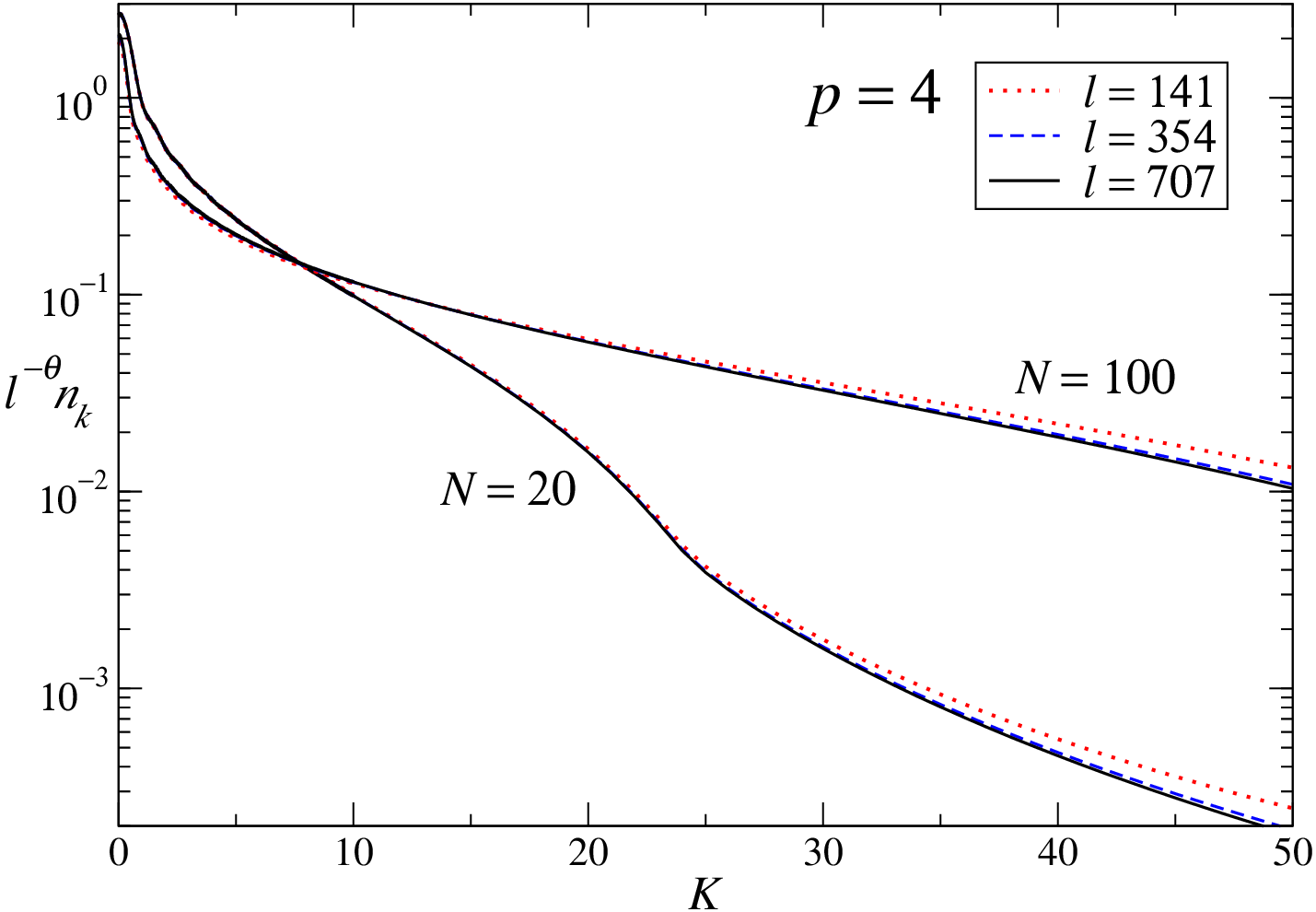}
\includegraphics*[scale=\graphicscale]{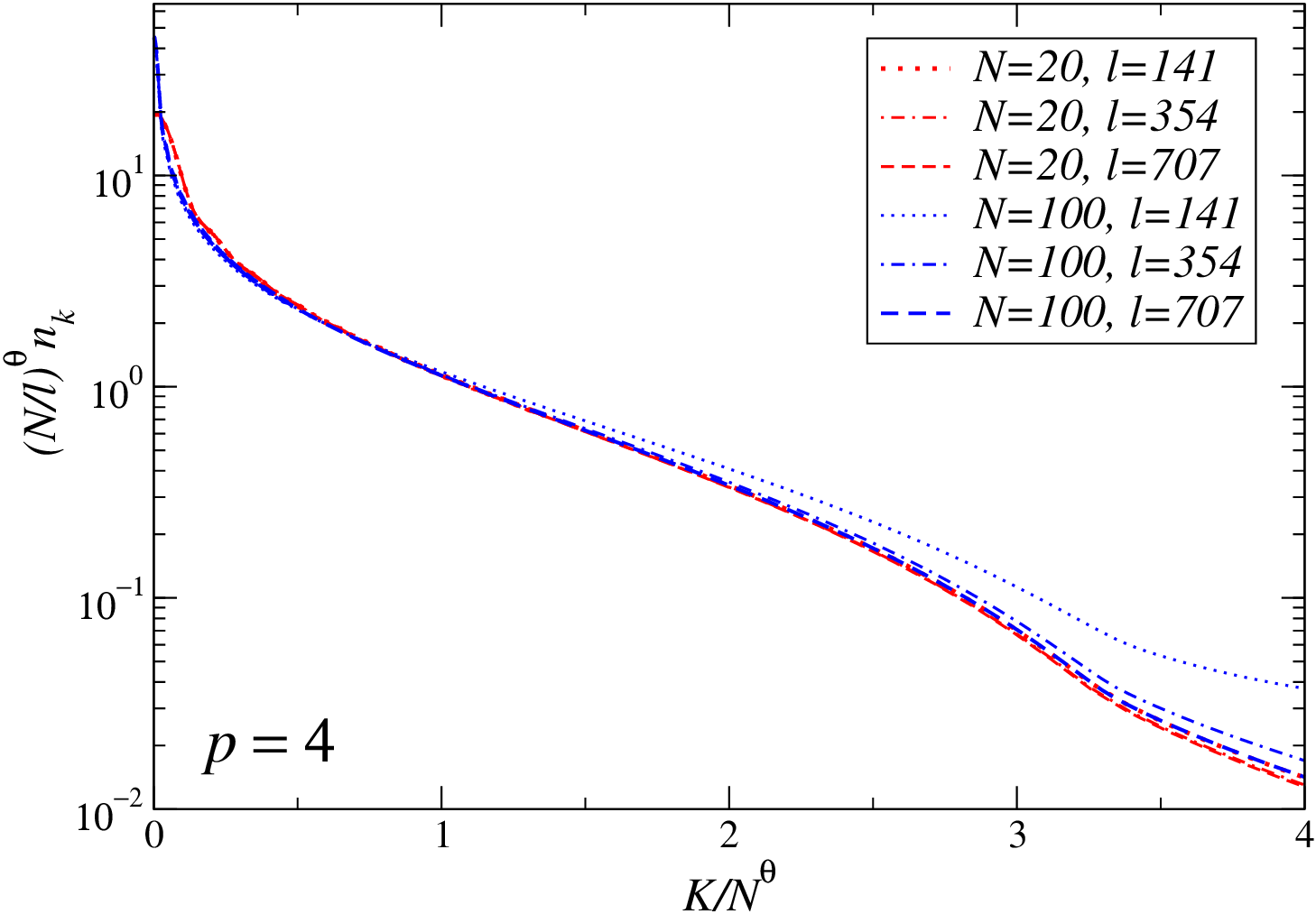}
\caption{ (Color online) Results for the momentum distribution $n_k$
for $p=4$, and several values of $N$ and $l$.  We plot $n_k$ vs $k$
(above), $l^{-\theta}n_k$ versus $K\equiv l^{\theta} k$ (middle), and
$(N/l)^{\theta}n_k$ vs $k/(Nl)^{\theta}$ (below).  We recall that
$\theta=2/3$ for $p=4$. }
\label{nkp4}
\end{figure}

Figures~\ref{nkp2} and \ref{nkp4} show results for the momentum
distribution $n_k$, cf. Eq.  (\ref{nkdef}), for $p=2$ and $p=4$
respectively, and several
values of $N$ and $l$.  We first note that the plots of $n_k$ vs $k$ show the
scaling behavior
\begin{equation}
n_k \approx f(N/l,k)
\label{abovenk}
\end{equation}
Moreover, as shown by the plots of $l^{-\theta}n_k$ versus $K\equiv
l^{\theta}k$, the data also support the TSS behavior (\ref{nkscal}), which is
expected to be approached with $O[(N/l)^{2\theta}]$ corrections.  Actually, as
shown by the bottom figures (\ref{nkp2}) and (\ref{nkp4}), the data 
for $k>0$ appear to scale as 
\begin{equation}
n_k = (N/l)^{-\theta} F(\widetilde{K}),\quad
\widetilde{K}\equiv (N/l)^{-\theta} k,
\label{bottomnk}
\end{equation}
for $N/l\ll 1$, which agrees with both Eq.~(\ref{abovenk}) and (\ref{nkscal}).
The zero component 
\begin{equation}
n_0={1\over N} \sum_{x,y} \rho_1(x,y)
\label{zerocompnk}
\end{equation}
scales differently, indeed $n_0=O(1)$ in the large-$N$ limit,
analogously to a gas of impenetrable bosons. See, e.g.,
Ref.~\cite{Papenbrock-03}.  At large $\widetilde{K}$,
$F(\widetilde{K}) \sim \widetilde{K}^{-4}$, which can be inferred from
the results of Refs.~\cite{MVT-02,OD-03} for a gas of impenetrable
bosons.

\subsection{Finite $U$ and universality of the low-density behavior}
\label{finiteU}

We now consider the BH model at finite values of the on-site repulsion
coupling $U$. We perform calculations using the DMRG method.
Specifically, we consider the BH model with $U=2$ in the presence of a
harmonic potential, up to trap sizes $l=O(10^3)$, and for several
values of $N$, up to $N=20$. The trap is again centered in the middle
site of a lattice of size $L$, which is taken sufficiently large to
make finite-$L$ effects negligible.  We set the cutoff on the number
of bosonic states per site $n_B = 5$, which turns out to be sufficient
to provide very accurate results; indeed the relative difference from the
results using $n_B = 4$ is at most $O(10^{-7})$.  We keep the maximum
eigenvalue truncated from the density matrix below $10^{-10}$; this
requires retaining up to 200 states.

The issue that we want to investigate is the universality of the low-density
TSS behavior with respect to variations of the on-site repulsion coupling $U$,
i.e., how it depends on $U$. If there is universality, then the low-density
asymptotic TSS at finite values of $U$ must be the same as that found in the
HC limit.  Actually, a rescaling of the trap size which depends on $U$ may be
allowed, although the data show that this is not necessary when one uses the
same definition of trap size with respect to the kinetic term, as we have done
in Eqs.~(\ref{bhmN}) and (\ref{potential}).

\begin{figure}[tbp]
\includegraphics*[scale=\graphicscale]{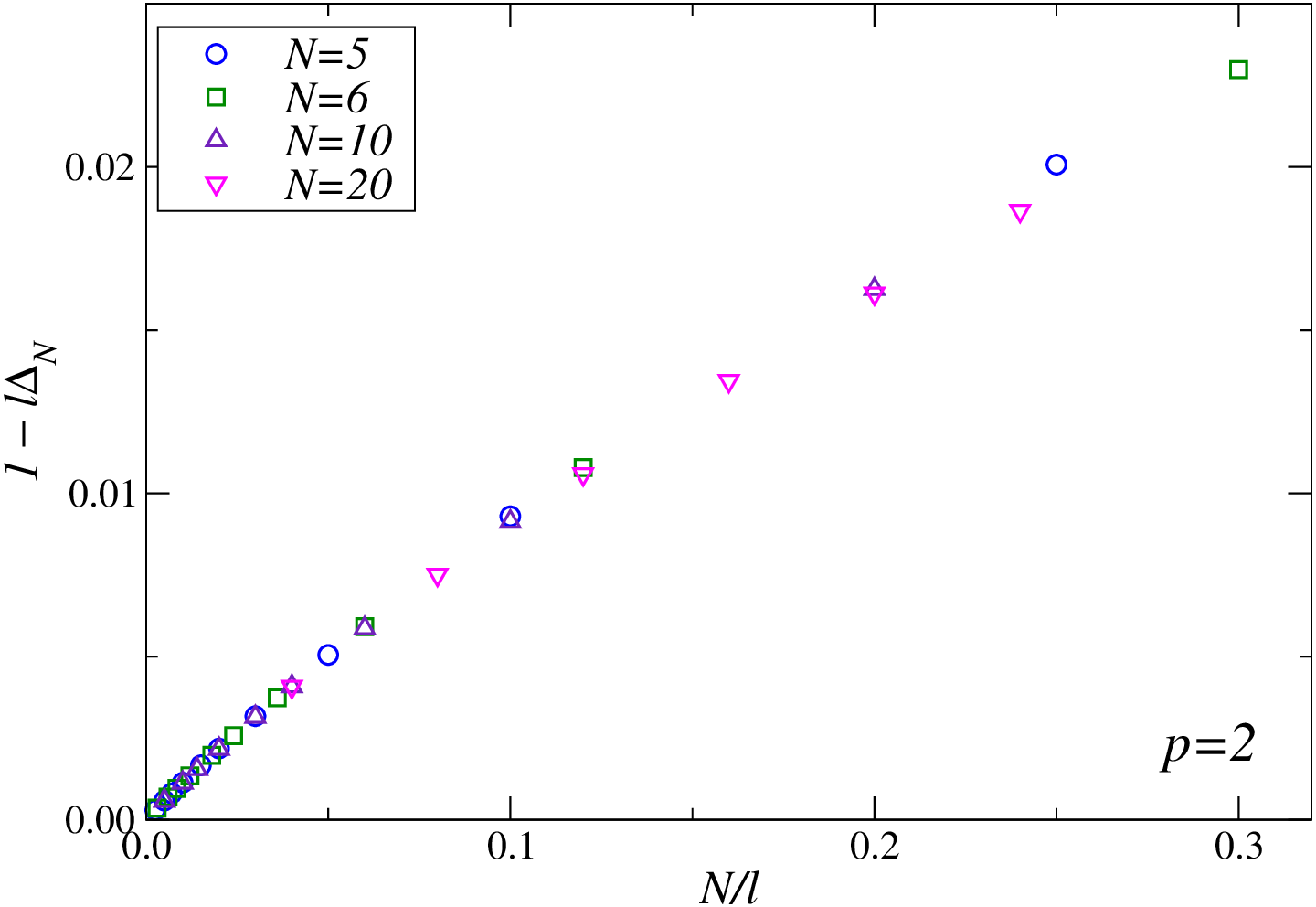}
\caption{ (Color online)
  Some results for the gap of the BH model with $U=2$ and $N=5,\,6,\,10,\,20$.
  They show that the data approach the asymptotic value $l\Delta_N=1$ with
  increasing $l$, with $O(l^{-1})$ corrections.  }
\label{gaphb}
\end{figure}

Fig.~\ref{gaphb} shows results for the gap, i.e., the difference between the
energy of the lowest states. The data show a behavior analogous to that found
analytically in the HC limit, see Sec.~\ref{tssobs}, i.e., $l\Delta_N =
1 + O(l^{-1})$ independently of the particle number $N$ [the large-$l$
extrapolation to get the leading behavior is checked within an accuracy of
$O(10^{-6})$]. This shows that there is no need of a $U$-dependent
normalization of the trap size.

\begin{figure}[tbp]
\includegraphics*[scale=\graphicscale]{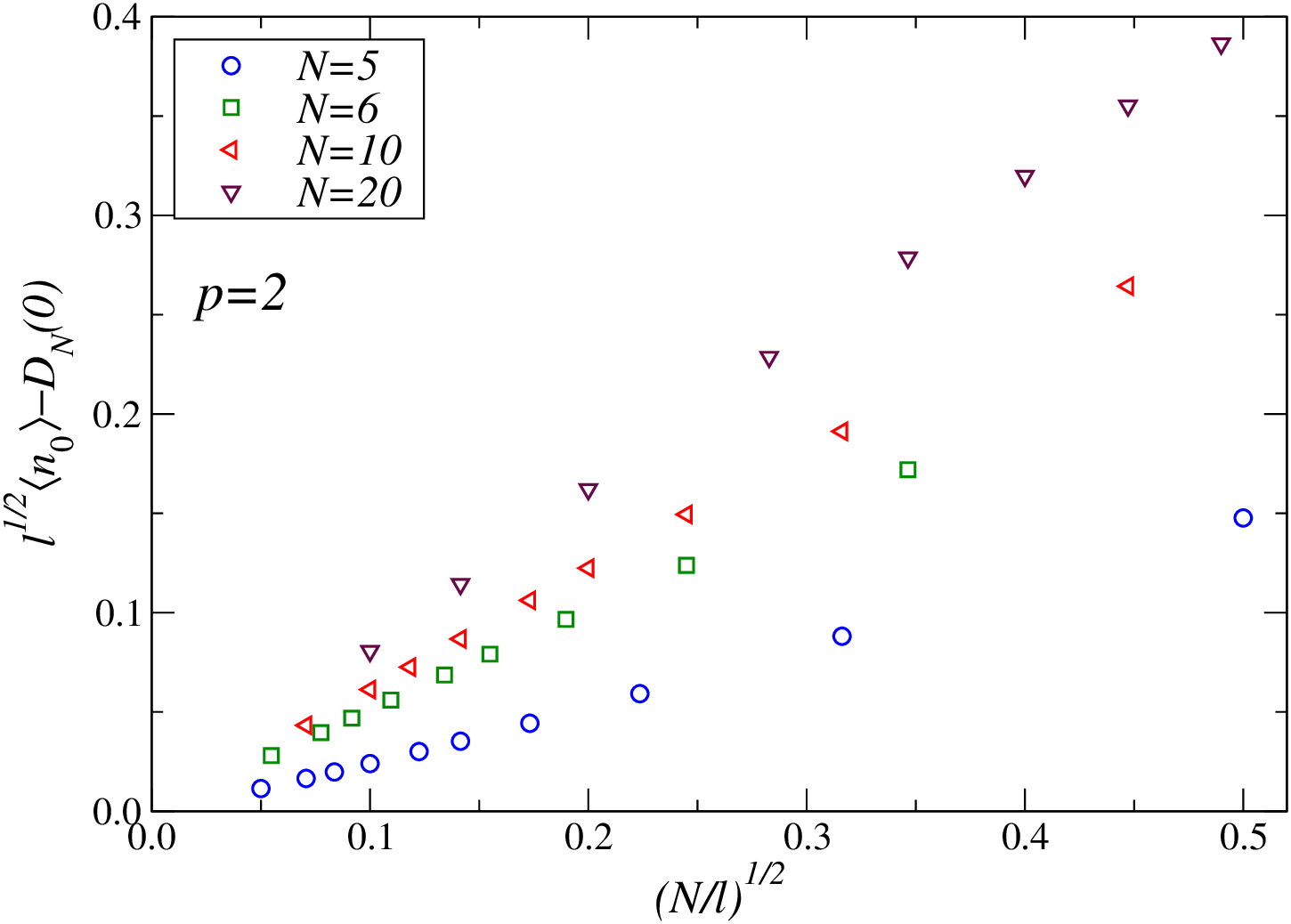}
\caption{ (Color online) The difference $l^{1/2} \langle n_0
  \rangle-{\cal D}_N(0)$ vs $(N/l)^{1/2}$, for the BH model with $U=2$ and
  $N=5,\,6,\,10,\,20$.  These results provide a clear evidence that the
  asymptotic value is consistent with ${\cal D}_N(0)$, and it is approached
  with $O(l^{-1/2})$ corrections.}
\label{rho0hb}
\end{figure}

\begin{figure}[tbp]
\includegraphics*[scale=\graphicscale]{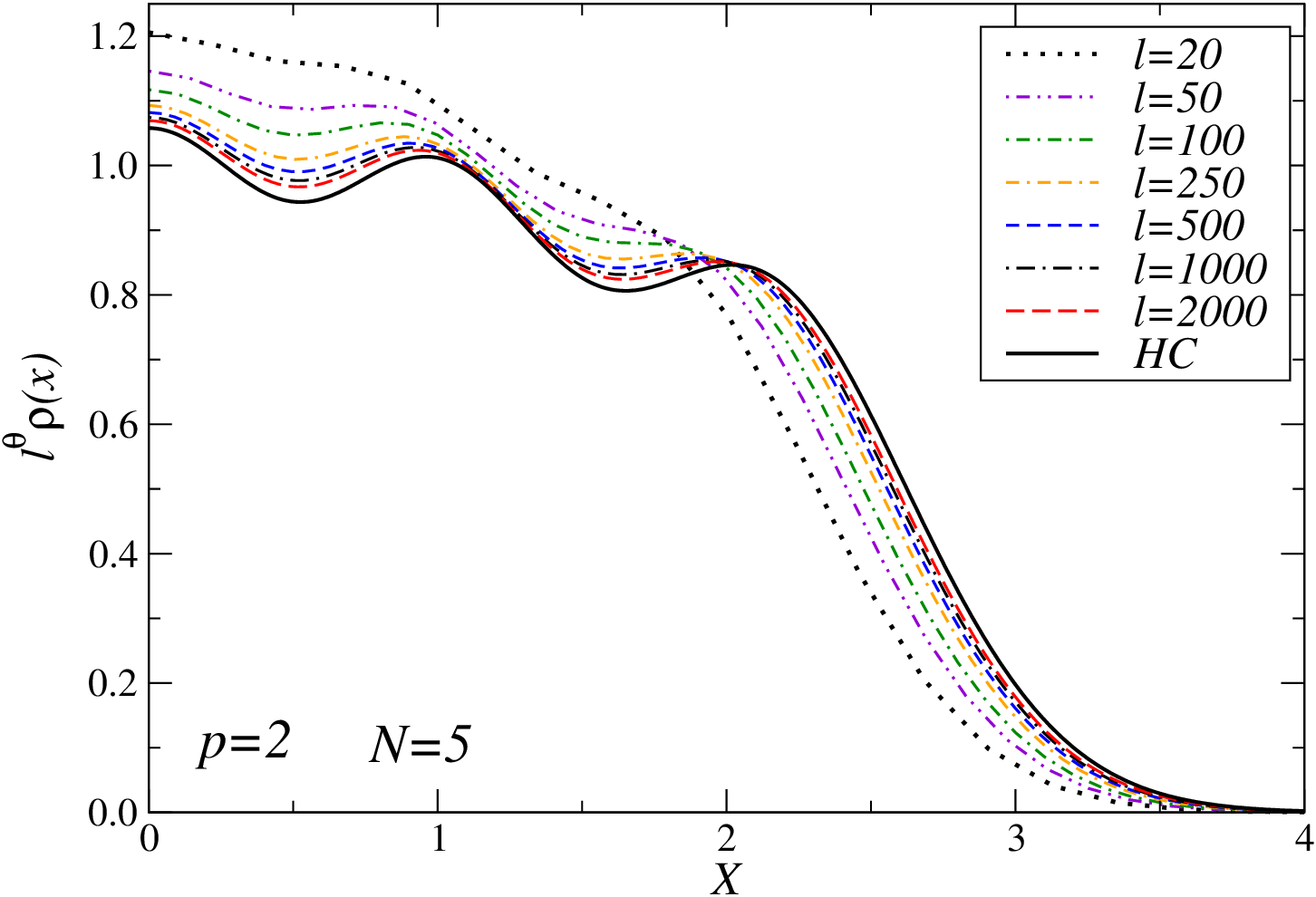}
\includegraphics*[scale=\graphicscale]{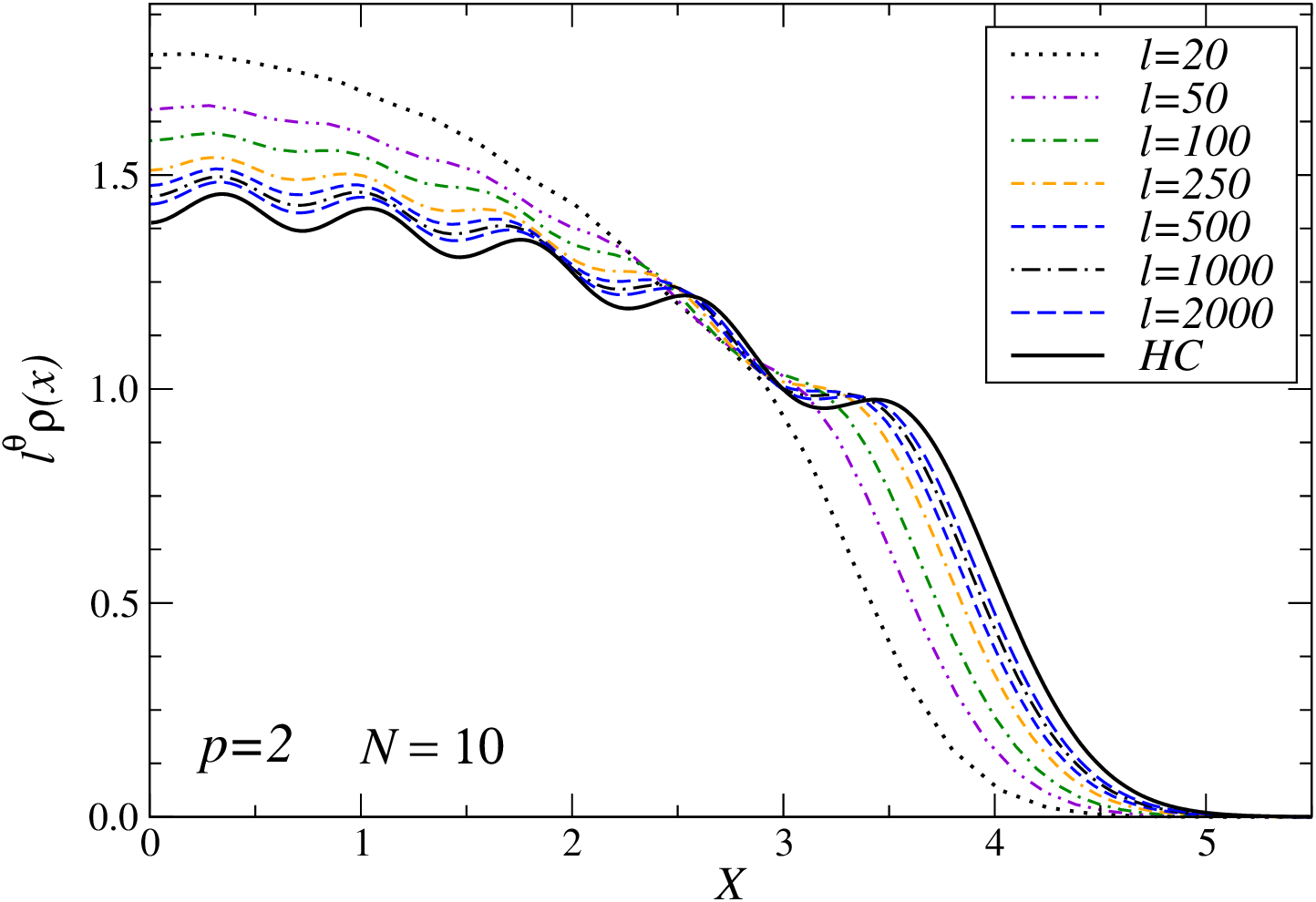}
\caption{ (Color online)
  Some results for the spatial depencence of the particle density in the BH
  model with $U=2$ and for $N=5$ (above) and $N=10$ (below).  They clearly
  approach the large-$l$ limit of the HC model, although their
  convergence appears significantly slower than that found in the HC limit.
}
\label{dnxp2hb}
\end{figure}

\begin{figure}[tbp]
\includegraphics*[scale=\graphicscale]{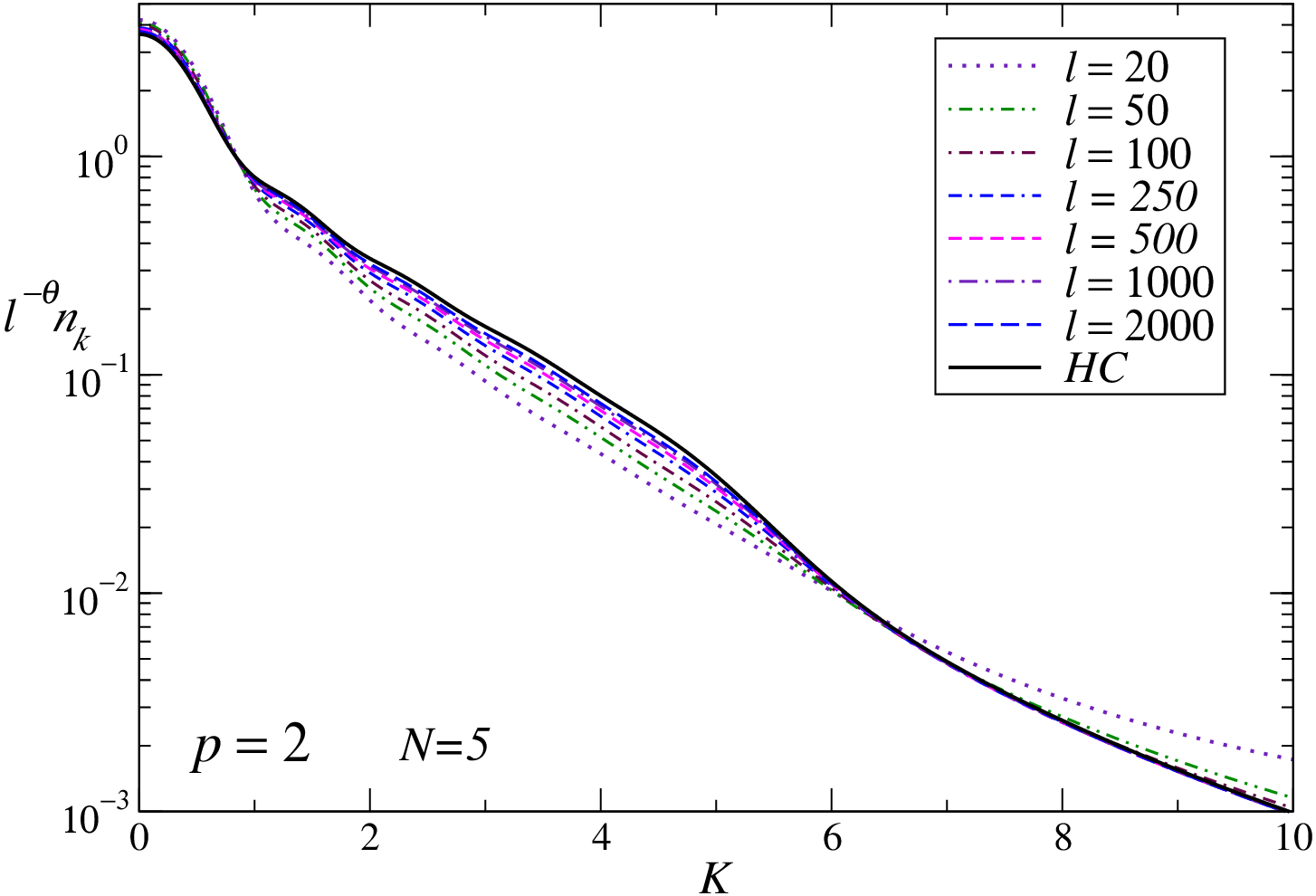}
\includegraphics*[scale=\graphicscale]{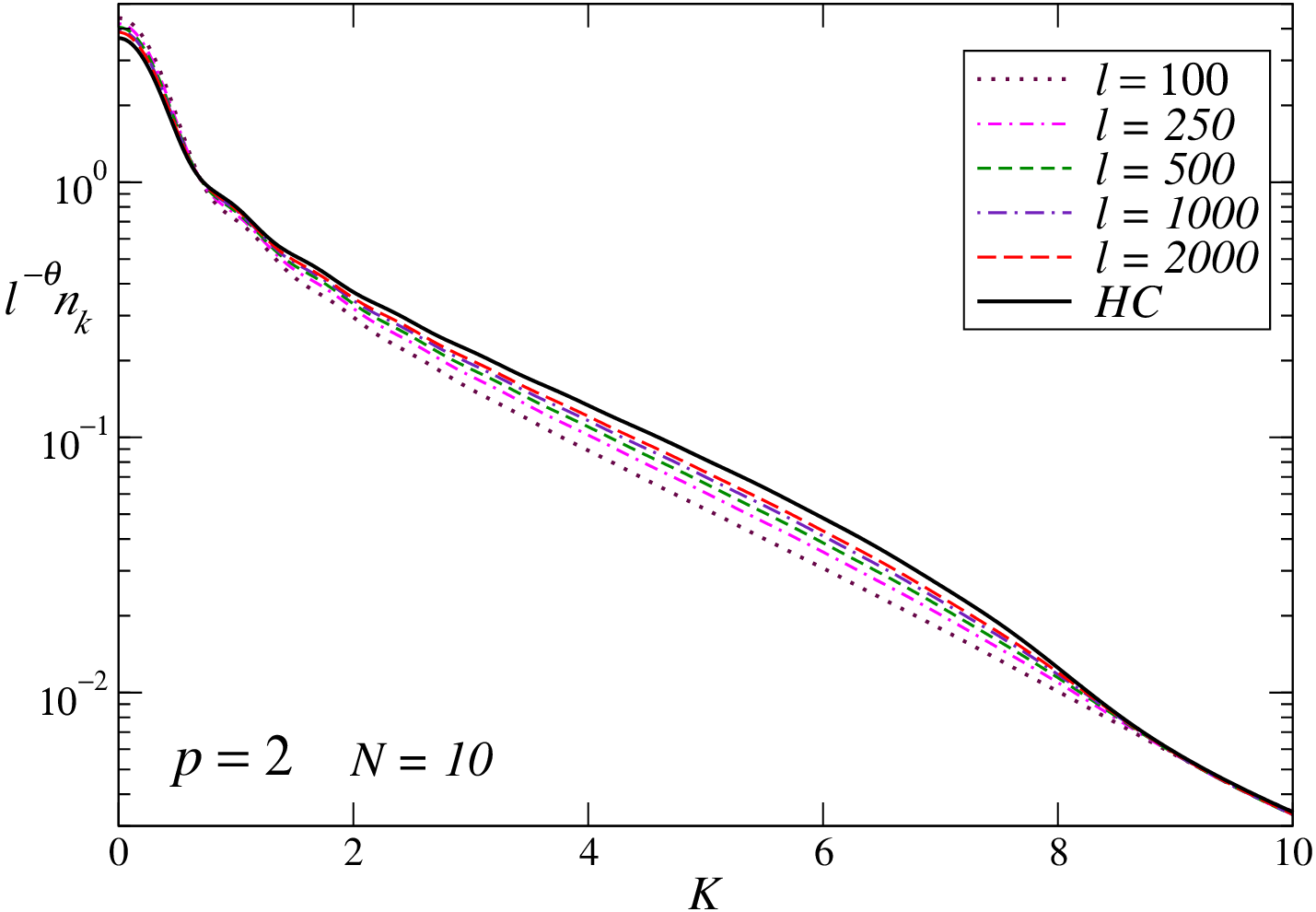}
\caption{ (Color online)
Some results for the BH model with $U=2$ for $N=5$ (above) and
$N=10$ (below).
}
\label{nktsshb20}
\end{figure}

Fig.~\ref{rho0hb} shows data for the particle density at the origin.
They are consistent with
\begin{eqnarray}
l^{1/2} \rho(0) = {\cal D}_N(0) \left[1+ O(l^{-1/2})\right],
\label{nxnhb}
\end{eqnarray}
where ${\cal D}_N(0)=\sqrt{2N}/\pi$, using Eqs.~(\ref{dnphi}) and
(\ref{eq:p2eig}).  Note that the power law of the scaling corrections
differs from that found in the HC limit, which was $O(l^{-2\theta})$,
thus $O(l^{-1})$ for $p=2$.  Therefore the approach to the asymptotic
behavior is significantly slower than that for the HC limit.  This is
also found for other observables.  Results for the spatial dependence
of the particle density are shown in Fig.~\ref{dnxp2hb} for $N=5$ and
$N=10$.  They clearly approach the scaling function ${\cal D}_N(X)$
with increasing the trap size.  Finally, in Fig.~\ref{nktsshb20} we show
results for the momentum distribution at several values of $N$, and
compare them with the asymptotic behavior computed in the HC limit.

In conclusion, these results confirm that the low-density TSS of $N$ particles
described by the BH model is universal with respect to the on-site
repulsion coupling $U$.  However, scaling corrections at finite $U$ appear
generally larger than those of the HC limit, $O(l^{-\theta})$ for generic
values of $U$ against $O(l^{-2\theta})$ in the $U\to\infty$ HC limit.  The
only exception was the gap where we have not found evidence of $O(l^{-1/2})$
corrections.

Actually, $O(l^{-\theta})$ corrections are generally expected, because the RG
dimension of the parameter $U$ has RG dimension $y_U=-1$ with respect to the
low-density scaling. This implies that it generally leads to $O(\xi^{-1})$
scaling corrections, which become $O(l^{-\theta})$ in terms of the trap
size.~\footnote{ The RG dimension of $U$ can be derived from the
  $\beta$-function associated with the quartic term of the corresponding
  bosonic continuum theory (\ref{lb}), which reads $\beta(u) = u - u^2/2$
  exactly~\cite{Sachdev-book}. It has a nontrivial fixed point for $u^*=2$,
  thus $y_U=\beta^\prime(u^*)=-1$.}  These corrections vanish in the HC limit.
Thus, within generic 1D BH models, the HC limit represents a RG-improved
model~\cite{PV-02} where the leading scaling corrections are absent.

\section{The 1D bosonic gas at low density}
\label{bosgas}

We now consider a system of 1D boson particles interacting through a repulsive
contact term in the presence of a confining potential such as
(\ref{potential}), described by the LL model, cf.  Eq.~(\ref{LLHam}). In the
low-density regime the system can be effectively described by the limit of
infinitely strong repulsive interaction~\cite{PSW-00}, i.e., a 1D gas of
impenetrable bosons (TG model).

The wave function for a 1D system of $N$ impenetrable bosons in a confining
potential is essentially defined by the one-particle Hamiltonian
\begin{equation}
H = {p^2\over 2m} + {1\over p} m \omega^p x^p
\label{oh}
\end{equation}
and the impenetrability condition, i.e., the fact that the wave function of
the $N$ particles  vanishes if two spatial variables coincide.  The
low-density condition to realize a 1D TG gas of impenetrable bosons
is~\cite{PSW-00,BDZ-08} $N a_s^2/l_{\rm osc}^2<< 1$ where $a_s$ is the 1D
scattering length, related to the quartic coupling by $g=-{4\hslash^2/(m^2
  a_s)}$, and $l_{\rm osc}\equiv \hslash^{1/2}/(m\omega)^{1/2}$ is the
oscillator length.  In the following we set $\hslash=1$ and $m=1$.

The ground-state wave function of the TG model can be written in terms of the
ground state wave function of $N$ free fermion
particles~\cite{Girardeau-60,Girardeau-65} described by the Hamiltonian
(\ref{oh}), which is
\begin{equation}
\Psi(x_1,...,x_N) = {1\over \sqrt{N!}} {\rm det} [\phi_i(x_j)],
\label{fpsi}
\end{equation}
where $\phi_i(x)$ are the lowest $N$ eigensolutions of the one-particle
Schr\"odinger equation 
$H\phi_i=E_i \phi_i$.
The wave function $\Phi$ of
$N$ impenetrable bosons is obtained by {\em symmetrizing} the fermion wave
function $\Psi$, i.e.,
\begin{eqnarray}
&&\Phi(x_1,...,x_N) = {\cal A}(x_1,...,x_N)  \Psi(x_1,...,x_N), 
\label{phieqbos}\\
&&{\cal A}(x_1,...,x_N) = \prod_{1\le i<j\le N}{\rm sign}(x_i-x_j) .
\end{eqnarray}
The ground-state wave function allows us to derive the the one- and
two-particle density matrices by
\begin{equation}
\rho_1(x,y) =  N \int \Phi(x,x_2,...,x_N)^* \Phi(y,x_2,...,x_N) 
dx_2...dx_N
\label{rhonbos}
\end{equation}
and
\begin{eqnarray}
&&\rho_2(x_1,x_2;y_1,y_2) = \label{twopmd} \\
&&N^2 \int \Phi(x_1,x_2,x_3,...,x_N)^* 
\Phi(y_1,y_2,x_3,...,x_N)  dx_3...dx_N.
\nonumber
\end{eqnarray}

In the case of the harmonic potential, we have
\begin{eqnarray}
&&E_k = \omega (k + 1/2), \quad k\ge 0, \label{Ekphi}\\
&&\phi_k(x) = \omega^{1/4}{H_k(\omega^{1/2} x)\over
 \pi^{1/4} 2^{k/2} (k!)^{1/2}} \, e^{-\omega x^2/2}, 
\nonumber 
\end{eqnarray}
thus leading to~\cite{GWT-01}
\begin{eqnarray}
&&\Phi(x_1,...,x_N) = c_N \omega^{N^2/4} B(x_1,...,x_N)
e^{-\sum_i \omega x_i^2/2}, \nonumber\\
&&B(x_1,...,x_N) = \prod_{1\le i<j\le N}|x_i-x_j|, 
\label{phisolu} 
\end{eqnarray}
where $c_N$ is the appropriate normalization constant 
\begin{equation}
c_N = \pi^{-N/4} \left[ N! \prod_{k=0}^{N-1} 2^{-k} k!\right]^{-1/2},
\label{cnco}
\end{equation}
so that $\int \prod_{i=1}^N dx_i |\Phi|^2=1$.
Some useful analytical developments, to evaluate the one-particle density
matrix, can be found in 
Refs.~\cite{LGW-02,FFGW-02,Papenbrock-03,FFGW-03,Gangardt-04,GFF-05}.

We now note that, after appropriate rescalings, the above results for the TG
model concide with the low-density TSS of the BH model, see Sec.~\ref{TSSlim},
which was derived by taking the continuum TSS limit of its HC limit.  This
implies that the trap-size dependence of the TG model exactly gives the
asymptotic TSS of BH model, after replacing~\footnote{Restoring the
  dependences on $J$, $a$ and $m$ in the BH and TG models, the trap-size
  correspondence between the {\em trap sizes} of the BH and TG models is
\begin{equation}
a^{2/p} l = {a^{2/p} J^{1/p}\over v} \leftrightarrow  {\hslash^{2/p}\over 
m^{2/p} \omega},
\label{tpcorr}
\end{equation}
thus $l=\omega^{-1}$ setting $J,a,m$ to one.  Therefore, the trap size of the
TG model in a harmonic potential is essentially given by $l\sim
\hslash/(m\omega)$, thus $l\sim l_{\rm osc}^2$ where $l_{\rm osc}\equiv
\sqrt{\hslash/(m\omega)}$ is the characteristic length scale of an oscillator
of frequency $\omega$.}  
$l=\omega^{-1}$.
This can be verified by explicit calculations, see below.

Straightforward calculations lead to the following expressions for the
particle density and its correlator:
\begin{eqnarray}
\rho(x)  \equiv \rho_1(x,x) = l^{-\theta} {\cal D}_N(X) 
\label{dnbos}
\end{eqnarray}
and
\begin{eqnarray}
&&G_n(x,y) = \langle n_{x} n_{y} \rangle_c  = \label{gnbos}\\
&&= \rho_2(x,y;x,y) -   
\rho_1(x,x)\rho_1(y,y) \nonumber \\
&&=l^{-2\theta} {\cal G}_N(X,Y)
\nonumber
\end{eqnarray}
where $\theta=p/(p+2)$ is the trap exponent already introduced in the
low-density TSS of the BH model, $X=x/l^{\theta}$ and $Y=y/l^{\theta}$, and
the TSS functions ${\cal D}_N$ and ${\cal G}_N$ are exactly given by
Eqs.~(\ref{dnphi}) and (\ref{eq:GnTSS}).

The one-particle density matrix can be written as
\begin{equation}
\rho_1(x,y) = l^{-\theta} {\cal M}_N(X,Y).
\label{rhonresc}
\end{equation}
Again, this exactly provides the large-$l$ TSS of the BH model, as
shown by results for the TG model and the HC-BH model in
Fig.~\ref{gbp2}.  In particular, its large-$N$ limit is given by
Eq.~(\ref{rho1lnscalbeh}), which implies that the rescaled density
matrix $(N/l)^{-\theta} \rho_1(x,y)$ has a nontrivial large-$N$ limit
$B(\zeta,\delta)$ keeping $\zeta\equiv x N^{-1+\theta} l^{-\theta}$
and $\delta\equiv (y-x) N^{-\theta} l^{-\theta}$ fixed.  This scaling
behavior was already noted in Ref.~\cite{Papenbrock-03}.

We can also compute the energy difference $\Delta_N$ between the two lowest
states.  The lowest excited state above the ground state is obtained by {\em
  exciting} only the {\em fermion} particle with the highest energy in the
ground state.  One can easily check that $\Delta_N$ is exactly given by the
asymptotic TSS behavior found for the HC-BH model, cf. Eq.~(\ref{deltaN}),
without corrections. In particular, $\Delta_N = 1/l$ for the harmonic
potential.

Summarizing, we have shown that the trap-size dependence in a 1D
trapped gas of $N$ impenetrable bosons coincides with the asymptotic
TSS of $N$ particles described by the 1D BH model, if appropriate
definitions of the trap size are considered. As already argued within
the BH model, the critical exponents associated with this TSS are
related to the {\em nonrelativistic} $\Phi^4$ theory (\ref{lb}). We
expect that the low-density TSS is also universal with respect to the
strength of the short-ranged repulsive interaction. Therefore, it
should exactly provide the asymptotic low-density trap-size dependence
of $N$ boson particles described by the LL model, when $Na_s^2/l_{\rm
osc}^2\ll 1$.

Note that the power-law TSS does not have corrections in trapped systems of
impenetrable bosons, while within the BH model it is only expected
asymptotically in the large-$l$ limit, i.e., it is approached with
$O(l^{-2\theta})$ corrections in the HC limit and $O(l^{-\theta})$ corrections
for finite $U$.  In a sense, in the language of the RG
theory~\cite{Wilson-82}, the TG model represents a fixed-point Hamiltonian,
i.e., a model where scaling corrections are totally absent, with respect to
the low-density behavior of the BH model and the LL gas. Using the same RG
arguments reported at the end of Sec.~\ref{finiteU}, we predict that scaling
corrections are $O(l^{-\theta})$ in the LL model.

\section{Trap-size scaling in a time-dependent trap}
\label{dynTSS}

The off-equilibrium dynamics is a quite complicated issue, more subtle than
issues related to the equilibrium behavior.  This is not a prerogative of the
quantum evolution only, but it is also found in classical systems, see, e.g.,
Refs.~\cite{JSS-89,CG-05}.

In this section we discuss the trap-size dependence of the off-equilibrium
time evolution of 1D bosonic gases in time-dependent traps, in the limit of
instantaneous variations and for a power-law time dependence, starting from
the equilibrium ground state for a initial trap size $l_0$.  We derive scaling
Ansatz for the asymptotic TSS with respect to the
initial trap size $l_0$ in the large-$l_0$ limit.

\subsection{Instantaneous variation of the confining potential}
\label{tssisqu}

Let us first discuss the case of an instantaneous change of the confining
potential. In particular, we assume that at $t=0$ the $N$-particle system is
at equilibrium, in the ground state with a confining potential of trap size
$l_0$.  Then, the trap is instantaneously changed to a larger trap size, $l_f
> l_0$, or dropped completely, corresponding to $l_f=\infty$. We are
interested in the asymptotic trap-size dependence of the quantum time
evolution after the instantaneous quench for large initial trap size $l_0$.

In this case we expect that the trap-size dependence of the off-equilibrium
dynamics after the quench is essentially determined by the trap-size
dependence of the inizial state at equilibrium, and by the ratio $l_f/l_0$ of
the final and initial trap sizes.  Thus, the simplest scaling Ansatz
for the large-$l_0$ behavior may be
\begin{equation}
\langle O\rangle_N(x;t) \approx
l_0^{-y_o \theta}
{\cal O}_N(x l_0^{-\theta}, t l_0^{-z\theta},l_f/l_0),
\label{scalbehNqu}
\end{equation}
where $z=2$ and $\theta=p/(p+2)$ is the equilibrium trap exponent.
For example, this would imply that the particle density 
of 1D and 2D bosonic gases behaves as
\begin{eqnarray}
\rho(x;t) \approx l_0^{-d\theta} 
{\cal D}_N(x l_0^{-\theta}, t l_0^{-z\theta},l_f/l_0).
\label{dnphitssqu}
\end{eqnarray}

This scaling Ansatz will be confirmed by the time evolution of a 1D gas of
impenetrable bosons, see Sec.~\ref{instdun}.  We also expect that, like the
equilibrium behavior, 1D impenetrables boson gases and 1D BH models of $N$
particles share the same TSS of the off-equilibrium dynamics after
instantaneous variations of the trap.

\subsection{Power-law time dependence of the confining potential}
\label{tsspltd}

A non trivial time dependence of the confining potential makes the issue more
complicated.  In the following, we consider a power-law time dependence, such
as that given by Eqs.~(\ref{tidpo}) and (\ref{ktaudef}).  We are again
interested in the asymptotic trap-size dependence of the quantum time
evolution for large initial trap size $l_0$.

We use RG scaling arguments to infer the scaling behavior of the time
dependence of generic observables under a change of the confining potential.
For this purpose, we write the perturbation associated with the
time-dependent confining potential to the continuum theory (\ref{lb}), i.e.,
\begin{equation}
\int d\tau d^dx \, u^p \,\tau^q \,|x|^p \,|\phi(x,\tau)|^2,
\label{pertV}
\end{equation}
where $\tau$ indicates a time variable. We are interested in the scaling
behavior at fixed $N$, large $l_0$, with $N/l_0\ll 1$.
The RG arguments of Sec.~\ref{TSSeq}, see also Ref.~\cite{CV-10}, may be
extended to allow for the presence of a time-dependent perturbation
(\ref{pertV}), and derive an off-equilibrium scaling Ansatz.  A standard
analysis of the RG dimensions of the coupling $u$ leads to~\cite{CV-10,CK-10}
$y_u = (2+p+zq)/p$.  It is convenient to introduce the initial trap size at
$t=0$, $l_0=1/u$, with the corresponding RG dimension
\begin{equation}
\theta_0 = {p\over 2+p+zq}.
\label{theta0}
\end{equation}

Let us consider an operator $O$ whose low-density critical behavior of its
matrix elements is described by the RG dimension $y_o$ in the homogeneous
system.  In the presence of a chemical potential, the simplest Ansatz for the
large-$l_0$ off-equilibrium behavior, which may be derived from the above RG
scaling arguments, is
\begin{equation}
\langle O\rangle (\mu,x;t) \approx l_0^{-y_o \theta_0}
A_0(x l_0^{-\theta_0}, \tau l_0^{-z\theta_0},\bar{\mu} l_0^{y_\mu\theta_0}).
\label{scalbeh}
\end{equation}
The corresponding Ansatz for the low-density TSS at fixed particle number $N$
is
\begin{equation}
\langle O\rangle_N(x;t) \approx l_0^{-y_o \theta_0}
{\cal O}_N(x l_0^{-\theta_0}, \tau l_0^{-z\theta_0}).
\label{scalbehN}
\end{equation}
For example, the application to the one-particle density matrix reads
\begin{equation}
\rho_1(x_1,x_2;t)\approx 
l_0^{-\theta_0}
{\cal M}_N(x_i l_0^{-\theta_0}, \tau l_0^{-z\theta_0}).
\label{opd}
\end{equation}

We warn that these scaling behaviors neglect possible relevant effects related
to the initial conditions, which may not allow us to take the $l_0\to\infty$
limit of the scaling functions ${\cal O}_N$, more precisely of the product
$l_0^{y_o\theta_0} \langle O\rangle_N(x;t)$ after the variable rescalings
$X\equiv x l_0^{-\theta_0}$ and $Z\equiv \tau l_0^{-z\theta_0}$.

The above scaling Ansatz can be reexpressed in terms of the instantaneous
trap size
\begin{equation}
l(t) = l_0 \tau^{-q/p}.
\label{itp}
\end{equation}
Replacing it in Eq.~(\ref{scalbehN}), we can write
\begin{equation}
\langle O\rangle_N(x;t) \sim  
l(t)^{-y_o \theta}
\widetilde{{\cal O}}_N(x l(t)^{-\theta}, \tau l(t)^{-z\theta}).
\label{scalbehNb}
\end{equation}

The RG arguments leading to the scaling Ansatz (\ref{scalbeh}) and
(\ref{scalbehN}) are quite general and can be applied to other models.  In
Ref.~\cite{CK-10} they were applied to the XY chain in a space- and
time-dependent trasverse field.  In the following we challenge them against
the off-equilibrium evolution of 1D bosonic particle systems.

\section{$N$ particles in a slowly time-dependent trap}
\label{slowdyn}

We here discuss the behavior of $N$ particles in a time-dependent confining
potential, which varies slowly, i.e., with a large parameter $t_q$ in
Eq.~(\ref{tidpo}), and for sufficiently large trap sizes to be in the
low-density regime.  More precisely, we assume that the external potential
is slowly varied so that the trap size slowly increases, corresponding to
the limit $t_q\to -\infty$ in Eq.~(\ref{ktaudef}), thus $l(t)\to \infty$ for
$t \to |t_q|$.

\subsection{Adiabatic evolution}
\label{addyn}

In the case of slow changes of the Hamiltonian parameters, the system
undergoes a quasi-equilibrium dynamics, i.e., starting from the ground state
at $t=0$, the evolution of the system passes through the instantaneous ground
states of the BH Hamiltonian with the confining potential $V(r,t)$ and trap
size $l(t)$.  We write the solution of the Schr\"odinger equation, 
\begin{equation}
i\partial_t \Psi(t) = {\cal H}(t)\Psi(t),
\label{seqpsi}
\end{equation}
in terms of the instantaneous eigenstates $\phi_n$ of the time-dependent
Hamiltonian (whose spectrum is discrete for any finite trap size), 
where $\phi_n(t)$ are solutions of 
\begin{equation}
{\cal H}(t)\phi_n(t) = E_n(t) \phi_n(t).
\label{iexp}
\end{equation}
Starting at $t=0$ from the ground state of the Hamiltonian at $t=0$, i.e.,
$\Psi(0)=\phi_0(0)$, and writing 
\begin{equation}
\Psi(t) = e^{-i\Theta_0(t)} \sum_n \alpha_n(t) \phi_n(t)
\label{redefpsi}
\end{equation}
where 
\begin{equation}
\Theta_n(t) = \int_0^t E_n(t^\prime)dt^\prime,
\label{thetadef}
\end{equation}
the zero-order adiabatic approximation gives 
\begin{equation}
\alpha_{n}(t) = \delta_{n0}. 
\label{0ordan}
\end{equation}
Note that the adiabatic quasi-equilibrium evolution requires the
absence of degeneracies and level crossings during the process, but
the {\em instantaneous} gap vanishes when $\tau\equiv 1+t/t_q\to 0$.
Thus we may already expect that, approaching the time corresponding to
$\tau=0$, the adiabiatic condition breaks down at some point of the
evolution.  We return to this point later.

Under the quasi-equilibrium dynamics due to the slow increasing of the
trap size, the particle number $N$ is conserved because the particle
number operator commutes with the time-dependent Hamiltonian. Since
the system passes through equilibrium ground states, we can use the
results obtained for the equilibrium TSS, see Secs.~\ref{TSSeq},
\ref{static}, and \ref{bosgas}.  The adiabatic evolution of a generic
observable $O$ can be obtained by computing its expectation values
over the {\em instantaneous} ground states. After a sufficiently large
time, when $N/l(t)\ll 1$, we are in the low-density regime, thus the
adiabatic time-dependence of the observables is obtained from the
static low-density behaviors, such as (\ref{deltaN}) and (\ref{nxn}),
by replacing $l$ with the instantaneous trap size (\ref{itp}), i.e.,
\begin{equation}
\langle O\rangle_{\rm adiab}(x;t) 
\sim  l(t)^{-y_o \theta} {\cal O}_N(x l(t)^{-\theta}),
\label{adsca}
\end{equation}
whee $l(t)$ is the instantaneous trap size (\ref{itp}).
Note that this is compatible with the dynamic TSS derived in
Sec.~\ref{dynTSS}, cf. Eqs.~(\ref{scalbehNb}) and (\ref{scalbehN}).

At large $N$, we may use the relation (\ref{cmu}) to define a time-dependent
chemical potential $\mu(t)$ at any $t$, by replacing $l$ with $l(t)$, along
the quasi-equilibrium evolution.  For example in the 1D HC-BH model, since
$l(t)\to\infty$ for $t\to\infty$ and therefore $N/l(t)\to 0$, we have that
$\mu(t)\to 1$, which is the location of the low-density to empty-state
transition.  Within the adiabatic dynamics, the time behavior of the
observables related to the ground state can be read from that at equilibrium,
by replacing the instantaneous trap size $l(t)$ and chemical potential $\mu(t)$,
obtained from Eq.~(\ref{cmu}), in the
corresponding TSS formulae obtained in Ref.~\cite{CV-10-2}.  It is then
convenient to define
\begin{equation}
\bar{\mu}(t)\equiv \mu(t)-1,
\label{mubdef}
\end{equation}
which tends to zero from below in the large-$t$ limit (after $t_q\to -\infty$).
Asymptotically, when $|\bar{\mu}(t)|\ll 1$, the time dependence corresponds to
varying the trap size $l(t)$ so that 
\begin{equation}
|\bar{\mu}(t)| l(t)^{2\theta} = b N^{2\theta} \left\{
1 + O[(N/l)^{2\theta)}]\right\}, 
\label{ap}
\end{equation}
where $b$ is a $p$-dependent constant
which can be easily derived from Eq.~(\ref{cmu}), for example $b=1$ for $p=2$.  
Note that the l.h.s. of Eq.~(\ref{ap}) corresponds to the rescaled
chemical potential $\mu_r\equiv l^{2\theta} \bar{\mu}$,
cf. Eq.~(\ref{muresc1}), and that it remains constant during the
adiabiatic changes since the r.h.s. remains fixed, apart from
suppressed corrections.

\subsection{First-order adiabatic perturbation theory
and breaking of the adiabatic condition}
\label{foapt}

We may also consider the first-order correction to Eq.~(\ref{0ordan}) within
the adiabatic perturbation theory, see, e.g., Refs.~\cite{Schiff-book,DP-10}.
The first-order approximation of the coefficients for $n>0$ of the expansion
(\ref{redefpsi}) over instantaneous bases is
\begin{eqnarray}
\alpha_n(t) \approx - e^{i\Delta\Theta_{n0}(t)}
\int_0^t dt' \langle n | \partial_{t'} |0\rangle 
e^{-i\Delta\Theta_{n0}(t')},
\label{foac}
\end{eqnarray}
where $\Delta\Theta_{nm} \equiv \Theta_n - \Theta_m$,
and, assuming
nondegenerate states,
\begin{equation}
\langle n | \partial_{t} |0\rangle = -
{\langle n | \partial_t {\cal H}  |0\rangle \over E_n(t) - E_0(t)}.
\label{icalc}
\end{equation}

In the low-density regime, we can use the equilibrium TSS developed in the
previous sections, to evaluate the first-order adiabatic approximation of the
coefficients $\alpha_n(t)$.  Energy differences behave as
\begin{equation}
E_n(t)-E_0(t) = e_n l(t)^{-z\theta},
\label{deltabeh}
\end{equation}
where $e_n$ generally depends on the instantaneous eigenstate
$|n\rangle$.  The scaling behavior of the matrix element $\langle n |
\partial_t {\cal H} |0\rangle$ is computed considering $t$ as a
parameter.  We evaluate the matrix element between the ground state
$|0\rangle$ and one of the excited states $|n\rangle$ in the
low-density and TSS limit.  We generally expect
\begin{equation}
\langle n | \partial_t \sum_i [x_i/l(t)]^p b_i^\dagger b_i | 0 \rangle 
  = g_n \partial_t [l(t)^{-z\theta}],
\label{partHme}
\end{equation}
where $g_n$ is a (eigenstate-dependent) constant. Some of the lowest
excited states, and in particular the lowest one, are obtained by
{\em exciting} only the particle with the highest energy in the ground
state, from the one-particle state $k=N-1$ to $k=N-1+2j$ with $j>0$.
In the case of 1D HC model we have
 \begin{eqnarray}
&&\langle e | \sum_i V(x_i) b_i^\dagger b_i | 0 \rangle 
= \langle e | \sum_i V(x_i) \phi_{ki}\phi_{qi} \eta_k^\dagger \eta_q | 0
\rangle  \nonumber \\
&& \approx
l^{-z\theta} \int dX (X^p/p) \varphi_{N-1+2j}(X) \varphi_{N-1}(X), 
\label{llco}
\end{eqnarray} 
where $N$ is the number of particles, and the functions $\varphi_k(X)$ are the
solutions of Eq.~(\ref{trapscaleqxx}). Thus the integral is finite. It
increases as $N^{2\theta}$ at large $N$.  In particular, for $p=2$
\begin{equation}
\int dX (X^2/2) \varphi_{N+1}(X) \varphi_{N-1}(X) = {N\over 4} [1 + 
O(N^{-1})].
\label{p2xp}
\end{equation}
The matrix element $\langle e | \partial_t {\cal H} | 0 \rangle$ is
then obtained by perfoming the time derivative of the r.h.s. of
Eq.~(\ref{llco}), in agreement with Eq.~(\ref{partHme}).  These
results apply also to a 1D gas of impenetrable bosons.

Inserting Eq.~(\ref{deltabeh}) and Eq.~(\ref{llco}) in the first-order
adiabatic expansion (\ref{foac}) of the coefficients $\alpha_n(t)$,
and defining the scaling variable $Z\equiv \tau l_0^{-z\theta_0}$,
where $\theta_0$ is the off-equilibrium trap exponent (\ref{theta0}),
we obtain
\begin{eqnarray}
\alpha_n(t) \approx e^{ie_n t_q Z^{b+1}/b}
\int_{Z_0}^Z {d\zeta\over \zeta} {bg_n\over e_n}   e^{-it_qe_n\zeta^{b+1}/b}
\label{alphaapprox}
\end{eqnarray} 
where $b=zq\theta/p$.  This expression agrees with the scaling Ansatz
(\ref{scalbeh}).

Note that it diverges logarithmically when $Z\to 0$,
\begin{equation}
\alpha_n(t) \approx {b g_n\over e_n} \ln (Z/Z_0)
\sim  \ln[l(t)/l_0] \sim \ln \tau.
\label{dtss}
\end{equation}
Since the adiabatic perturbative expansion requires $|\alpha_n(t)|\ll 1$, it
fails when $\tau$ becomes too small.  This is not unexpected because when
$\tau\to 0$ the spectrum tends to be degenerate.

A simple example of the breaking of the adiabatic evolution when approaching a
Hamiltonian with vanishing instantaneous gap is provided by a quantum
oscillator with a time-dependent frequency, see App.~\ref{oscill}.

\section{1D impenetrable bosons in a
  time-dependent harmonic trap}
\label{bosgasdyn}

In this section we determine the trap-size dependence of the off-equilibrium
evolution of a 1D gas of impenetrable bosonic particles in a time-dependent
confining harmonic potential, i.e., $p=2$, starting from an equilibrium ground
state configuration with initial trap size $l_0$.  We consider  instantaneous
changes to a confining potential with different trap size $l_f$, and also
power-law time dependences such as
\begin{equation}
V(x,t) = {1\over 2} \kappa(t) x^2 ,
\label{vxt}
\end{equation}
where  
\begin{equation}
\kappa(t) = \kappa_0 \tau^q \equiv {1\over l(t)^2},\quad \tau\equiv 1+t/t_q,
\quad \kappa_0 \equiv 1/l_0^{2},
\label{kappat}
\end{equation}
and $t_q$ is a time rate.  In the following analyses of the power-law time
dependence, we set $t_q=1$ for simplicity.

\subsection{Off-equilibrium time evolution}
\label{tddm}

As shown in Ref.~\cite{KSS-96}, see also \cite{MG-05}, the time-dependent wave
function of the system can be derived from the solutions $\psi_j(x,t)$
of the one-particle Schr\"odinger equation
\begin{eqnarray}
i\partial_t \psi_j(x,t) = 
\left[ -{1\over 2}\partial_x^2 + V(x,t)\right] \psi_j(x,t),\label{scoeq}
\end{eqnarray}
with the initial condition $\psi_j(x,0)=\phi_j(x)$ where $\phi_j(x)$ are the
eigensolutions of the Hamiltonian at $t=0$, characterized by a trap size
$l_0$, with eigenvalue $E_j = (j+1/2)/l_0$, cf. Eqs.~(\ref{Ekphi}).  The
solution can be obtained introducing a time-dependent function $s(t)$,
writing~\cite{PP-70,KSS-96}
\begin{eqnarray}
\psi_j(x,t) = && s^{-1/2} \phi_j(x/s) \times \label{psijsol}\\
&&\times {\rm exp}\left( 
i {\dot{s}x^2\over 2 s} - i E_j\int_0^t s^{-2} dt' \right),
\nonumber
\end{eqnarray}
where $\phi_j(x)$ is the $j^{\rm th}$ eigenfunction of the Schr\"odinger
equation of the one-particle Hamiltonian at $t=0$, thus with trap size
$l_0$, and $s(t)$ satisfies the nonlinear differential equation
\begin{equation}
\ddot{s} + \kappa(t) s = \kappa_0 s^{-3}
\label{fdef}
\end{equation}
with initial conditions $s(0)=1$ and $\dot{s}(0)=0$.

The time-dependent wave function $\Phi$ of $N$ impenetrable bosons, with
$\Phi(x,0)$ given by the ground state of the Hamiltonian at $t=0$, can be
obtained following the same steps as at equilibrium, see Sec.~\ref{bosgas},
obtaining~\cite{GW-00}
\begin{eqnarray}
&&\Phi(x_1,...,x_N;t) = {\cal A}(x_1,...,x_N) \Psi(x_1,...,x_N;t), 
\nonumber\\
&& \Psi(x_1,...,x_N;t)= 
{1\over \sqrt{N!}} {\rm det}[\psi_i(x_j;t)],
\label{phieqbosdyn} 
\end{eqnarray}
where the determinant involves the $N$ lowest eigensolution a fixed $t$.
Then, using Eq.~(\ref{psijsol}), one can write the wave function of the
ground state of an $N$-particle system as
\begin{eqnarray}
&&\Phi(x_1,...,x_N;t) =  s^{-N/2}\Phi(x_1/s,...,x_N/s;0)\times
\nonumber \\
&&\times {\rm exp}\left( {i\dot{s}\over 2s}\sum_j x^2_j -
i \sum_j E_j \int_0^t s^{-2} dt' \right),
\label{phisoldyn}
\end{eqnarray}
where $\Phi(x_1,...,x_N;0)$ is the wave function of the ground state for
the Hamiltonian at $t=0$.  The time-dependent one-particle density matrix
reads~\cite{MG-05}
\begin{eqnarray}
&&\rho_1(x,y;t) = \label{rhonbosdyn}\\
&&=N\int \Phi(x,x_2,...,x_N;t)^* 
\Phi(y,x_2,...,x_N;t) dx_2...dx_N
\nonumber \\
&&= s^{-1} \rho_1(x/s,y/s;0) 
{\rm exp}\left[ i{\dot{s}\over 2s}(y^2-x^2)\right],
\nonumber
\end{eqnarray}
where $\rho_1(x,y;0)$ is the equilibrium one-particle density matrix at a
trap size $l=l_0$, i.e.,
\begin{equation}
\rho_1(x,y;0)=\rho_1(x,y)|_{l=l_0},
\label{rho1rhol0}
\end{equation}
computed in Sec.~\ref{bosgas}, cf.  Eq.~(\ref{rhonbos}).

Examples of explicit solutions of the function $s(t)$, cf.  Eq.~(\ref{fdef}),
are the following.

(i) Instantaneous drop of the trap, so that $\kappa(t)=0$ for $t>0$,
\begin{equation}
s(t) = \sqrt{1+\kappa_0 t^2}.
\label{ftqinf}
\end{equation}

(ii) Instantaneous change to a confining potential with trap size $l_f$, so
that $\kappa(t)=l_f^{-2}$ for $t>0$,
\begin{equation}
s(t) = \sqrt{ 1 + (r^2-1) \left[{\rm sin}(\kappa_0^{1/2} t/r)\right]^2} ,
\label{stsol}
\end{equation}
where $r=l_f/l_0$.

\begin{figure}[tbp]
\includegraphics*[scale=\graphicscale]{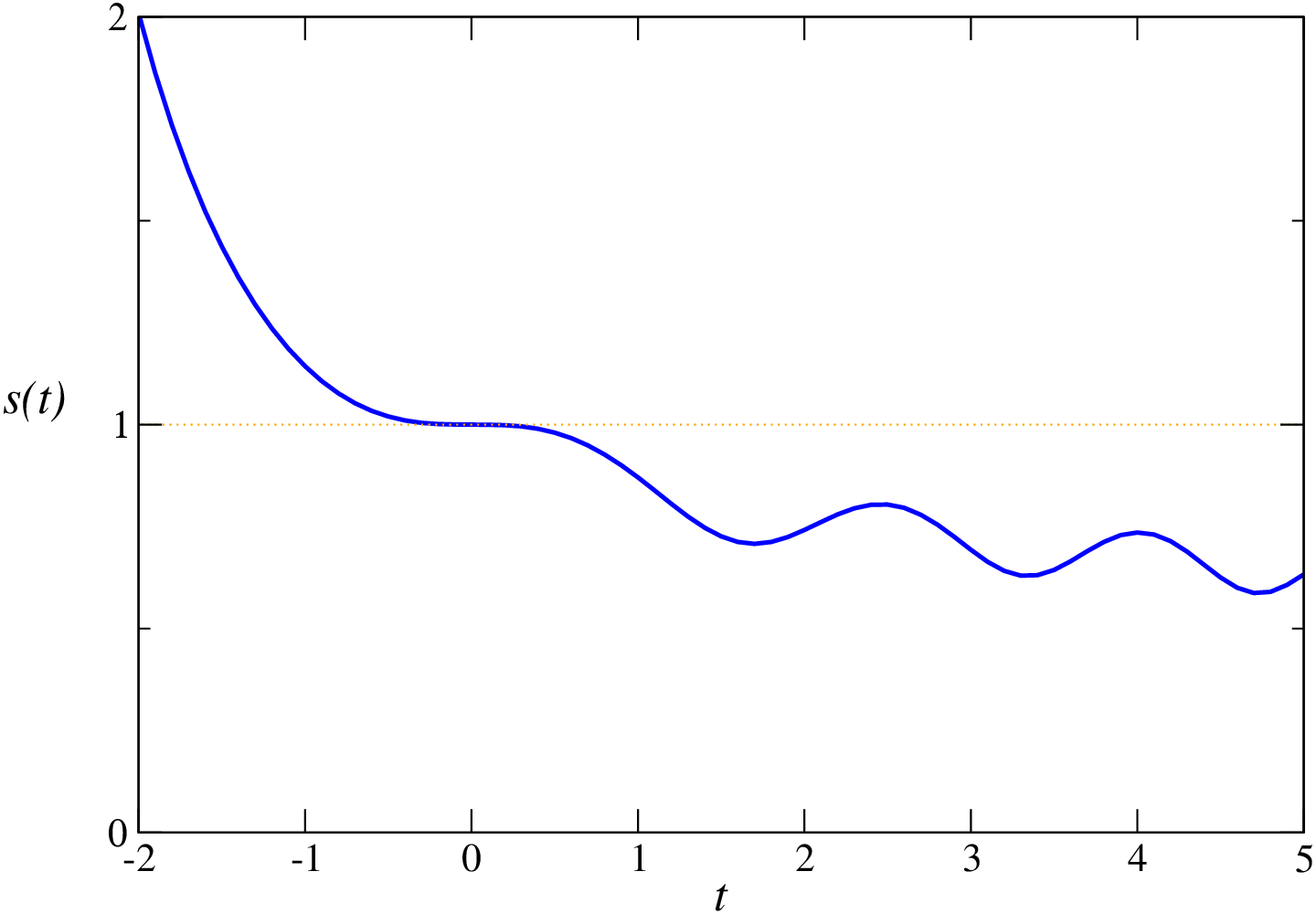}
\caption{ (Color online)
The function $s(t)$ for $\kappa_0=1$, cf. Eq.~(\ref{exsolq1}).  
}
\label{st}
\end{figure}

(iii) Linear time dependence of the trapping potential, i.e., $q=1$ in
Eq.~(\ref{kappat}),
\begin{eqnarray}
s(t) = [{\rm Re} W(\tau)]^{-1/2}, \;\;
\dot{s}(t) = -{\kappa_0^{1/2}{\rm Im} W(\tau)\over [ {\rm Re} W(\tau)]^{1/2}},
\label{exsolq1}
\end{eqnarray}
where the complex function $W(\tau)$ is the solution of the
differential equation
\begin{equation}
i W' = \kappa_0^{1/2} (W^2 - \tau) \label{www} 
\end{equation}
with $W(1)=1$,~\footnote{
In the general case, i.e., when $V(x;t)=\kappa(t)x^2/2$, the replacement (\ref{exsolq1})
leads to the differential equation 
$i W' = \kappa_0^{1/2} W^2 - \kappa_0^{-1/2} \kappa(\tau)$ with $W(1)=1$.}
which can be written as a combination of Airy
functions,~\cite{PG-08}
\begin{eqnarray}
&& W(\tau) =  i \kappa_0^{-1/6}
{{\rm Bi}^\prime(-\kappa_0^{1/3}\tau) + 
c {\rm Ai}^\prime(-\kappa_0^{1/3}\tau)
\over {\rm Bi}(-\kappa_0^{1/3}\tau) + c {\rm Ai}(-\kappa_0^{1/3}\tau)},
\nonumber \\
&& c = - {\kappa_0^{1/6} {\rm Bi}(-\kappa_0^{1/3}) - i 
{\rm Bi}^\prime(-\kappa_0^{1/3})\over {\kappa_0^{1/6}
\rm Ai}(-\kappa_0^{1/3}) - i 
{\rm Ai}^\prime(-\kappa_0^{1/3})}.\label{wtsol} 
\end{eqnarray}
A plot of $s(t)$ is shown in Fig.~\ref{st}.

\subsection{TSS at instantaneous quenches}
\label{instdun}

We now show that the time evolution after instantaneous changes of the trap
size is consistent with the scaling Ansatz put forward in Sec.~\ref{tssisqu}
in terms of the equilibrium trap exponent $\theta=1/2$.

Let us first consider an instantaneous drop of the trap.  The energy after the
quench can be computed within 1D BH model in the TSS limit, by evaluating the
expectation value of the unconfined BH Hamiltonian ${\cal H}_u$ over the
ground state $|0_c \rangle$ of the confined BH Hamiltonian
\begin{equation}
{\cal H}_c = {\cal H}_u + \sum_i V(x_i) n_i 
\label{cuh}
\end{equation}
in the low-density region $N/l\ll 1$.  We have
\begin{equation}
E_i\equiv \langle 0_c | {\cal H}_u |0_c \rangle = 
\langle 0_c | {\cal H}_c - \sum_i V(x_i) n_i |0_c \rangle .
\label{hucm}
\end{equation}
For $p=2$, we have
\begin{eqnarray}
&&E_i= l_0^{-1} \Big[ \sum_{k=0}^{N-1} (k+1/2) -\nonumber\\
&&-\int dX (X^2/2) \sum_{k=0}^{N-1}\varphi_{k}(X)^2\Big]=
{N^2\over 4l_0}.
\label{hucmp2}
\end{eqnarray}
For generic values of $p$, we have $E_i\sim N^{2\theta+1}/l^{2\theta}$
Therefore, for large initial trap size $l_0$, thus $N/l_0\ll 1$, only 
low-energy states are involved.

We again expect that in the low-density regime the asymptotic trap-size
dependence is that of the gas of impenetrable bosons, and that the lattice
structure of the BH model gives only rise to suppressed power-law corrections.
Therefore, in the case of the harmonic trap we can use the general solutions
reported in the previous subsection to derive the TSS behavior at a quench.

\begin{figure}[tbp]
\includegraphics*[scale=\graphicscale]{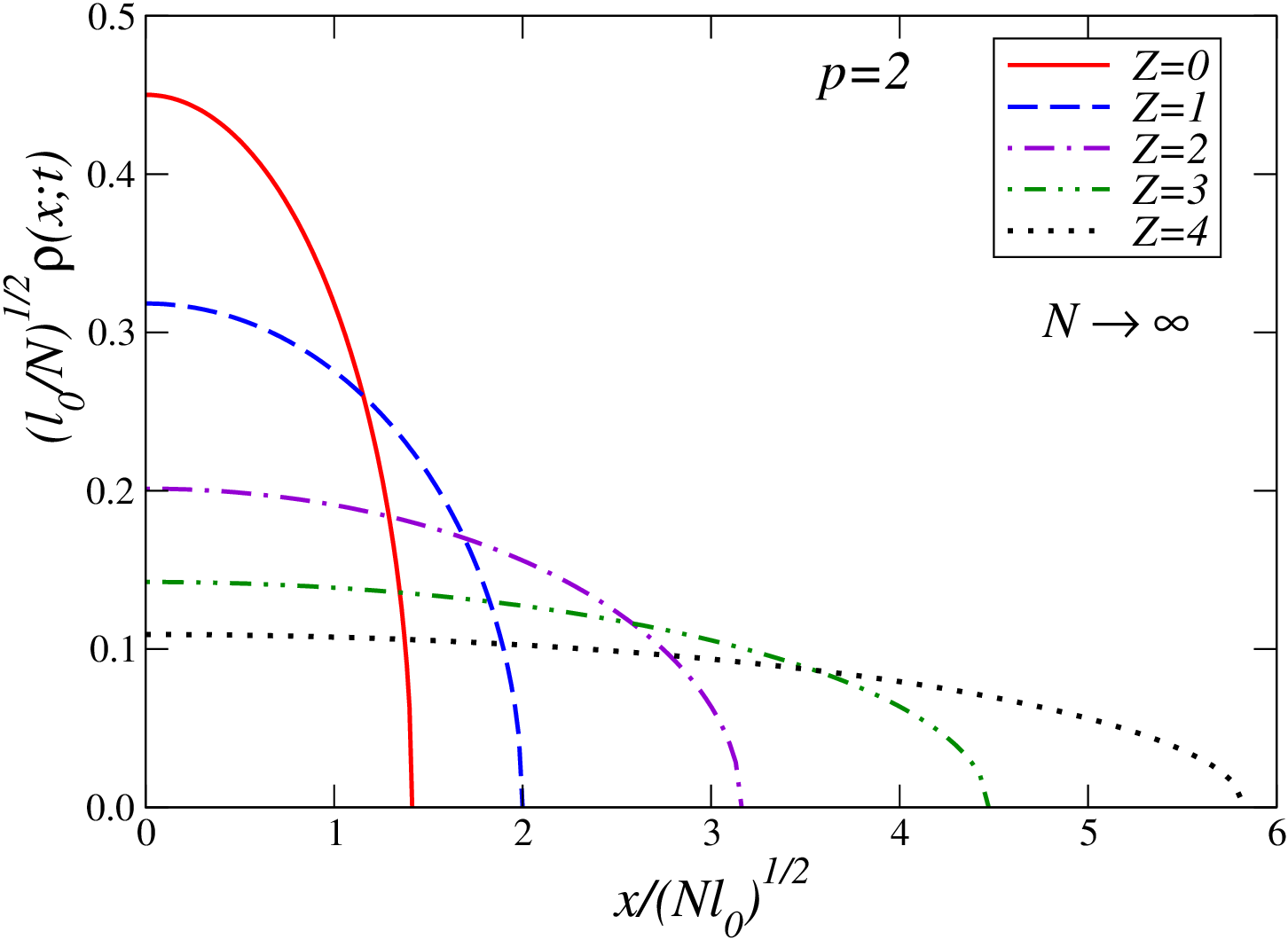}
\includegraphics*[scale=\graphicscale]{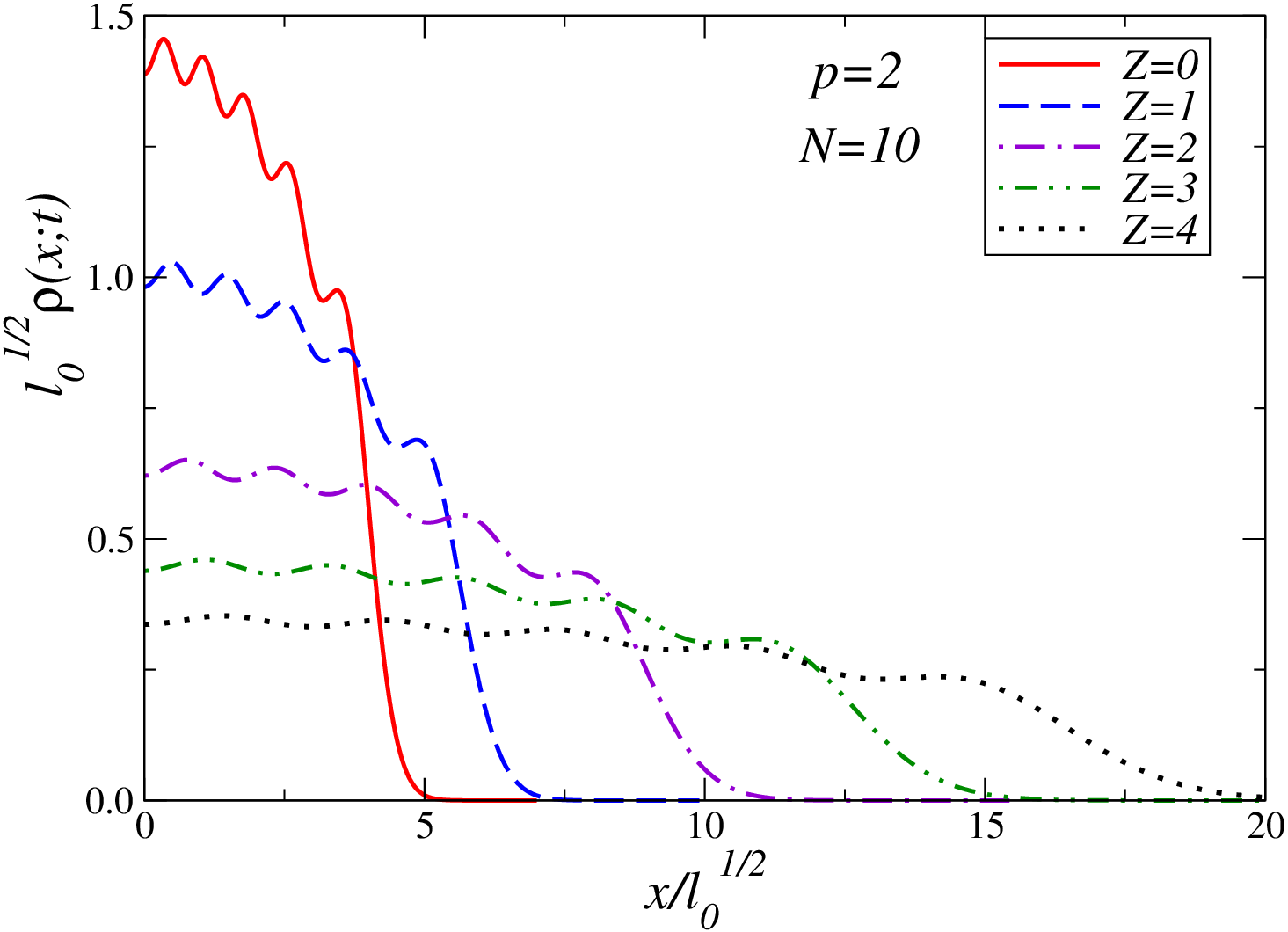}
\caption{ (Color online)
  $l_0^{\theta}\rho(x;t)$ for
  some values of $Z\equiv l_0^{-2\theta} t$, for $N=10$ (below) and in the
  limit $N\to\infty$ (above), in the case of a quench to the unconfined
  Hamiltonian.  }
\label{tidnnxqu}
\end{figure}

\begin{figure}[tbp]
\includegraphics*[scale=\graphicscale]{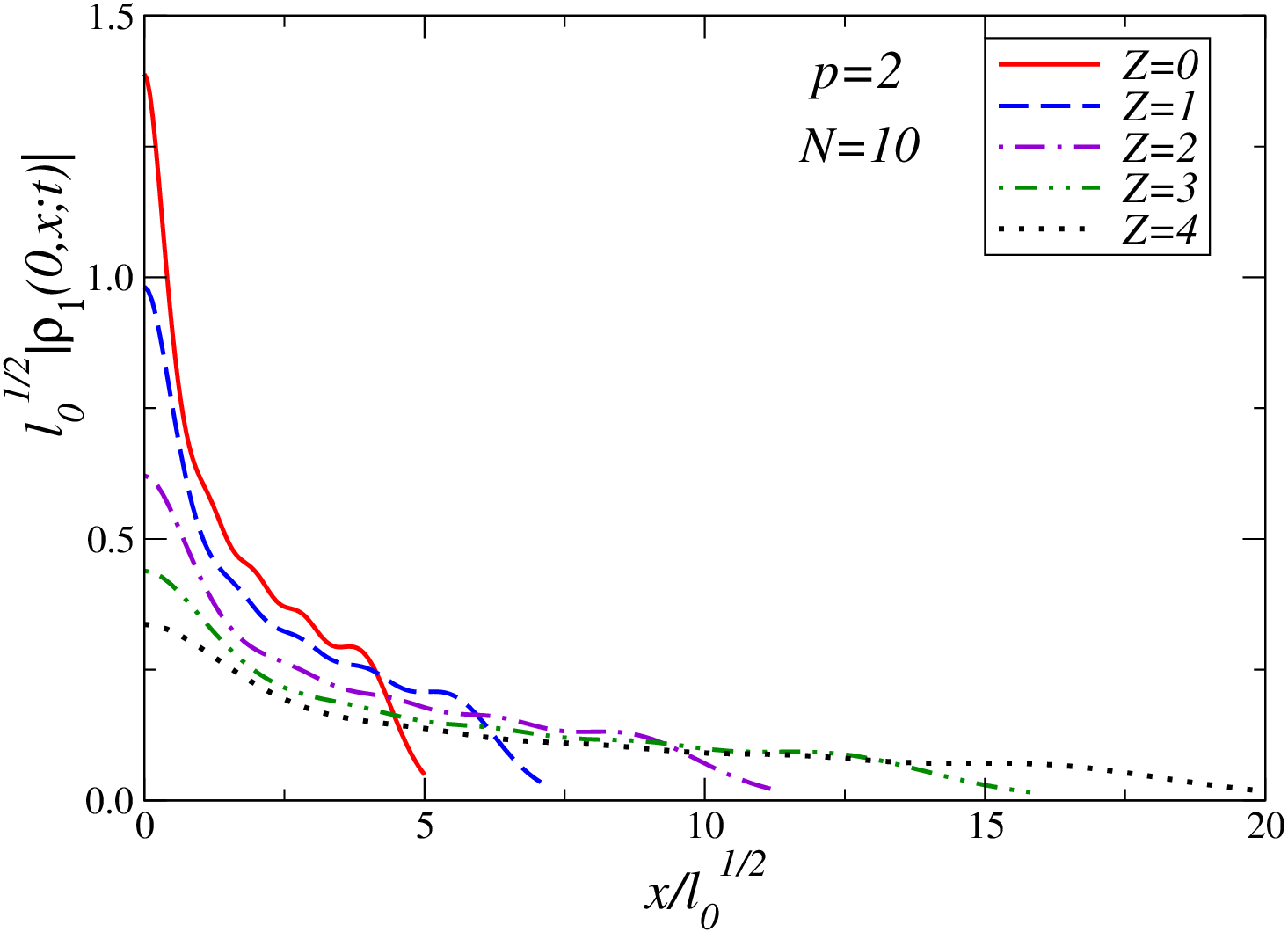}
\caption{ (Color online)
$l_0^{\theta}|\rho_1(0,x;t)|$ for some values of 
$Z\equiv l_0^{-2\theta} t$,
for $N=10$ and in the case of a quench to the unconfined
Hamiltonian.
}
\label{tidngbqu}
\end{figure}

The time-dependence of the one-particle density matrix, after turning the trap
off, is obtained by inserting the function $s(t)$ of Eq.~(\ref{ftqinf}) into
Eq.~(\ref{rhonbosdyn}).  Then, using equilibrium relation
\begin{equation}
\rho_1(x,y;0) \approx l_0^{-\theta} {\cal M}(x/l_0^\theta,y/l_0^\theta),
\label{rhonresc2}
\end{equation}
and defining
\begin{eqnarray}
&&X=x/l_0^{\theta},\;\; Y=y/l_0^{\theta},\;\; Z=t/l_0^{z\theta}, 
\label{rescwh}\\ 
&& Q(Z) = \sqrt{1 + Z^2}, \label{qtsol}
\end{eqnarray}
where $\theta=1/2$ is the equilibrium trap exponent, 
we write 
\begin{eqnarray}
&&\rho_1(x,y;t) =  l_0^{-\theta} Q^{-1} \times \label{dyntssrhoqu}\\
&&\times {\cal M}_N(X/Q,Y/Q)
{\rm exp}\left[i{Q'\over 2 Q} (Y^2-X^2)\right].
\nonumber
\end{eqnarray}
The particle density is given by
\begin{equation}
\rho(x;t) = \rho_1(x,x;t) = l_0^{-\theta} Q^{-1}{\cal D}_N(X/Q),
\label{rhoxtqu}
\end{equation}
where ${\cal D}_N(X)$ can be derived from Eqs.~(\ref{eq:p2eig}) and
(\ref{dnphi}).  Analogously, one can derive the particle-density
correlation $G_n$, cf. Eq.~(\ref{gnbos}), obtaining
\begin{equation}
G_n(x,y;t) = l_0^{-2\theta} Q^{-2}{\cal G}_N(X/Q,Y/Q).
\label{gmqu}
\end{equation}
Results for the particle density and the one-particle density matrix are shown
in Figs.~\ref{tidnnxqu} and \ref{tidngbqu}.

\begin{figure}[tbp]
\includegraphics*[scale=\graphicscale]{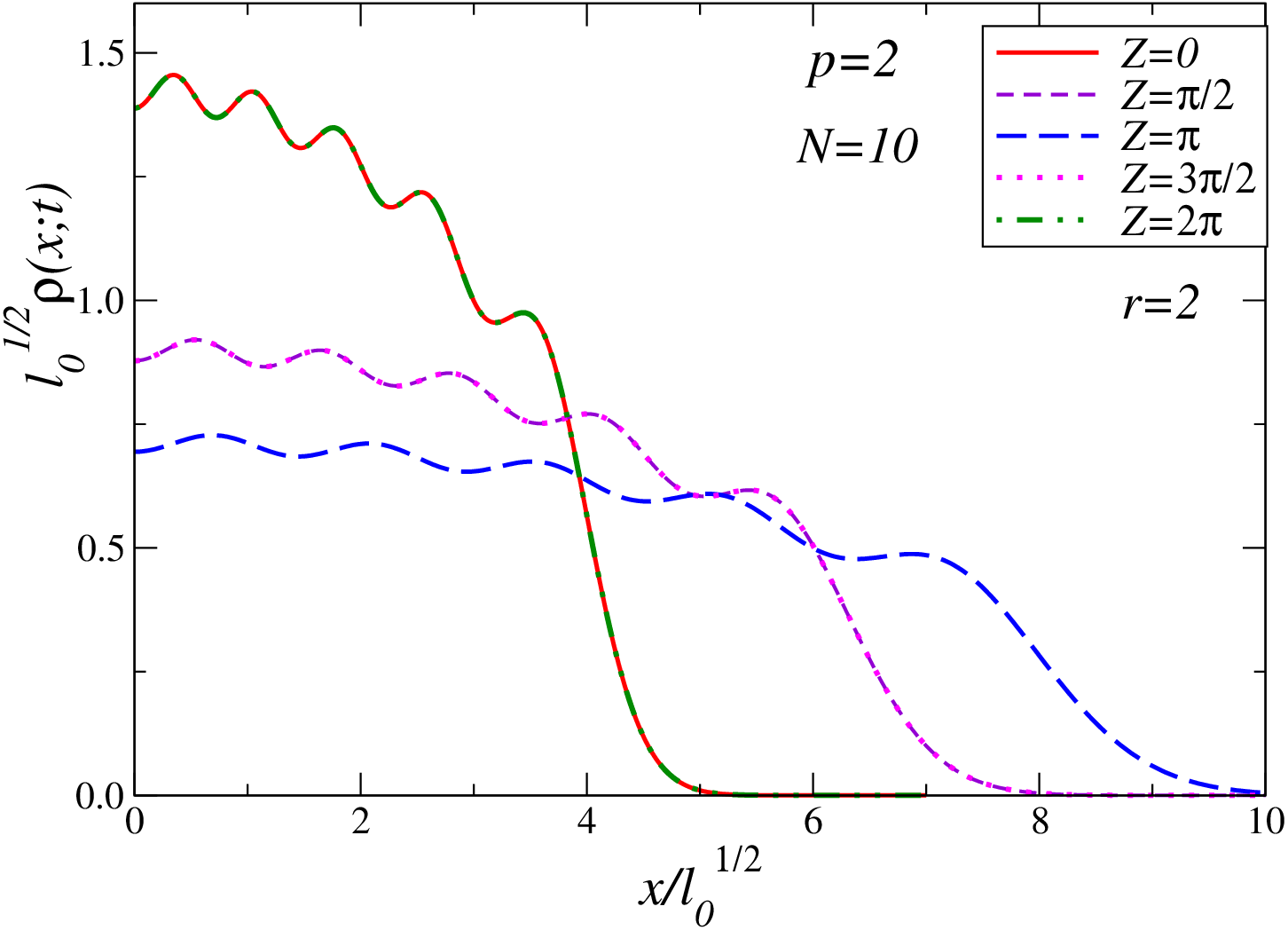}
\caption{ (Color online) Time dependence of the rescaled particle density
  $l_0^{\theta}\rho(x;t)$ for some values of $Z\equiv l_0^{-2\theta} t$,
  for $N=10$ and in the case of a quench to a trap with size $l_f=2l_0$.  It
  oscillates between the $Z=0$ and $Z=\pi$ curves.  }
\label{tidnnxqu2}
\end{figure}

In the case of a quench to a larger trap size $l_f>l_0$, replacing
Eq.~(\ref{stsol}) into Eq.~(\ref{rhonbosdyn}), we again obtain the expression
(\ref{dyntssrhoqu}), but
\begin{eqnarray}
Q(Z) = \sqrt{ 1 + (r^2-1) \left[{\rm sin}(Z/r)\right]^2}, 
\label{qtsol2}
\end{eqnarray}
where $r \equiv l_f/l_0$.  Therefore, we have a periodic time evolution with
period $Z_p=r \pi$.  Fig.~\ref{tidnnxqu2} shows results for the periodic
time evolution of the particle density for $N=10$ and $r=2$.

For a large number of particles in a harmonic potential, we can derive the TSS
using the asymptotic behavior given by Eqs.~(\ref{dnto1on}) and
(\ref{ry}). We obtain
\begin{equation}
l_0^{\theta} \rho(x;t) \approx  {(2N)^{1/2}\over \pi Q}  
\sqrt{1 - {X^2\over 2NQ^2}}.
\label{rhoxtp2lnq}
\end{equation}

The above results show that, after instantaneous changes of the
trap size of the harmonic confining potential, the trap-size
dependence satisfies the scaling Ansatz put forward in
Sec.~\ref{tssisqu}

It is worth noting that the time-dependence of the particle density
$\rho(x;t)$ and its correlation $G_n(x,y;t)$ can be reexpressed as
their equilibrium TSS with an effective time-dependent trap size
\begin{eqnarray}
\tilde{l}(t) = l_0 s(t)^{1/\theta} = l_0 Q(Z)^{1/\theta}, 
\label{itpnqu}
\end{eqnarray}
so that
\begin{eqnarray}
\rho(x;t) = \tilde{l}(t)^{-\theta} {\cal D}_N(\widetilde{X}),
\qquad \widetilde{X} \equiv x/\tilde{l}(t)^\theta.
\label{reexrho}
\end{eqnarray}

We finally mention that the case of a gas of impenetrable bosons in a
hard-wall trap, and its expansion after the drop of the trap, was
considered in Refs~\cite{OS-02,CM-06}.  A hard-wall trap of size $L$
corresponds to the $p\to\infty$ limit of the confining potential, cf.
Eq.~(\ref{potential}), with trap size $l=L/2$.  One can easily check
that the time evolution of the particle density after the drop of the
trap, computed in Ref.~\cite{CM-06}, is consistent with the scaling
Ansatz (\ref{dnphitssqu}) taking into account that the $p\to\infty$
limit of the trap exponent (\ref{thetaexp}) is $\theta=1$ and
$l_0=L/2$.

\subsection{Power-law time dependence of the trapping potential}
\label{offtsse}

We now consider the case of a power-law time dependence of the confining
potential, cf. Eq.~(\ref{vxt}).

Let us define the quantities
\begin{eqnarray}
&&S(Z) \equiv l_0^{q\theta_0/2} s(t),
\quad Z\equiv l_0^{-2\theta_0} \tau,
\label{ftr} 
\end{eqnarray}
where 
\begin{equation}
\theta_0 = {1\over 2+q}
\label{theta0q}
\end{equation} 
is the  off-equilibrium trap exponent
obtained by replacing $z=2$ and $p=2$ in Eq.~(\ref{theta0}).
$S(Z)$ satisfies the equation
\begin{equation}
S'' + Z^q S = S^{-3}
\label{fde}
\end{equation}
where $S''\equiv \partial_Z^2 S$.
Then, using the equilibrium relation (\ref{rhonresc2}),
we rewrite Eq.~(\ref{rhonbosdyn}) as
\begin{eqnarray}
&&\rho_1(x,y;t) =  l_0^{-\theta_0} S^{-1} 
\times \label{dyntssrho}\\
&&\times {\cal M}_N(X/S,Y/S) \times
{\rm exp}\left[i{S'\over 2S} (Y^2-X^2)\right],
\nonumber
\end{eqnarray}
where $X=x/l_0^{\theta_0}$, $Y=y/l_0^{\theta_0}$, and ${\cal M}_N$ is the same
scaling function appearing in Eq.~(\ref{rhonresc}).  

The evolution of the particle density is easily obtained:
\begin{equation}
\rho(x;t) = \rho_1(x,x;t) = 
l_0^{-\theta_0} S^{-1} {\cal D}_N(X/S) ,
\label{rhoxt}
\end{equation}
where ${\cal D}_N$ is the scaling function (\ref{dnphi}).  For a large number
of particles we can derive the off-equilibrium TSS using the asymptotic
behavior given by Eqs.~(\ref{dnto1on}) and (\ref{ry}). We obtain
\begin{equation}
l_0^{\theta_0} \rho(x;t) \approx  {(2N)^{1/2}\over \pi S}  
\sqrt{1 - {X^2\over 2NS^2}}.
\label{rhoxtp2ln}
\end{equation}

The time dependence of $\rho(x;t)$ can be again reexpressed using the
equilibrium expression with an effective trap size
\begin{eqnarray}
\tilde{l}(t) = l_0^{\theta_0/\theta}  S^{1/\theta} = l_0 s(t)^{1/\theta}.
\label{itpn}
\end{eqnarray}
The function $\tilde{l}(t)$ increases monotonically with decreasing $t<0$. If
one prefers to invert the time evolution, so that the effective trap size
increases with increasing $t\ge 0$, it is sufficient to redefine $\tau=1-t$ in
Eqs.~(\ref{kappat}) and (\ref{exsolq1}).  Note that $\tilde{l}(t)$ remains
finite for $\tau=0$, i.e., when the external potential (\ref{vxt}) vanishes,
indeed $s(\tau=0)\simeq 1.14313$, then it diverges in the limit $\tau\to
-\infty$ (note that for $\tau<0$ and $q=1$ the potential (\ref{vxt}) changes
sign, so it does not trap the particles anymore).

The above scaling behaviors are apparently consistent with those predicted by
the scaling arguments of Sec.~\ref{tsspltd} for the off-equilibrium TSS in the
low-density regime.  However, the function $S(Z)$ maintains a residual
dependence on $l_0$, beside on $Z$, due to the initial condition of $s(t)$
which corresponds to $l_0^{-q\theta_0/2} S(l_0^{-2\theta_0})=1$.  Thus, the
scaling Ansatz can be actually considered as fully verified only if the
function $S(Z)$ has a nontrivial scaling limit for $l_0\to\infty$.

\begin{figure}[tbp]
\includegraphics*[scale=\graphicscale]{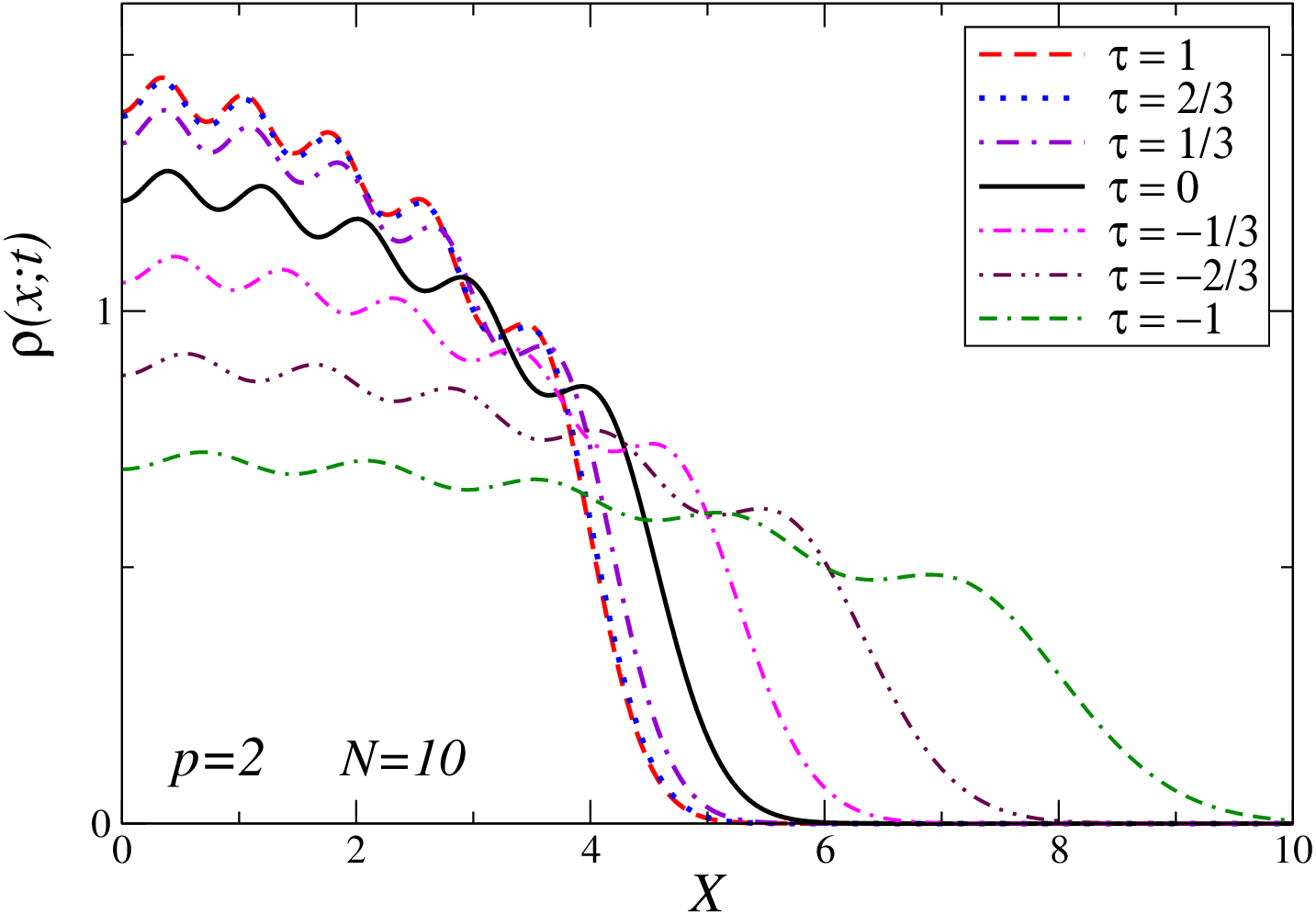}
\caption{ (Color online) 
Time dependence of the particle density
for $N=10$ in a time-dependent trap with $\kappa(t)= \tau$, 
corresponding to $l_0=1$.
}
\label{tidnnx}
\end{figure}

\begin{figure}[tbp]
\includegraphics*[scale=\graphicscale]{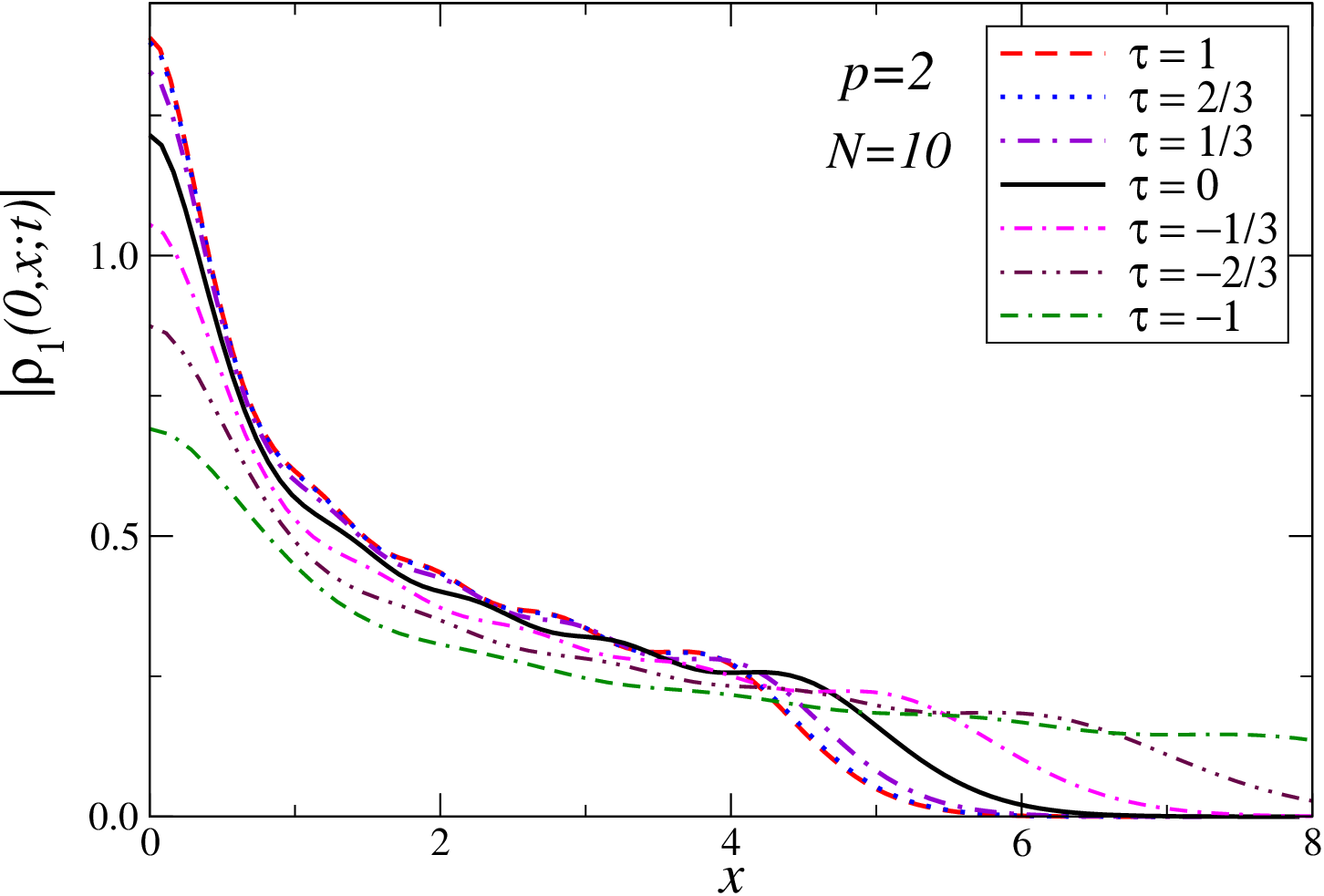}
\caption{ (Color online)
Time dependence of the absolute value of the one-particle density matrix 
for $N=10$ in a time-dependent trap with $\kappa(t)= \tau$, 
corresponding to $l_0=1$.
}
\label{tidngb}
\end{figure}

In the case of a linear time-dependence of $\kappa(t)$, i.e.,
$\kappa(t)=\kappa_0 \tau$, the function $S(Z)$ can be derived from the
corresponding solution $s(t)$, cf.  Eq.~(\ref{exsolq1}).  Then, using the
equilibrium results of Sec.~\ref{bosgas}, we obtain the time dependence of the
one-particle density matrix, the particle density, particle density
correlators, momentum distribution, etc....  Some results for the particle
density and one-particle density matrix are respectively shown in
Figs.~\ref{tidnnx} and \ref{tidngb}.

An important remark is in order. The analytical solution in the case of a
linear time dependence shows that the function $S(Z)$ does not have a
nontrivial scaling limit for $l_0\to\infty$, indeed it appears to diverge,
roughly as $(\ln l_0)^2$ at fixed $Z$. This may reflect the fact that the
initial conditions are somehow weakly relevant, leaving a residual weak
(logarithmic) dependence in the large-$l_0$ limit. We should further note that
$\widetilde{S}(Z)\equiv s(t)$ with $Z\equiv l_0^{-2\theta_0} \tau$ (with
$\tau=1+t$) has a nontrivial $l_0\to\infty$ limit satisfying the
differential equation $S''+ZS=0$. But this rescaling would not fit any scaling
behavior consistent with the dynamic exponent $z=2$.  This point deserves
further investigation.

\section{Summary and conclusions}
\label{conclusions}

We study the trap-size dependence of the quantum behavior of dilute gases of
bosonic particles in the presence of a confining potential trapping the
particles within a limited spatial region.  We consider systems of bosonic
particles constrained in an optical lattice, described by the Bose-Hubbard
(BH) model in the presence of a confining potential coupled to the particle
density, cf. Eq.~(\ref{bhmN}).  In the case of a harmonic potential $V(x) = v^2
x^2/2$, the corresponding trap size is defined as $l=\sqrt{J}/v$ where $J$ is
the hopping parameter.  We consider systems at equilibrium and off equilibrium
during the unitary time evolution arising from changes of the trapping
potential, at zero temperature, i.e., at a sufficient low temperature to
neglect its effects.  We investigate the trap-size dependence in the
low-density regime using the framework of the trap-size scaling (TSS)
theory~\cite{CV-10,CV-09}.

Using scaling arguments, we infer the power-law trap-size dependence
of observables related to the equilibrium lowest states of a dilute
gas of $N$ particles in the low-density regime.  The low-density
regime of the BH model (\ref{bhmN}), i.e., $Na^d/l^d<< 1$ where $a$ is
the lattice spacing, can be seen as the critical regime of a quantum
transition from low density to the empty state, which may be
considered as a $n=0$ Mott transition.  Mott transitions driven by the
chemical potential are described by the {\em nonrelativistic} $\Phi^4$
continuum theory (\ref{lb}), where the dynamic exponent is $z=2$ and
the RG dimension of the chemical potential is $y_\mu=2$, in one and
two spatial dimensions.  In the presence of a confining potential, the
power-law trap-size dependence is described by the equilibrium trap
exponent~\cite{CV-10,CV-10-2} $\theta=p/(p+2)$ where $p$ is the power
of the confining potential. This allows us to derive the universal
scaling features of the asymptotic power-law trap-size dependence
keeping fixed the particle number $N$. For a generic observable, whose
low-density critical behavior is described by the RG dimension $y_o$
in the homogeneous system, we obtain the scaling Ansatz $\langle
O\rangle_N(l,x) \approx l^{-y_o \theta} {\cal O}_N(x l^{-\theta})$,
see Sec.~\ref{TSSeq}.

The equilibrium TSS scenario is verified in 1D systems by analytical
and numerical calculations. We show analytically that the expected TSS
holds in the hard-core (HC) limit $U\to\infty$ of the BH model. We
compute the scaling functions of some observables,
such as the particle density and its correlators,
the one-particle density matrix, see Sec.~\ref{tssobs}.
The universality of
the low-density TSS with respect to the on-site repulsion coupling $U$
is supported by numerical calculations at a finite value of $U$, i.e.,
$U=2$, using DMRG methods.  We show that the asymptotic TSS of $N$
particles at equilibrium described by the 1D BH model, in the HC limit
and at finite $U$, is identical to that of a 1D gas of impenetrable
bosons (Tonks-Girardeau model), with appropriate definitions of the
trap size (in the case of harmonic traps the trap sizes are
proportional to the inverse frequency in both models). The lattice
structure gives rise to subleading $O(l^{-2\theta})$ scaling
corrections in the HC limit of the BH model.  The approach to
the asymptotic behavior is slower at finite values of $U$, in
agreement with the RG arguments which predict subleading
$O(l^{-\theta})$ scaling corrections.  We argue that the same
scenario applies to the Lieb-Liniger model with a finite contact
interaction in the low-density regime, i.e., it presents the same
universal asymptotic TSS with
$O(l^{-\theta})$ scaling corrections.

We investigate the trap-size dependence of the off-equilibrium
dynamics due to time-dependent confining potentials, such as $V(r,t)
\sim \left(1 + t/t_q \right)^q r^p$, or instantaneous changes of the
trap size, including the instantaneous drop of the trap.  We extend
the scaling Ansatz for the trap-size dependence at equilibrium to
off-equilibrium quantum evolutions, to describe the TSS with respect
to the initial trap size $l_0$, see Sec.~\ref{dynTSS}. 
We argue that $\langle O\rangle_N(x;t)
\approx l_0^{-y_o \theta} {\cal O}_N(x l_0^{-\theta}, t
l_0^{-z\theta},l_f/l_0)$ in the case of an instantaneous change of the
trap size from $l_0$ to $l_f$.  In the case of a power-law time
dependence, we introduce an off-equilibrium trap exponent, given by
$\theta_0=1/(2+q)$ in the case of a harmonic trapping potential, and
put forward the scaling Ansatz $\langle O\rangle_N(x;t) \approx
l_0^{-y_o \theta_0} {\cal O}_N(x l_0^{-\theta_0}, \tau
l_0^{-z\theta_0})$.  The above results are expected to be quite
general in the dilute regime of 1D bosonic gases, such the lattice BH
model and continuous Lieb-Liniger model.

We then analyze the trap-size dependence of the off-equilibrium
dynamics of 1D bosonic gases with respect to the initial trap size
$l_0$, using adiabatic approximations in the case of slow changes of
the parameters, and exact solutions of the Schr\"odinger equation of
$N$ impenetrable bosons in time-dependent traps or after instantaneous
changes of the trap size.  The evolution after instantaneous quenches
agrees with the corresponding off-equilibrium Ansatz
(\ref{scalbehNqu}), where the equilibrium trap exponent $\theta$
characterizes the power-law dependence on the initial trap size, see
Sec.~\ref{instdun}.  In the case of a power-law time dependence of the
potential, the evolution supports the scaling Ansatz in terms of the
off-equilibrium trap exponent $\theta_0$, see Sec.~\ref{offtsse}.
However, in the case of a linear time dependence, for which we have an
analytical solution, the simplest Ansatz (\ref{scalbehN}) does not
provide a complete description of the asymptotic large-$l_0$ behavior
because the resulting scaling functions maintain a weak logarithmic
dependence on $l_0$ in the large-$l_0$ limit, demonstrating that the
initial conditions are somehow relevant.  This point deserves further
investigation.

Our results are of experimental relevance for systems of cold atomic gases
trapped by a confining potential. Indeed, the easy tunability and long
characteristic time scales of these systems may allow a careful study of the
trap-size dependence of the zero-temperature properties of $N$-particle boson
gases, in the continuum and on optical lattices, at equilibrium and off
equilibrium in a time-dependent confining potential.

\medskip Helpful discussions with P. Calabrese, D. Giuliano, M. Mintchev and
G. Morchio are gratefully acknowledged.

\appendix

\section{TSS in traps induced by a spatial 
dependence of the hopping parameter}
\label{tdhp}

As suggested in Ref.~\cite{RBSMJ-10}, ultracold atomic systems in
optical lattices may be also get trapped by appropriate
spatially inhomogeneous hopping parameters in the BH model.  An
example is given by the model
\begin{eqnarray}
{\cal H}_{t_{ij}} = - {J\over 2}
\sum_{\langle ij\rangle} {t_{ij}\over 2}(b_j^\dagger b_i+ b_i^\dagger b_j)
+ {U\over2} \sum_i n_i(n_i-1) 
\label{bhmN2}
\end{eqnarray}
with 
\begin{eqnarray}
&&t_{ij} \equiv h(x_{ij}), \quad x_{ij} = {x_i + x_j\over 2}, 
\label{tijdef}\\
&&\quad h(x) = \left[ 1 + {1\over p} (x/l)^p\right]^{-1},
\nonumber
\end{eqnarray}
and $x_i$ are the positions of the sites of the lattice.
The rescaled hopping parameter $t_{ij}$ tends to one at the
middle of the trap and vanishes at large distance, giving
rise to an effective trap, with trap size $l$. 

In the HC $U\to\infty$ limit, the Hamiltonian can be diagonalized
exploiting the fermion quadratic representation (\ref{fermmod}) with
\begin{eqnarray}
h_{ij} = \delta_{ij} - {1\over 2} t_{ij} (\delta_{i,j-1} + \delta_{i,j+1})
\label{hijdef2}
\end{eqnarray}
following the procedure outlined in Sec.~\ref{TSSlim},
cf. Eqs.~(\ref{hcn}), (\ref{hijeq}) and (\ref{hcdia}).

In the dilute region, i.e., for sufficiently small $N/l$, we can
follow the same steps of Sec.~\ref{TSSlim} to arrive at a continuum
TSS limit.  We end up with the same Schr\"odinger-like equation
(\ref{trapscaleqxx}) after the same rescalings (\ref{xresc}),
(\ref{eresc}) and (\ref{phiresc}) and $\theta=p/(p+2)$ as well.  Thus,
the TSS arising from the spatial-inhomogeneity of the hopping
parameter, like Eq.~(\ref{tijdef}), is identical to that of model
(\ref{bhmN}), i.e., of a trap achieved by coupling an external
potential to the particle density. One can also infer that
scaling corrections are $O(l^{-2\theta})$ as well.

\section{Trap-size scaling for a large number of particles}
\label{lnbeh}

In this appendix we determine the large-$N$ behavior of the TSS
functions of the observables considered in Sec.~\ref{tssobs}.

\subsection{The large-$N$ behavior of the TSS functions}
\label{lnbeh1}

\begin{figure}[tbp]
\includegraphics*[scale=\graphicscale]{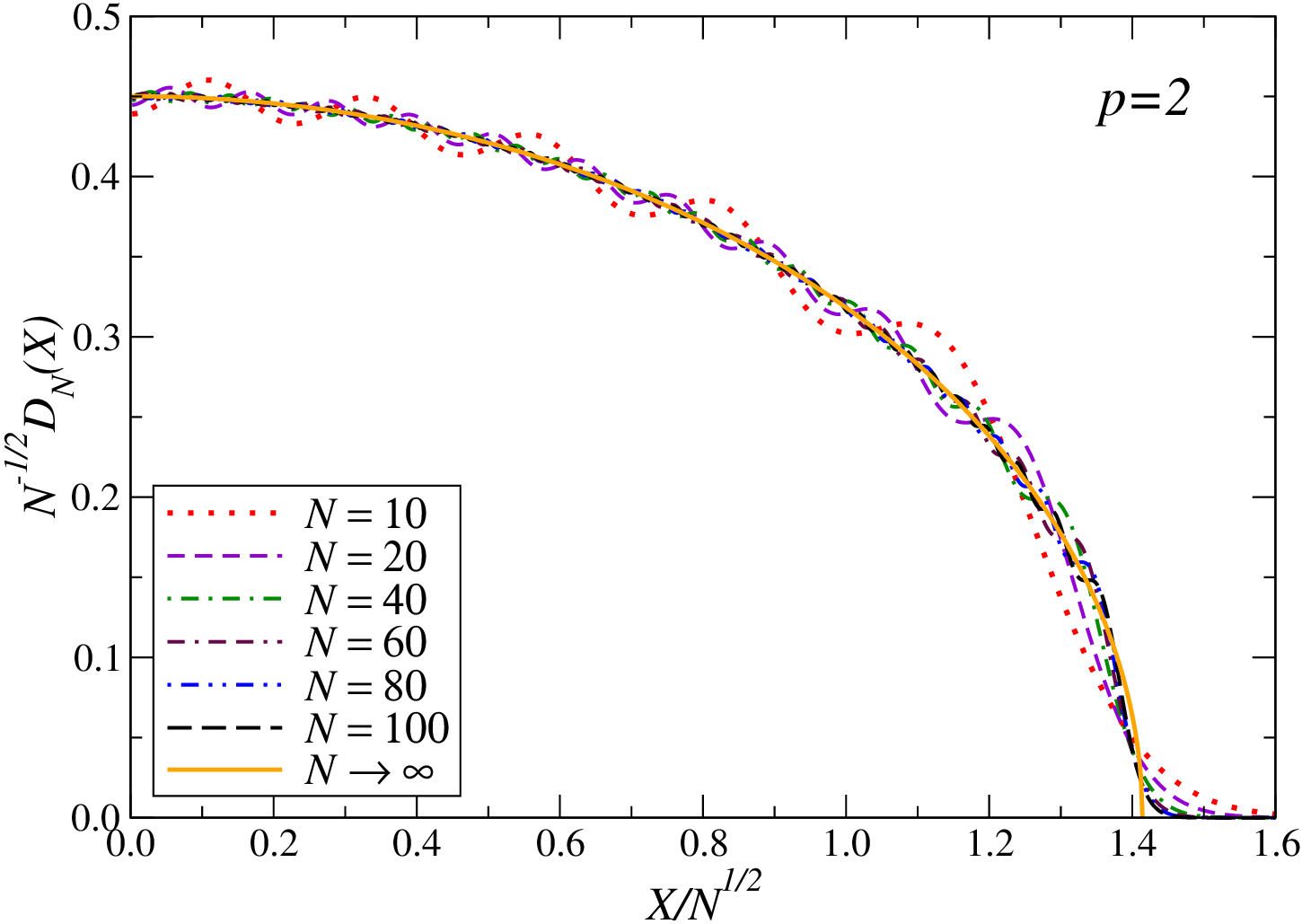}
\caption{ (Color online) The large-$N$ behavior of 
  ${\cal D}_N(X)$ for $p=2$. The full line shows its
  $N\to\infty$  limit, cf. Eq.~(\ref{ry})}
\label{dnxlnp2}
\end{figure}

\begin{figure}[tbp]
\includegraphics*[scale=\graphicscale]{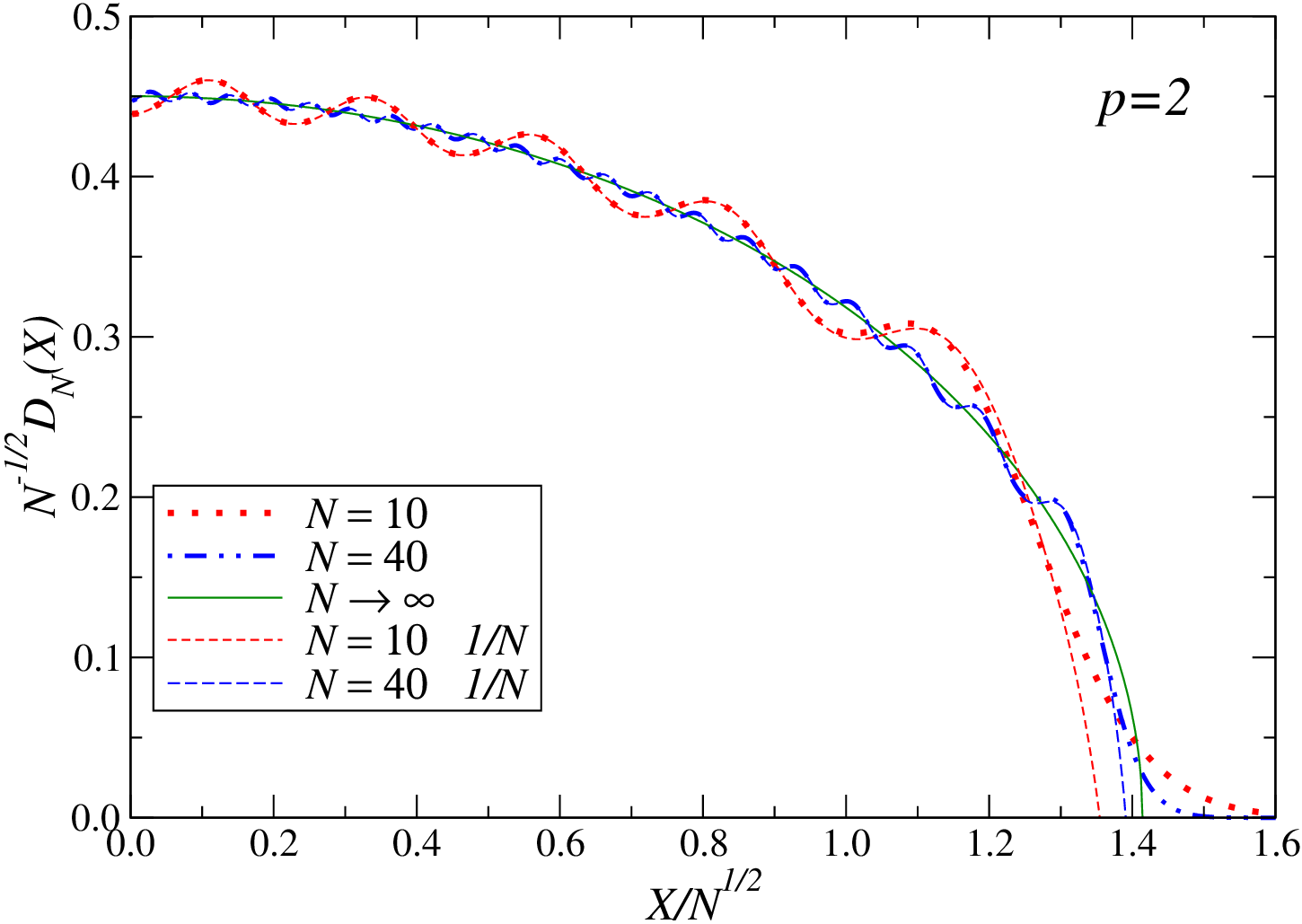}
\caption{ (Color online) Comparisons of ${\cal D}_N(X)$ for $p=2$ and $N=10$
  and $N=40$ with their next-to-leading approximations, 
cf. Eq.~(\ref{o1ncorr}), denoted by ``$1/N$'' in
  the figure.  }
\label{dnxlnp2b}
\end{figure}

In the case of the harmonic potential, the rescaled particle density,
cf.  Eq.~(\ref{dnphi}), behaves as
\begin{eqnarray}
{\cal D}_N(X) = N^{1/2} \left[ R_D(\widetilde{X}) 
+ {C_D(\widetilde{X})\over N} 
+ O(1/N^2)\right],\label{dnto1on} 
\end{eqnarray}
where
$\widetilde{X}\equiv N^{-1/2} X$.
The approach to the large-$N$ behavior is shown in
Fig.~\ref{dnxlnp2}, where $N^{-1/2}D_N(X)$ is plotted versus
$\widetilde{X}$ for several values of $N$.  The leading large-$N$
behavior is 
\begin{equation}
R_D(x) = {1\over \pi} \sqrt{2 - x^2},
\label{ry}
\end{equation}
for $x\le\sqrt{2}$, and $R_D(x)=0$ for $x>\sqrt{2}$. This is also the
$N\to\infty$ limit of the particle density in a bosonic gas of
impenetrable bosons~\cite{KB-02,GFF-05}.  Actually, since the
low-density limit of the HC-BH model matches the behavior of a gas of
impenetrable bosons, as discussed in Sec.~\ref{bosgas}, we  can use
results obtained for the TG gas~\cite{KB-02} to infer that
\begin{eqnarray}
&& C_D(x) = - {(-1)^N {\rm cos}[Nq(x)]\over \pi \sqrt{2} (2-x^2)},
\label{o1ncorr}\\
&& q(x) = x \sqrt{2-x^2} + 2{\rm arcsin}(x/\sqrt{2}).
\nonumber
\end{eqnarray}
A comparison of ${\cal D}_N(X)$ for $N=10$ and $N=40$ with their
next-to-leading large-$N$ approximations is shown in
Fig.~\ref{dnxlnp2b}.

\begin{figure}[tbp]
\includegraphics*[scale=\graphicscale]{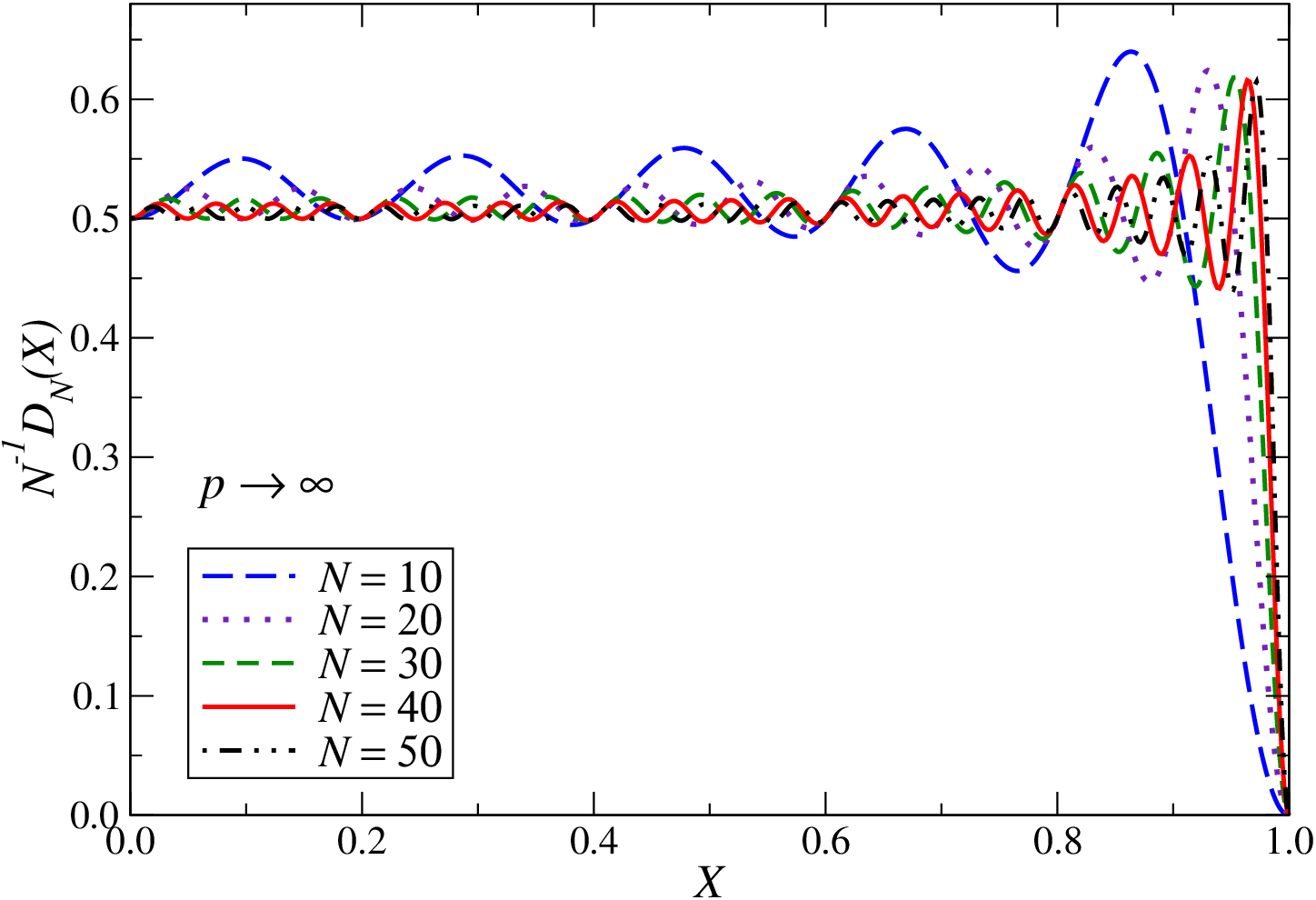}
\includegraphics*[scale=\graphicscale]{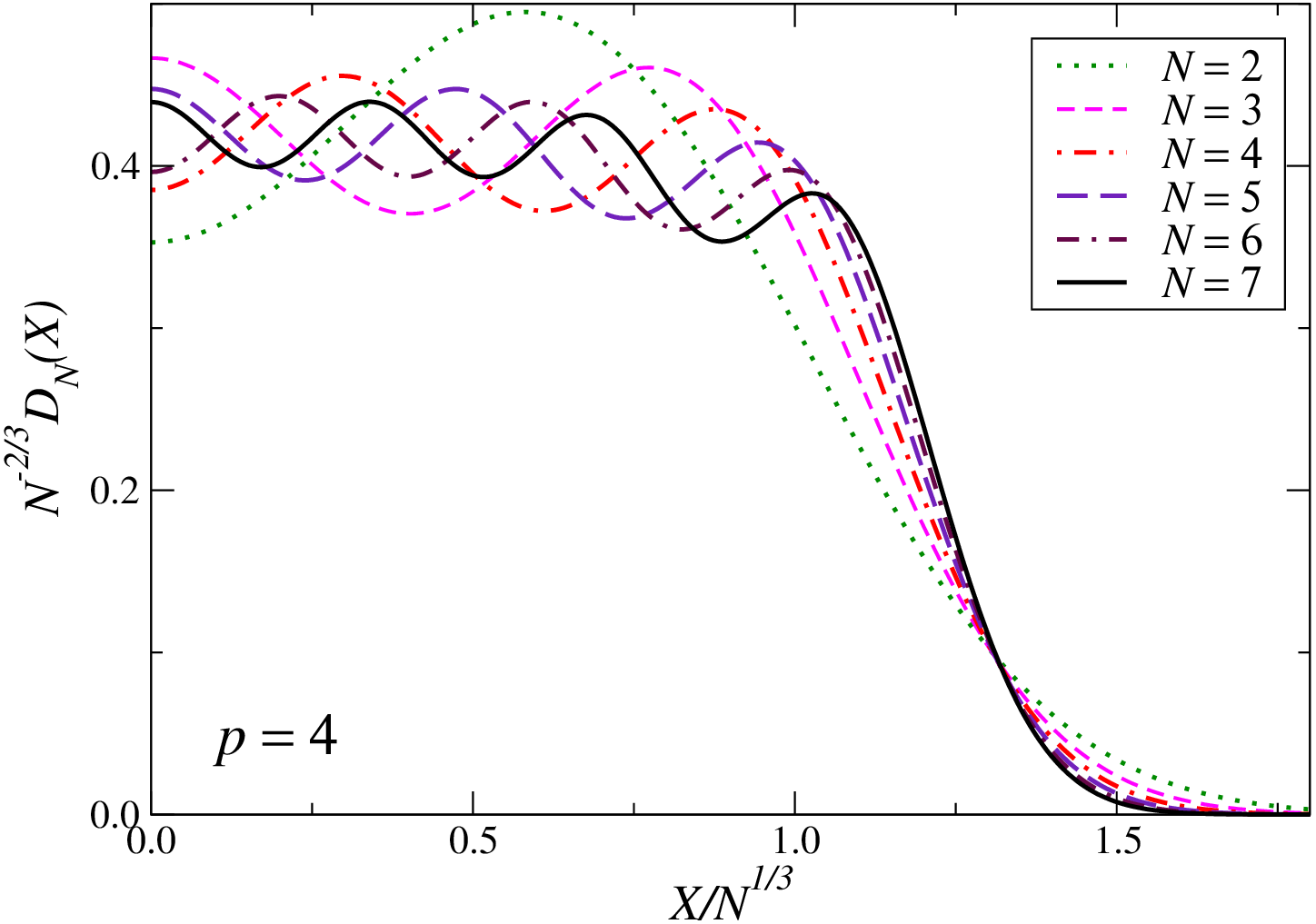}
\caption{ (Color online)
The scaling functions ${\cal D}_N(X)$, cf. Eq.~(\ref{nxn}), for 
$p=4$ (below) and $p\to\infty$ (above).
}
\label{dnxpr4i}
\end{figure}

Analogous results can be derived for other power laws of the confining
potential. Results for $p=4$ and $p\to\infty$ are shown in Fig.~\ref{dnxpr4i}.
The particle-density scaling functions show again $N$ peaks, with an
underlying structure scaling as ${\cal D}_N(X) \approx N^{2/3}
R_D(X/N^{1/3})$ for $p=4$ and ${\cal D}_N(X) = N/2 + O(1)$ for $p\to\infty$ (at
least not too close to $X=1$), with the oscillatory terms suppressed by $1/N$
with respect to the leading terms. These results suggest
the general behavior 
\begin{equation}
{\cal D}_N(X) = N^{\theta} \left[ R_D(X/N^{1-\theta})  + O(1/N)\right]
\label{rhoxsrgenp}
\end{equation}
for any power $p$,
where the function $R_D$ depends on $p$.

\begin{figure}[tbp]
\includegraphics*[scale=\graphicscale]{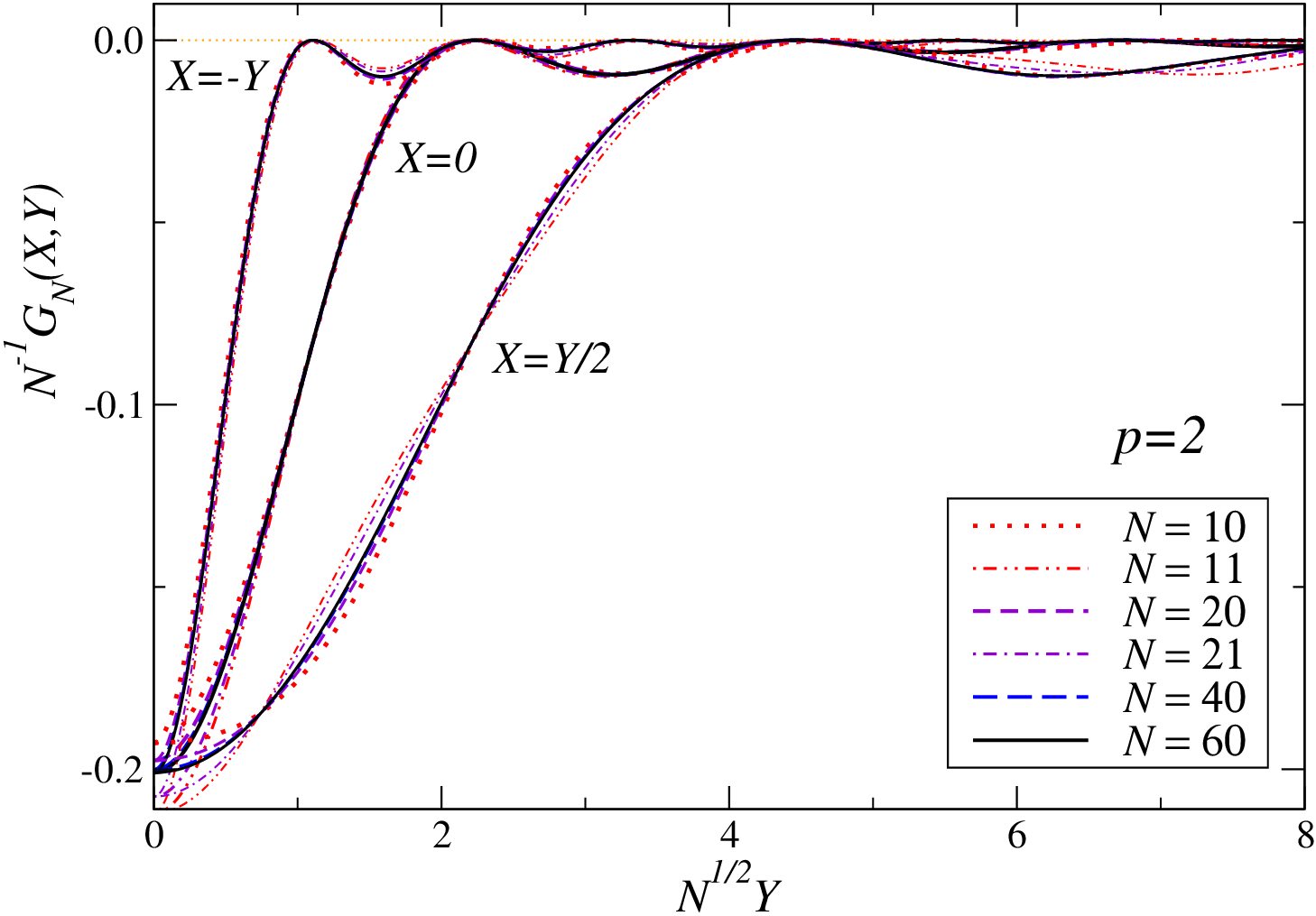}
\caption{ (Color online) Plot of $N^{-1}{\cal G}_N(X,Y)$ for $X=Y/2$, $X=0$,
  and $X=-Y$, vs $N^{1/2} Y$, for several values of $N$. They 
  approach unique curves with increasing $N$. Even and odd values of $N$
  converge from opposite sides.  }
\label{gnxp2ln}
\end{figure}

Fig.~\ref{gnxp2ln} shows results for the scaling function ${\cal
G}_N(X,Y)$ associated with the density-particle correlation,
cf. Eq.~(\ref{gncotss}), for several values of $N$, obtained from
Eq.~(\ref{eq:GnTSS}) for the harmonic potential.  These plots show
that the large-$N$ behavior is
\begin{equation}
{\cal G}_N(X,Y) \approx N R_G(N^{1/2}X,N^{1/2}Y)
\label{gnxlnbeh}
\end{equation}
Note the different $N$-rescaling of the spatial coordinates  with respect
to that of the particle density.

Concerning the one-particle density matrix, cf. Eq.~(\ref{rho1mu}), we
note that the scaling behaviors (\ref{rhoxsrgenp}) and
(\ref{bottomnk}), of the particle density and the momentum
distribution respectively, can be both derived from the following
nontrivial large-$N$ scaling behavior of the one-particle density
matrix:
\begin{eqnarray}
\rho_1(x,y) \approx (N/l)^\theta B[N^{-1+\theta}X,N^\theta(Y-X)],
\label{rho1lnscalbeh}
\end{eqnarray}
where $B$ is a scaling function, $X=x/l^\theta$ and $Y=y/l^\theta$
(note the different power of $N$ in the two arguments of the function
$B$; in the case of a harmonic potential they are $N^{-1/2}X$ and
$N^{1/2}(Y-X)$ respectively).  The above scaling behavior would imply
that, with increasing $N$, the region where the diagonal component
is significantly nonzero increases as
$N^{1-\theta}$, while the width around it decreases as $N^{-\theta}$.
The approach to this large-$N$ limit is
generally characterized by $O(N^{-1/2})$ oscillating corrections in
the case of the harmonic potential.

\subsection{
Power-law behavior at the boundaries where the particle-density vanishes
}
\label{lnbehb}

In this section we discuss the asymptotic large-$N$ power-law behavior
at the boundaries of the trap where the particle density
is suppressed.

The large-$N$ asymptotic behavior found in App.~\ref{lnbeh} holds for
$\widetilde{X} \equiv N^{-1/2}X<\widetilde{X}_c$ where
$\widetilde{X}_c=\sqrt{2}$ is the value where $R_D(\widetilde{X})=0$,
which is the point around which the particle density vanishes
in the large-$N$ limit.  However, around the spatial points
corresponding to $\widetilde{X}_c$ another power law behavior sets, as
suggested by the behavior around $\widetilde{X}_c$ of the curves shown
in Fig.~\ref{dnxlnp2b}.  This fact was already noted within the
Gaussian unitary ensembles of random
matrices~\cite{Forrester-93,GFF-05}, whose eigenvalue density
corresponds to the particle density in harmonically trapped systems of
impenetrable bosons.  An analogous change of power law is observed at
fixed chemical potential, thus $N\sim l$, at the boundaries of the
trap~\cite{CV-10-2}.

\begin{figure}[tbp]
\includegraphics*[scale=\graphicscale]{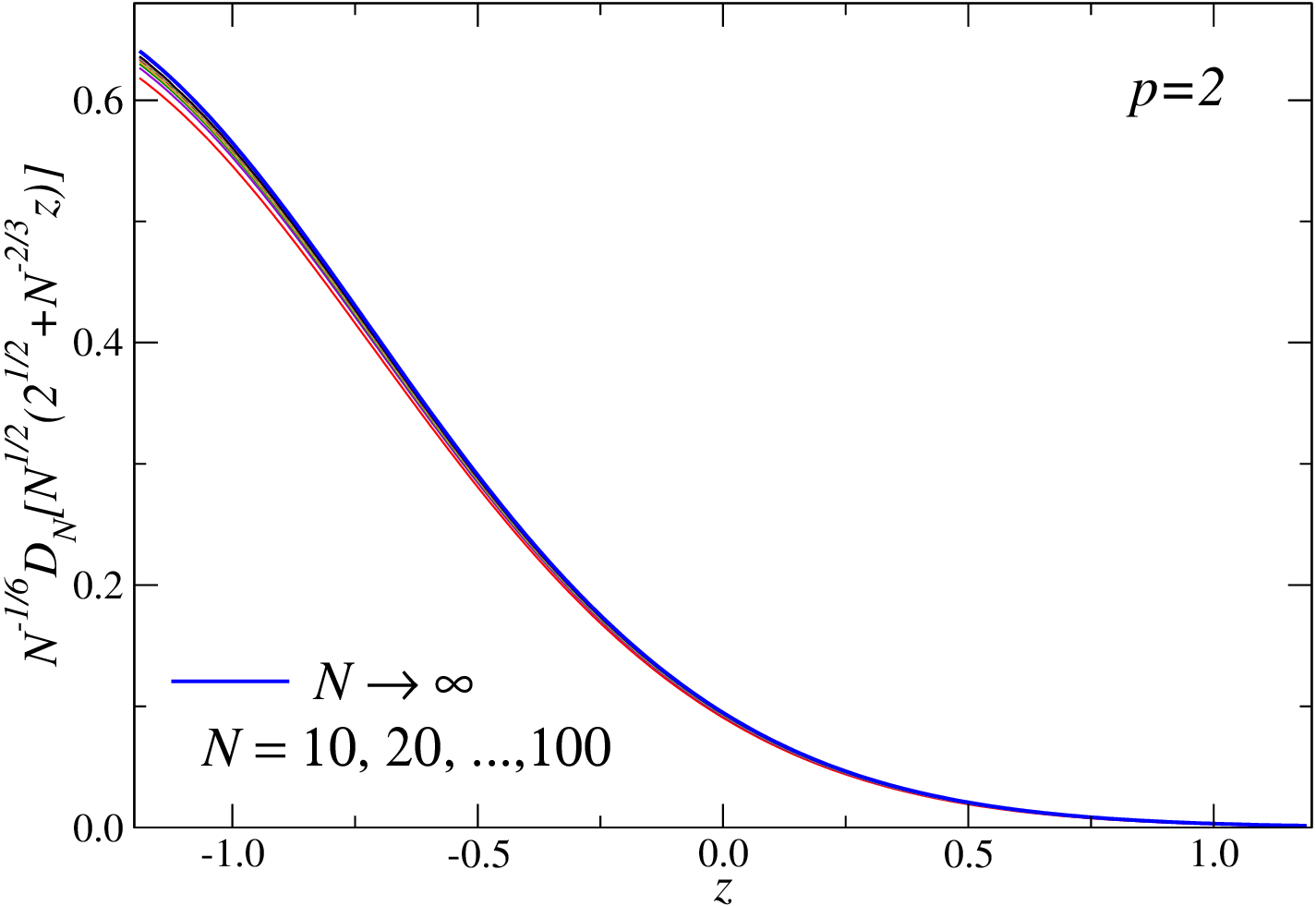}
\caption{ (Color online) Plot of $N^{-1/6}{\cal D}_N[N^{1/2}(\widetilde{X}_c +
  N^{-2/3}z)]$ versus $z$ for $p=2$, where $\widetilde{X}_c=2^{1/2}$,
for $N=10,20,...,100$.
The curves appear to converge toward the 
$N\to\infty$ limit (\ref{fez}).}
\label{dnxp2bo}
\end{figure}

This phenomenon is related to a {\em real-space transition} between
the low-density particle regime, for $\widetilde{X}\lesssim
\widetilde{X}_c$, and the empty state for $\widetilde{X}>
\widetilde{X}_c$, which occurs at the points $x_c$ corresponding to
$\widetilde{X}_c$.  Thus we expect that the region around $x=x_c$
develops {\em critical} modes related to a low-density Mott
transition. The effective external potential at $x_c$ can be obtained
by expanding the trapping potential around $x_c$, thus obtaining an
approximately linear potential $V_l(x) \sim x-x_c$.  Around
$x_c$, other critical modes develop with length scale $\xi\sim l^\sigma$,
where $\sigma$ is the exponent associated with a linear external
potential.  The value of $\sigma$ can be inferred by RG arguments
analogous to those leading to the determination of the trap exponent
$\theta$ at the low-density Mott transition~\cite{CV-10-2,CV-10},
which give $\sigma=1/3$.~\footnote{The exponent $\sigma$ can be determined by a RG
  analysis of the perturbation corresponding to a linear potential
  $V_l(x)=ux$, i.e., $\int {\rm d}^dx\,{\rm d} t\, V_l(x) |\phi(x)|^2$, at the
  fixed point 
  of the continuous theory describing the Mott transition~\cite{FWGF-89}.  The
  exponent $\sigma$ is related to the RG dimension $y_u$ of the parameter $u$,
  which can be obtained from the relations $y_u - 1 = d+z-y_{|\phi|^2} =
  y_\mu=2$, thus $y_u=3$, and therefore $\sigma\equiv 1/y_u=1/3$ for $d=1$ and
  $d=2$.} For example, this implies that
\begin{equation}
\widetilde{X}_c - \widetilde{X}_{\rm max} \sim N^{-2/3}
\label{ylp}
\end{equation}
where $\widetilde{X}_{\rm max}$ corresponds to the abscissa of the rightmost
maximum of ${\cal D}_N(X)$.  
More generally, we have the
scaling behavior
\begin{equation}
{\rm lim}_{N\to\infty} N^{-1/6}{\cal D}_N[N^{1/2}
(\widetilde{X}_c + N^{-2/3}z)] = f(z).
\label{fzsc}
\end{equation}
The scaling function $f(z)$ can be obtained from related computations within
the Gaussian unitary ensembles of random matrices~\cite{Forrester-93,GFF-05}:
\begin{eqnarray}
f(z) = 2^{1/2} |{\rm Ai}^\prime(2^{1/2}z)|^2 -  2z|{\rm Ai}(2^{1/2}z)|^2 .
\label{fez} 
\end{eqnarray}
Fig.~\ref{dnxp2bo} shows that the above asymptotic behavior is rapidly
approached in the large-$N$ limit.

\section{The quantum oscillator with a time-dependent frequency}
\label{oscill}

Let us consider a quantum oscillator described by the Hamiltonian
\begin{equation}
H = {p^2\over 2} + \kappa(t) {x^2\over 2}
\label{qosc}
\end{equation}
with a time dependent frequency
\begin{equation}
\kappa(t) = \omega_0^2 (1+t/t_q)^q\equiv \omega_0^2 \tau^q, 
\label{kt}
\end{equation}
where $t_q$ is the time rate of the time dependence. In the following we set
$t_q=1$ for simplicity; its dependence can be easily inferred by appropriate
rescalings of the results.  We assume that at $t=0$ the oscillator is in its
ground state, i.e., its wave function is
\begin{equation}
\psi_0(x) = (\omega_0/\pi)^{1/4} e^{-\omega_0 x^2/2}. 
\label{incond}
\end{equation}
The evolution equation 
\begin{equation}
i\partial_t \psi(x,t) = H\psi(x,t)
\label{seq}
\end{equation}
with $\psi(x,0)=\psi_0(x)$ preserves the Gaussian spatial dependence. 
We write the solution of Eq.~(\ref{seq}) as
\begin{eqnarray}
&&\psi(x,t) = (\omega_0/\pi)^{1/4}
e^{-w(t) \omega_0 x^2/2 + z(t)},\label{solosc}
\end{eqnarray}
where $w(t)$ and $z(t)$ satisfy the equations
\begin{eqnarray}
&& i \dot{w} = \omega_0 (w^2 - \tau^q), \label{wteq} \\
&& i \dot{z} = \omega_0 w/2 \nonumber,
\end{eqnarray}
with initial conditions $w(0)=1$ and $z(0) = 0$.
In the case of a linear dependence of $\kappa$, i.e.,
\begin{equation}
\kappa(t) = 1+t \equiv \tau,
\label{kappatau}
\end{equation}
the solution is $w(t)=W(\tau)$ where $W(\tau)$ is the
complex function given in Eq.~(\ref{wtsol}), and 
\begin{equation}
z(t) = - {i\omega_0\over 2} \int_0^t dt' \,w(t').
\label{zexp}
\end{equation}  
Note that $\psi(x,t)$ remains exponentially suppressed at
large $x$ even at $t=-1$ when $\kappa=0$, indeed
$w(-1)|_{\omega_0=1}=W(0)|_{\omega_0=1}=0.765265 + i 0.346358$.

\begin{figure}[tbp]
\includegraphics*[scale=\graphicscale]{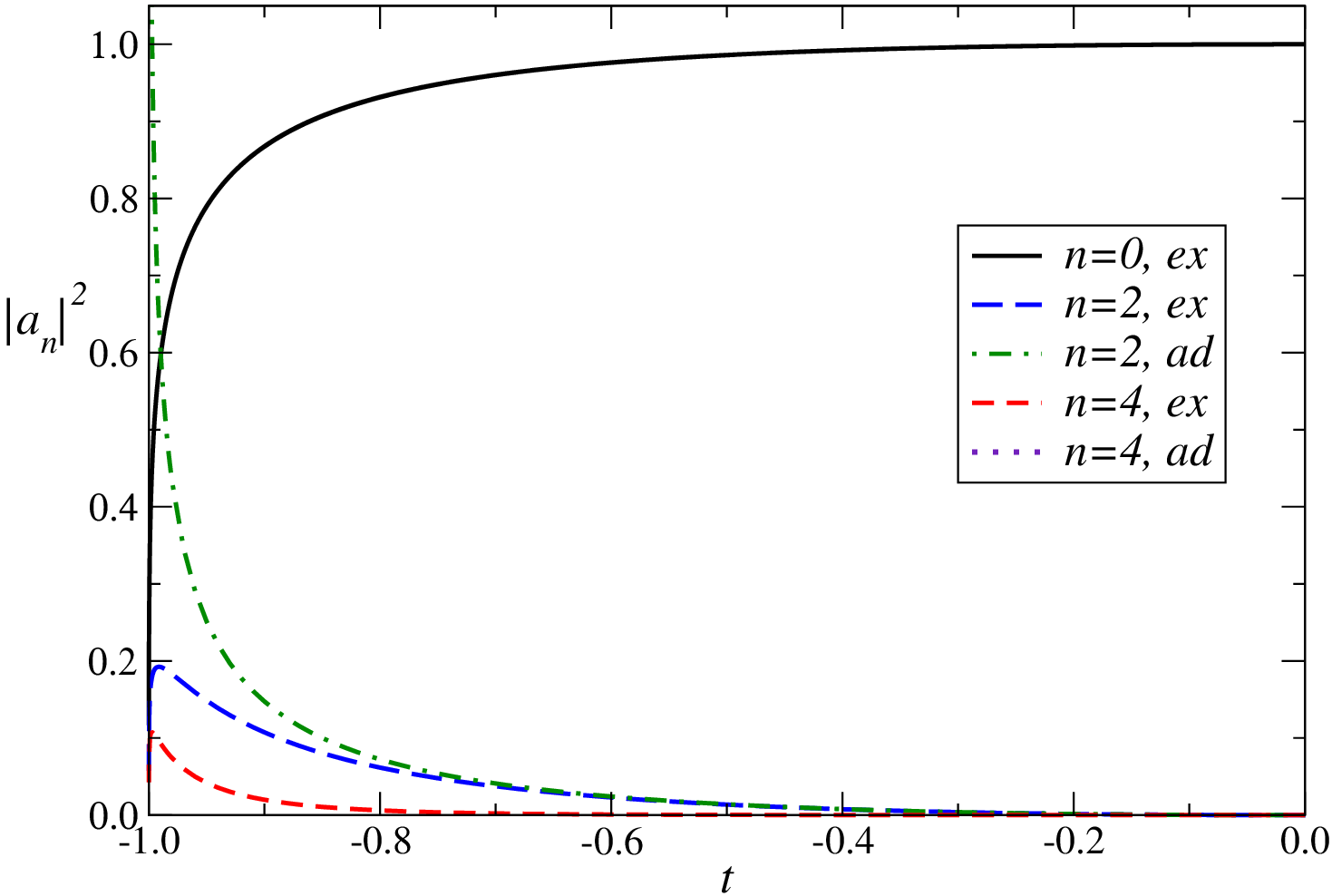}
\caption{ (Color online)
Some results for the coefficients of the expansion (\ref{exphixt})
versus $t$, using exact results from Eq.~(\ref{antex})
and from the adiabatic approximation (\ref{antad}). 
}
\label{oscillator}
\end{figure}

It is interesting to compare the exact solution (\ref{solosc}) with 
the evolution
predicted by the adiabatic perturbation theory.  In the following calculations
we set $\omega_0=1$. We expand the wave function as
\begin{equation}
\psi(x,t) = \sum_{n} a_n(t) \phi_n(x,t),
\label{exphixt}
\end{equation}
where $\sum_n |a_n(t)|^2=1$ and
$\phi_n(x,t)$ are {\em instantaneous} eigenstates, i.e.,
\begin{eqnarray}
&&H\phi_n(x,t) = E_n(t) \phi_n(x,t),\label{eigsol}\\
&&E_n=\omega(t)(n+1/2),\quad \omega(t)=\sqrt{\kappa(t)},\nonumber\\
&&\phi_n(x,t) = {1\over \pi^{1/4} (2^n n!)^{1/2}} 
H_n[x\omega(t)^{1/2}] e^{-\omega(t) x^2/2},
\nonumber
\end{eqnarray}
where $H_n(x)$ are the Hermite polynomials.  The coefficients $a_n(t)$ are
given by
\begin{equation}
a_n(t) = \int dx \,\psi(x,t)^* \phi_n(x,t).
\label{antex}
\end{equation}
Note that $a_{n}=0$ for odd values of $n$ (when the corresponding
eigenfunction $\phi_n$ is odd). Moreover, $a_n(t)\to 0$ for $t\to -1$.

Assuming a very slow variation of the Hamiltonian parameters, we expect an
adiabatic evolution, i.e., the system starting from the ground state at
$t=0$ evolves through the {\em instantaneous} ground states
$\phi_0(x,t)$. Thus, the leading behavior is given by
$a_{n}(t) = e^{i\Theta_0(t)} \delta_{n0}$ 
where $\Theta_n(t) = \int_0^t E_n(t)dt$.
The time dependent coefficients $a_n(t)$ can be computed to the
next-to-leading order of the adiabatic perturbative expansion, see, e.g.,
Refs.~\cite{Schiff-book,DP-10}, obtaining
\begin{eqnarray}
a_n(t) = -e^{i\Theta_n(t)} 
\int_0^t dt' {\langle n| \partial_t H|0\rangle \over E_n(t') - E_0(t')}
e^{-i\Delta\Theta_{n0}(t')} 
\label{antad}
\end{eqnarray}
where $\Delta\Theta_{nm}\equiv \Theta_n-\Theta_m$.  Beside
$a_0(t)$, only the coefficient $a_2(t)$ differs from zero in the first order
adiabatic approximation.  
Note that the limit $t\to -1$, thus $\tau\to 0$, is singular; indeed
$|a_2(t)| \sim |\ln\tau|$.
This shows that the adiabatic approximation fails approaching the {\em
  critical} point where the spectrum tends to become degenerate.

Some results for the coefficients $a_n(t)$ are shown in
Fig.~\ref{oscillator}, as obtained from Eq.~(\ref{antex}) and their
adiabatic approximation.  We find that $|a_0(t)|^2\gtrsim 0.99$ for
$\tau\equiv 1+t\gtrsim 0.5$, which is the region where the zero-order
adiabatic approximation works within 1\%, and
$|a_0(t)|^2+|a_2(t)|^2\gtrsim 0.99$ for $\tau\gtrsim 0.2$., which is
where the first-order adiabatic approximation is effective.


\begin{thebibliography}{99}

\bibitem{CWK-02} E.A. Cornell and C.E. Wieman, Rev. Mod. Phys. 74, 875 (2002);
  N. Ketterle, Rev. Mod. Phys. 74, 1131 (2002).
  
\bibitem{BDZ-08} I. Bloch, J. Dalibard, and W. Zwerger, Rev.\ Mod.\ Phys.\ 80,
  885 (2008).

\bibitem{HSBBD-06} Z. Hadzibabic, P. Kr\"uger, M. Cheneau, B. Battelier,
  and J. Dalibard, Nature 441, 1118 (2006).

\bibitem{DRBOKS-07} T. Donner, S. Ritter, T. Bourdel, A. \"Ottl, M. K\"ohl,
  and T. Esslinger, Science 315, 1556 (2007).
  
\bibitem{CRRHP-09} P. Clad\'e, C. Ryu, A. Ramanathan, K. Helmerson, and W.D.
  Phillips, Phys. Rev. Lett. 102, 170401 (2009).

\bibitem{GBMHS-02} M. Greiner, I. Bloch, M.O. Mandell, T. H\"ansch,
  and T. Esslinger, Nature 415, 39 (2002).

\bibitem{KWW-05} T. Kinoshita, T. Wenger, and D.S. Weiss, Science 305, 1125
  (2004); Phys. Rev. Lett.  95, 190406 (2005).

\bibitem{FWMGB-06} S. F\"olling, A. Widera, T. M\"uller, F. Gerbier, and I.
  Bloch, Phys.\ Rev.\ Lett.\ 97, 060403 (2006).
  
\bibitem{SPP-07} I.B. Spielman, W.D. Phillips, and J.V. Porto, Phys.\ Rev.\ 
  Lett.\ 98, 080404 (2007); Phys.\ Rev.\ Lett.\ 100, 120402 (2008).

\bibitem{CFFFI-09} D. Cl\'ement, N. Fabbri, L. Fallani, C. Fort, 
and M. Inguscio,
  Phys. Rev. Lett. 102, 155301 (2009).

\bibitem{MCA-10}
O. Morsch, D. Ciampini, and E. Arimondo,
Europhysicsnews 41/3, 21 (2010).

\bibitem{Greiner-etal-02} M. Greiner, O. Mandel,
T.W. H\"ansch, and I. Bloch, 
Nature 419, 51 (2002).

\bibitem{KWW-06} T. Kinoshita, T. Wenger, and D.S. Weiss, 
Nature 440, 900 (2006).

\bibitem{Sadler-etal-06} L.E. Sadler, M. Higbie, S.R. Leslie, M. Vengalattore,
  and D.M. Stamper-Kurn, Nature 443, 312 (2006).

\bibitem{KWW-04} T. Kinoshita, T. Wenger, and D.S. Weiss, Science 305, 1125
  (2004); Phys.\ Rev.\ Lett.\ 95, 190406 (2005).

\bibitem{SMSKE-04} T. St\"oferle, H. Moritz, C. Schori, M. K\"ohl, and T.
  Esslinger, Phys.\ Rev.\ Lett.\ 92, 130403 (2004).
      
\bibitem{PWMMFCSHB-04} B. Paredes, A. Widera, V. Murg, O. Mandel, S.
F\"olling, I. Cirac, G. Shlyapnikov, R.W. H\"ansch, and I. Bloch, Nature
429, 277 (2004).


\bibitem{THHPRP-04}
B. Laburthe Tolra, K.M. O'Hara, J.H. Huckans, W.D. Phillips,
S.L. Rolston, and J.V. Porto,
Phys. Rev. Lett. 92, 190401 (2004).


\bibitem{HLFSS-07}
S. Hofferberth, I. Lesanovsky, B. Fischer, T. Schumm, and J. Schmiedmayer,
Nature 449, 324 (2007).

\bibitem{JBCGZ-98} D. Jaksch, C. Bruder, J.I. Cirac, C.W. Gardiner, and P.
  Zoller, Phys.\ Rev.\ Lett.\ 81, 3108 (1998).

\bibitem{FWGF-89} M.P.A. Fisher, P.B. Weichman, G. Grinstein, and D.S.
  Fisher, Phys.\ Rev.\ B 40, 546 (1989).
  
\bibitem{BSSD-04}
V. Bretin, S. Stock, Y. Seurin, and J. Dalibard,
Phys. Rev. Lett. 92, 050403 (2004).

\bibitem{Sachdev-book}
S. Sachdev, {\em Quantum Phase Transitions}
(Cambridge Univ.\ Press, 1999).

\bibitem{RM-04} M. Rigol and A. Muramatsu, Phys.\ Rev.\ A 70, 031603 (2004);
  Phys.\ Rev.\ A 72, 013604 (2005).

\bibitem{CV-10} M. Campostrini and E. Vicari, Phys.\ Rev.\ A 81, 023606
  (2010); J. Stat. Mech.: Theory Exp.  P08020 (2010).

\bibitem{PGS-04}
D.S. Petrov, D.M. Gangardt, and G.V. Shlyapnikov,
J. Phys. IV France 116, 3-44 (2004).

\bibitem{LL-63} E.H. Lieb and W. Liniger, Phys. Rev. 130, 1605 (1963); E.H.
  Lieb, Phys. Rev. 130, 1616 (1963).

\bibitem{Girardeau-60}
M. Girardeau, J. Math. Phys. (N.Y.) 1, 516 (1960)

\bibitem{Girardeau-65} M.D. Girardeau, Phys. Rev. 139, B500 (1965).

\bibitem{PSW-00} D.S. Petrov, G.V. Shlyapnikov, and J.T.M. Walraven, Phys.
  Rev. Lett. 85, 3745 (2000).

\bibitem{KSS-96} Yu. Kagan, E.L. Surkov, and G.V. Shlyapnikov, Phys. Rev. A
  54, R1753 (1996).

\bibitem{GW-00} M.D. Girardeau and E.M. Wright, Phys. Rev. Lett. 84, 5691
  (2000).

\bibitem{KNSX-00}
E.B. Kolomeisky, T.J. Newman, J.P. Straley, and Xiaoya Qi,
Phys. Rev. Lett. 85, 1146 (2000).  


\bibitem{DLO-01}
V. Dunjko, V. Lorent, and M. Olshanii, Phys. Rev. Lett. 86, 5413 (2001).

\bibitem{GWT-01}
M.D. Girardeau, E.M. Wright, and J.M. Triscari,
Phys. Rev. A 63, 033601 (2001).

\bibitem{LGW-02}
G.J. Lapeyre, M.D. Girardeau, and E.M. Wright,
Phys. Rev. A 66, 23606 (2002).


\bibitem{BRSRMDT-02} G.G. Batrouni, V. Rousseau, R.T. Scalettar, M. Rigol, A.
  Muramatsu, P.J.H. Denteneer, and M. Troyer, Phys.\ Rev.\ Lett.\ 89, 117203
  (2002).

\bibitem{MS-02}
C. Menotti, S. Stringari, Phys. Rev. A 66, 043610 (2002).

\bibitem{PSG-02}
A. Polkovnikov, S. Sachdev, and S.M. Girvin, Phys. Rev. A 66, 053607 (2002).

\bibitem{FFGW-02} P.J. Forrester, N.E. Frankel, T.M. Garoni, and N.S. Witte,
Commun. Math. Phys. 238, 257 (2003); 
Phys. Rev. A 67, 043607 (2003).

\bibitem{KB-02} F. Kalish and D. Braak, J. Phys. A 35, 9957 (2002).

\bibitem{Papenbrock-03}
T. Papenbrock, Phys. Rev. A 67, 041601 (2003).

\bibitem{FFGW-03}
P.J. Forrester, N.E. Frankel, T.M. Garoni, and N.S. Witte,
Phys. Rev. A 67, 043607 (2003).

\bibitem{GS-03}
D.M. Gangardt and G.V. Shlyapnikov, Phys. Rev. Lett. 90, 010401 (2003).

\bibitem{Gangardt-04}
D.M. Gangardt, J. Phys. A 37, 9335 (2004).

\bibitem{KSDZ-04} C. Kollath, U. Schollw\"ock, J. von Delft, 
and W. Zwerger, Phys.\ 
  Rev.\ A 69, 031601 (2004).

\bibitem{PRHD-04} L. Pollet, S. Rombouts, K. Heyde, and J. Dukelsky, Phys.\ 
  Rev.\ A 69, 043601 (2004).

\bibitem{WATB-04} S. Wessel, F. Alet, M. Troyer, and G.G. Batrouni, Phys.\ 
  Rev.\ A 70, 053615 (2004).



\bibitem{MG-05} A. Minguzzi and D.M. Gangardt, Phys. Rev. Lett. 94, 240404
  (2005).

\bibitem{GFF-05} T.M. Garoni, P.J. Forrester, and N.E. Frankel,
J. Math. Phys. 46, 103301 (2005). 

\bibitem{CDEO-08} M. Cramer, C.M. Dawson, J. Eisert, and T.J. Osborne, Phys.
  Rev. Lett. 100, 030602 (2008).


\bibitem{PG-08} A. Polkovnikov and V. Gritsev, Nature Phys. 4, 477 (2008).

\bibitem{BKMRS-08} G.G. Batrouni, H.R. Krishnamurthy, K.W. Mahmud, V.G.
  Rousseau, and R.T. Scalettar, Phys. Rev. A 78, 023627 (2008).
  
\bibitem{RBRS-09} M. Rigol, G.G. Batrouni, V.G. Rousseau and R.T. Scalettar,
  Phys.\ Rev.\ A 79, 053605 (2009).


\bibitem{CV-10-2} M. Campostrini and E. Vicari, Phys.\ Rev.\ A 81, 063614
  (2010).

\bibitem{CK-10} M. Collura and D. Karevski, Phys. Rev. Lett. 104, 200601
  (2010).

\bibitem{Roux-10} G. Roux, Phys. Rev. A 81, 053604 (2010).

\bibitem{PPS-10}
L. Pollet, N.V. Prokof'ev, and B.V. Svistunov,
Phys. Rev. Lett. 104, 245705 (2010).

\bibitem{CV-09} M. Campostrini and E. Vicari, Phys.\ Rev.\ Lett.\ 102, 240601
  (2009).

\bibitem{book-numerov} E. Hairer, S.P. Norsett, and G. Wanner, {\em
Solving ordinary differential equations I: Nonstiff problems} (Berlin,
New York: Springer-Verlag, 1993).

\bibitem{Landau-book} L.D. Landau and E.M. Lifshitz, {\em Quantum
Mechanics: Non-Relativistic Theory}, Vol. 3 (3rd ed.)  (Pergamon Press
1977).

\bibitem{MVT-02}
A. Minguzzi, P. Vignolo, and M.P. Tosi, Phys. Lett. A 294, 222 (2002).

\bibitem{OD-03} M. Olshanii and V. Dunjko, Phys. Rev. Lett. 91, 090401 (2003).

\bibitem{PV-02} A. Pelissetto and E. Vicari, Phys. Rep. 368, 549 (2002).

\bibitem{Wilson-82} K.G. Wilson, in {\em Nobel Lectures in Physics 1981-1990}, 
G.~Ekspong Ed., World Scientific Publ., Singapore, 1993.

\bibitem{JSS-89}
H.K. Janssen, B. Schaub and B. Schmittmann, 
Z. Phys. B 73, 539 (1989).

\bibitem{CG-05} P. Calabrese and A. Gambassi, J. Phys. A 38, R133 (2005).

\bibitem{Schiff-book} L.I. Schiff, {\em Quantum Mechanics}, McGraw-Hill
  Kogakusha, LTD (1977).

\bibitem{DP-10} C. De Grandi and A. Polkovnikov,
 {\em Quantum Quenching, Annealing
  and Computation}, Eds. A. Das, A. Chandra and B. K. Chakrabarti, Lect. Notes
  in Phys., vol. 802 (Springer, Heidelberg 2010)

\bibitem{PP-70} V.S. Popov and A.M. Perelomov, Sov. Phys. JETP 30, 910 (1970);
A.M. Perelomov and Y.B. Zel'dovich,
{\em Quantum mechanics} (World Scientific, Singapore, 1998).

\bibitem{OS-02}
P. \"Ohberg, L. Santos,
Phys. Rev. Lett. 89, 240402 (2002).

\bibitem{CM-06}
A. del Campo and J.G. Muga,
Europhys. Lett. 74, 965 (2006).

\bibitem{RBSMJ-10}
V.G. Rousseau, G.G. Batrouni, D.E. Sheehy, J. Moreno, and M. Jarrell,
Phys. Rev. Lett. 104, 167201 (2010).



\bibitem{Forrester-93} P.J. Forrester, Nucl. Phys. B 402, 709 (1993).  

\end{thebibliography}
\end{document}